\documentclass[12pt]{article}
\usepackage{amsmath,amsfonts}
\usepackage[nosort]{cite}
\usepackage{graphicx}
\input epsf

\makeatletter\@addtoreset{equation}{section}\makeatother

\def\bR {\mathbb{R}}

\def\bI {\mathbb{I}}

\newcommand{\be}{\begin{equation}}
\newcommand{\ee}{\end{equation}}
\newcommand{\bea}{\begin{eqnarray}}
\newcommand{\eea}{\end{eqnarray}}
\newcommand{\half}{\frac{1}{2}}
\newcommand{\vev}[1]{{\left< {#1} \right>}}

\newcommand{\Tr}{{\rm Tr\,}}

\newcommand{\cA}{{\mathcal A}}
\newcommand{\cC}{{\mathcal C}}

\newcommand{\cJ}{{\mathcal J}}

\newcommand{\cL}{{\mathcal L}}
\newcommand{\cN}{{\mathcal N}}
\newcommand{\cP}{{\mathcal P}}
\newcommand{\cQ}{{\mathcal Q}}
\newcommand{\cS}{{\mathcal S}}

\renewcommand{\title}[1]{\vbox{\center\LARGE{#1}}\vspace{5mm}}
\renewcommand{\author}[1]{\vbox{\center#1}\vspace{5mm}}
\newcommand{\address}[1]{\vbox{\center\em#1}}
\newcommand{\email}[1]{\vbox{\center\tt#1}\vspace{5mm}}
\begin{document}
\begin{titlepage}
\begin{center}
\hfill {\tt HU-EP-07/58}\\
\hfill{\tt Imperial-TP-RR-05/2007}

\title{Supersymmetric Wilson loops on $S^3$}

\author{Nadav Drukker$^{1,a}$,
Simone Giombi$^{2,b}$,
Riccardo Ricci$^{3,4,c}$,
Diego Trancanelli$^{5,d}$}

\address{$^1$Humboldt-Universit\"at zu Berlin, Institut f\"ur Physik,\\
Newtonstra{\ss}e 15, D-12489 Berlin, Germany\\
\medskip
$^2$Jefferson Physical Laboratory, Harvard University, Cambridge, MA 02138, USA \\
\medskip
$^3$Theoretical Physics Group, Blackett Laboratory,\\
Imperial College, London, SW7 2AZ, U.K.\\
\medskip
$^4$The Institute for Mathematical Sciences, \\
Imperial College, London, SW7 2PG, U.K.\\
\medskip
$^5$Department of Physics, University of California,\\
Santa Barbara, CA 93106-9530, USA}

\email{$^a$drukker@physik.hu-berlin.de,
$^b$giombi@physics.harvard.edu,
$^c$r.ricci@imperial.ac.uk,
$^d$dtrancan@physics.ucsb.edu}

\end{center}

\abstract{ \noindent This paper studies in great detail a family
of supersymmetric Wilson loop operators in $\cN=4$ supersymmetric
Yang-Mills theory we have recently found. For a generic curve on
an $S^3$ in space-time the loops preserve two supercharges but we will
 also study special cases which preserve 4, 8 and 16 supercharges.
 For certain loops we find the string
theory dual explicitly and for the general case we show that
string solutions satisfy a first order differential equation. This
equation expresses the fact that the strings are
pseudo-holomorphic with respect to a novel almost complex
structure we construct on $AdS_4\times S^2$. We then discuss loops
restricted to $S^2$ and provide evidence that they can be
calculated in terms of similar observables in purely bosonic YM in
two dimensions on the sphere. }

\vfill

\end{titlepage}

{\addtolength{\parskip}{-1ex}
\tableofcontents
}

\section{Introduction}
\label{Introduction}
The $AdS$/CFT correspondence \cite{Maldacena:1997re,Gubser:1998bc,Witten:1998qj}
relates $\cN=4$ supersymmetric Yang-Mills (SYM) theory in four
dimensions and string theory on $AdS_5\times S^5$. One
calculates quantities at weak coupling using the gauge theory description
and at strong coupling using string theory techniques, but usually the
ranges of validity of the two calculations do not overlap and one
cannot compare the perturbative results with those derived from string
theory.

Some notable exceptions to this last statement do exist. For
example the Bethe-ansatz techniques for calculating the anomalous
dimensions of local operators have allowed to interpolate from
weak to strong coupling. One particularly striking example are the
recent results on the cusp anomalous dimension
\cite{Beisert:2006ez,Bern:2006ew,Roiban:2007jf,Benna:2006nd,Alday:2007qf,
Basso:2007wd}. An older example of such an interpolation is the
circular Wilson loop operator, whose expectation value calculated
from the gauge theory point of view seems to be captured by a
matrix model \cite{Erickson:2000af,Drukker:2000rr}. These results
agree with string calculations including an infinite series of
corrections in $1/N$
\cite{Drukker:2005kx,Drukker:2006zk,Yamaguchi:2006tq}\footnote{For
 these probe brane computations see also \cite{Gomis:2006sb,Okuyama:2006jc,Hartnoll:2006is,Chen:2006iu,Giombi:2006de}, while fully  back-reacted geometries dual to Wilson loops are studied in \cite{Yamaguchi:2006te,Lunin:2006xr,D'Hoker:2007fq,Okuda:2007kh}.}
as well as some proposed string calculations
valid to all orders in $1/\sqrt{g_{YM}^2N}$ \cite{Drukker:2006ga}.

Finding such examples is a subtle art-form, and one has to
progress by tiny incremental steps from trivial quantities to more
complicated ones. For the spectrum of local operators the starting
point were long supersymmetric operators and their small
excitations \cite{Berenstein:2002jq}. Later it was understood that
this problem is related to the existence of certain integrable
spin-chains \cite{Minahan:2002ve}. Bethe-ansatz techniques to
calculate the spectrum to all loop order in perturbation theory
were then developed and their predictions  matched to the
computation of quantum corrections to the semiclassical string
result, see
\cite{Beisert:2003tq,Beisert:2004ry,Beisert:2005fw,string1,string2,string3}
and references therein.

While the understanding of Wilson loops is much more fractured, the
cases that are understood have again been obtained by starting with
simple examples and generalizing on them. In the case of the circular
loop, it can be related by a conformal transformation to the trivial
straight line, where the difference between them is due to a subtle
change in the global properties of the loop. Then if one considers two
local deformations of the line or circle they can be analyzed again
using spin-chain techniques \cite{Drukker:2006xg}. Another family of Wilson loops that is well understood was
constructed by Zarembo \cite{Zarembo:2002an}, and can also be
considered as a generalization of the straight line. Like the line,
these loops have trivial expectation values, and we will review them
shortly.

In this paper we elaborate on the family of supersymmetric Wilson loops
introduced in \cite{Drukker:2007dw,Drukker:2007yx} and on some
techniques we can use to compute their
expectation values. These loops are similar to the ones constructed by
Zarembo, but their expectation values, in general, are complicated
functions of $g_{YM}$ and $N$. Instead of generalizing on the straight
line, they may be viewed as generalizations of the circle. As we will
show, despite their complexity, in many cases there are natural guesses
for what these
functions are. We do not have yet the full solution for all the
loops in this class, but we are optimistic that these loops reside
precisely in that regime where exact calculations are within reach
of current technology. It is also our hope that this construction
will lead to further developments that will allow to calculate
more Wilson loop operators and derive more exact results in the
$AdS$/CFT correspondence.

As further motivation for the study of Wilson loops we would like
to mention that there are some interesting connections between
local operators and Wilson loops. One example is the relation
between the cusp anomaly of a light-like Wilson line and the
anomalous dimension of large spin twist-2 operators
\cite{grossgeorgi,korchemsky1,korchemsky2,GKP,Kruczenski,makeenko,frolovlog,kruczenski2,roibanstrong,alday}.
Quite remarkably light-like Wilson loops with cusps have also been
conjectured by Alday and Maldacena to compute gluon scattering
amplitudes \cite{Alday:2007hr}.

In the rest of the introduction we will review the construction of
our Wilson loop operators and provide more details on the proof
that they are supersymmetric.

In Section~\ref{example-sec} we will go over some specific examples
of families of operators with enhanced supersymmetry. The most
general case in our class will preserve two supercharges, but we will
show some cases with four, eight and sixteen unbroken supersymmetries.
Some of the information there has already been anticipated in
\cite{Drukker:2007dw}, but we go over it in much more detail and include
many new results.

Section~\ref{J-section} contains the basic characterization of the
string duals of our Wilson loops. Beyond the standard claim that
they should be described by semi-classical string solutions, we
find a first-order differential equation satisfied by the strings.
This equation is derived by considering a novel almost complex
structure on an $AdS_4\times S^2$ subspace of $AdS_5\times S^5$.
Requiring that the strings are pseudo-holomorphic with respect to
this almost complex structure leads to the correct boundary
conditions on the strings  and to preservation of the expected
supersymmetry. The string world-sheets will be interpreted as
calibrated surfaces and their expectation values computed in terms
of the integral of the calibration form on the world-sheet. The
results in this section have not been published before.

In Section \ref{S2-section} we discuss Wilson loops restricted to an $S^2$
subspace of space-time and provide some evidence, both from the gauge
theory and from string theory, that those loops can be evaluated by a
perturbative prescription for two-dimensional bosonic YM expanding
on \cite{Drukker:2007yx}.

We complete the paper with a series of appendices.
In Appendix~\ref{algebra_conv} we collect our conventions for the
superconformal algebra while in Appendix~\ref{supergroups} we
provide all the details for the computation of the various supergroups
preserved by the loops introduced in Section~\ref{example-sec}.
Appendix~\ref{solutions-appendix} is dedicated to obtaining the
explicit string surfaces in $AdS_5\times S^5$ corresponding to
some of the loops presented in the text.
In Appendix~\ref{S6app} we review the construction of the almost
complex structure for $S^2$ and $S^6$ as a warm-up for the discussion of
the almost complex structure relevant to our loops presented in
Section~\ref{acs section}.
Finally, in Appendix~\ref{WML-appendix} we present a sample
computation in the two-dimensional Yang-Mills theory for our
loops restricted to an $S^2$.

\subsection{The loops}
\label{the loops}
The gauge multiplet of $\cN=4$ SYM includes all fields
in the theory: One gauge field, six real scalars and four complex spinor fields and it is then
natural to incorporate them into the Wilson loop operator. We
will consider the extra coupling of the scalars
$\Phi^I$ (with $I=1,\cdots,6$) so the Wilson loop is
\cite{Rey,Maldacena-wl}
\begin{equation}
W=\frac{1}{N}\,\Tr\,\cP\exp \oint dt
(iA_\mu\dot x^\mu(t) + |\dot x|\Theta^I(t)\Phi^I)\,,
\label{Wilson-loop}
\end{equation}
where $x^\mu(t)$ is the path of the loop and $\Theta^I(t)$ are arbitrary
couplings. A necessary requirement for SUSY is that the norm of $\Theta^I$ be
one. But that alone leads only to ``local'' supersymmetry. If one considers
the supersymmetry variation of the loop, then at every point along the loop
one finds another condition for preserved supersymmetry. Only if all those
conditions commute, will the loop be globally supersymmetric.

A simple way to satisfy this is if at every point one finds the same equation.
This happens in the case of the straight line, where $\dot x^\mu$ is a
constant vector and one takes also $\Theta^I$ to be a constant. This
idea was generalized in a very ingenious way by Zarembo
\cite{Zarembo:2002an}, who assigned for every tangent vector in
$\bR^4$ a unit vector in $\bR^6$ by a $6\times 4$ matrix $M^I{}_\mu$
and took $|\dot x|\Theta^I=M^I{}_\mu\dot x^\mu$. That construction
guarantees that if a curve is contained within a one-dimensional linear
subspace of $\bR^4$ it preserves half of the super-Poincar\'e symmetries
generated by $Q$ and $\bar Q$ (see the notations in
Appendix~\ref{algebra_conv}).
Inside a 2-plane it will preserve $1/4$,
inside $\bR^3$ $1/8$ of them, and for a generic curve $1/16$.
In special cases the loops might also preserve some of the superconformal
symmetries, generated by $S$ and $\bar S$. We will refer to these
loops often throughout the paper and call them ``$Q$-invariant loops''.

An amazing
fact about those loops is that their expectation values seem to be trivial,
with evidence both from perturbation theory, from $AdS$ and from
a topological argument \cite{Zarembo:2002an,Guralnik:2003di,
Guralnik:2004yc,Dymarsky:2006ve,Kapustin:2006pk}.
This construction can be associated to a topological twist of $\cN=4$
SYM, where one
identifies an $SO(4)$ subgroup of the $SO(6)$ $R$-symmetry group with
the Euclidean Lorentz group. Under this twist
four of the scalars become a space-time vector
$\Phi_\mu\equiv M^I{}_\mu\Phi_I$ and in the Wilson loop we use
a modified connection $A_\mu\to A_\mu+i\Phi_\mu$.

The construction we will discuss in the rest of this
paper is quite similar to this, but the expectation value of the Wilson
loops will in general be non-trivial.
A simple way to motivate our construction is by considering a different
twist, where {\em three} of the scalars are transformed into a self-dual
tensor
\begin{equation}
\Phi_{\mu\nu}=\sigma^i_{\mu\nu}M^i{}_I\Phi^I\,,
\label{phi-twist}
\end{equation}
and the Wilson loop will involve the modified connection
\begin{equation}
A_\mu\to A_\mu+i\Phi_{\mu\nu}x^\nu\,.
\end{equation}

The important ingredient in this construction are the tensors
$\sigma^i_{\mu\nu}$. They can be defined by the decomposition of the
Lorentz generators in the anti-chiral spinor representation
($\gamma_{\mu\nu}$) into Pauli matrices $\tau_i$
\begin{equation}
\frac{1}{2}(1-\gamma^5)\gamma_{\mu\nu}
=i\sigma^i_{\mu\nu}\tau_i\,,
\label{gamma-sigma}
\end{equation}
where we included the projector on the anti-chiral representation
($\gamma^5=-\gamma^1\gamma^2\gamma^3\gamma^4$).
The matrix $M^i{}_I$ appearing in (\ref{phi-twist}) is $3\times6$
dimensional and is norm preserving, {\it i.e.}
$M M^{\top}$ is the $3\times 3$ unit matrix.
When we need an explicit choice of $M$ we take
$M^1{}_1=M^2{}_2=M^3{}_3=1$ and all other entries zero.

These $\sigma$'s are also essentially the same as 't Hooft's $\eta$
symbols used in writing down the instanton solution, which is not
surprising, since there the gauge field is self-dual. Finally another
realization of them is in terms of the invariant one-forms on $S^3$
\begin{equation}
\begin{aligned}
\sigma_1^{R,L}& = 2 \left[\pm(x^2 dx^3-x^3 dx^2) +
(x^4 dx^1-x^1 dx^4) \right] \\
\sigma_2^{R,L} &= 2 \left[\pm(x^3 dx^1-x^1 dx^3) +
(x^4 dx^2-x^2 dx^4) \right] \\
\sigma_3^{R,L} &= 2 \left[\pm(x^1 dx^2-x^2 dx^1) +
(x^4 dx^3-x^3 dx^4) \right],
\label{one-forms}
\end{aligned}
\end{equation}
where $\sigma_i^R$ are the right (or left-invariant) one-forms and
$\sigma_i^L$ are the left (or right-invariant) one-forms
(adhering to the conventions of \cite{Gauntlett:1995fu}). We chose
our construction to rely on the right-forms (and the anti-chiral spinors)
so
\begin{equation}
\sigma_i^R=2\sigma^i_{\mu\nu}x^\mu dx^\nu\,.
\label{one-forms-decompose}
\end{equation}
These two realizations of $\sigma^i_{\mu\nu}$ will be important in our
exposition. The relation to the spinor representation of the Lorentz group
will be crucial for the proof of supersymmetry and the relation to
the one-forms on $S^3$ will be important for the geometric understanding
and classifications of our loops.

The Wilson loops we study in this paper can then be written in the following
two ways, first in form notation and then explicitly%
\footnote{It is tempting to couple the three remaining scalars
$\Phi^4$, $\Phi^5$ and $\Phi^6$ with the left-forms $\sigma_i^L$,
however this in general does not yield a supersymmetric loop.}
\begin{equation}
W=\frac{1}{N}\Tr\,\cP\exp \oint
\left( i A + \frac{1}{2} \sigma_i^R M^i{}_I\Phi^I \right)
=\frac{1}{N}\Tr\,\cP\exp \oint dx^\mu
\left( i A_\mu -\sigma^i_{\mu\nu}x^\nu M^i{}_I\Phi^I \right).
\label{susy-loop}
\end{equation}
One can of course also package the last expression in terms of the
modified connection $A_\mu+i\Phi_{\mu\nu}x^\nu$.

Note that this construction involves introducing a length-scale,
which can be seen by the fact that the tensor (\ref{phi-twist})
has mass dimension one instead of two. So this construction would
seem to make sense only when we fix the scale of the Wilson loop.
Indeed the operator (\ref{susy-loop}) will be supersymmetric only
if we restrict the loop to be on a three dimensional sphere. This
sphere may be embedded in $\bR^4$, or be a fixed-time slice of
$S^3\times\bR$. We will always take it to be of unit radius, but
it is simple to generalize to other radii by putting the radius
factors where they are required by dimensionality.

\subsection{Supersymmetry}
\label{susy-subsection}
We can now show that our ansatz (\ref{susy-loop}) leads to a
supersymmetric Wilson loop. The supersymmetry
variation of the Wilson loop will be proportional to
\begin{equation}
\label{deltaW}
\delta W \simeq \left(i\dot x^\mu\gamma_\mu
-\sigma^i_{\mu\nu}\dot x^\mu x^\nu M^i{}_I\rho^I\gamma^5
\right) \epsilon(x)\,,
\end{equation}
where $\gamma_\mu$ and $\rho^I$ are respectively the gamma
matrices of $SO(4)$ and $SO(6)$, the Poincar\'e and $R$-symmetry
groups and they are taken to commute with each-other. Note that later in
Section~\ref{acs section}, where we discuss the strings in
$AdS_5\times S^5$ that describe our loops, we will use
10-dimensional notations, where all gamma matrices anti-commute.
This is achieved by the simple replacement
$\rho^I\gamma^5\to\rho^I$. In (\ref{deltaW}) $\epsilon(x)$ is a
conformal-Killing spinor given in $\bR^4$ by two arbitrary constant
16-component Majorana-Weyl spinors as
\begin{equation}
\label{conformal spinor}
\epsilon(x)=\epsilon_0 +
x^\mu\gamma_\mu\epsilon_1\,.
\end{equation}
$\epsilon_0$ is related to the Poincar\'e supersymmetries while
$\epsilon_1$ is related to the super-conformal ones.

To simplify the expressions we eliminate the matrix $M$ so there
is an implicit choice of three scalars (using the index $i=1,2,3$). Then,
using the fact that
$x^\mu x^\mu=1$, we rearrange the variation of the loop as
\begin{equation}
\delta W \simeq i\dot x^\mu x^\nu\left(\gamma_{\mu\nu}\epsilon_1
+i\sigma^i_{\mu\nu}\rho^i\gamma^5\epsilon_0\right)
-i\dot x^\mu x^\nu x^\eta\gamma_\eta\left(\gamma_{\mu\nu}\epsilon_0
+i\sigma^i_{\mu\nu}\rho^i\gamma^5\epsilon_1\right).
\label{susy-variation}
\end{equation}
Requiring that this variation vanishes for arbitrary curves on $S^3$ leads to the
two equations
\begin{equation}
\label{susy-fieldtheory}
\begin{aligned}
\gamma_{\mu\nu}\epsilon_1
+i\sigma^i_{\mu\nu}\rho^i\gamma^5\epsilon_0&=0\,,\\
\gamma_{\mu\nu}\epsilon_0
+i\sigma^i_{\mu\nu}\rho^i\gamma^5\epsilon_1&=0\,.
\end{aligned}
\end{equation}
These equations are not hard to solve, since $\sigma^i_{\mu\nu}$ are
related to $\gamma_{\mu\nu}$ in the anti-chiral representation
(\ref{gamma-sigma}). We just need to decompose $\epsilon_0$
and $\epsilon_1$ into their chiral and anti-chiral components
(labeled respectively by a $+$ and $-$ superscript) and impose
\begin{equation}
\tau^i\epsilon^-_1=\rho^i\epsilon^-_0\,,\qquad
\epsilon_1^+=\epsilon_0^+=0\,.
\label{constraints}
\end{equation}

To solve this set of equations
we can eliminate for example $\epsilon_0^-$ from
(\ref{constraints}) to get
\begin{equation}
i\tau_1\epsilon_1^-=-\rho_{23}\epsilon_1^-\,,\qquad
i\tau_2\epsilon_1^-=-\rho_{31}\epsilon_1^-\,,\qquad
i\tau_3\epsilon_1^-=-\rho_{12}\epsilon_1^-\,.
\label{newconstraints}
\end{equation}
This is a set of constraints that are consistent with each other.
However it is easy to see that only two of them are independent
since the commutator of any two give the remaining equation. With
two independent projectors, we are thus left with two independent
components of $\epsilon^-_1$, while $\epsilon^-_0$ depends
on $\epsilon^-_1$. So we conclude that for a generic curve on
$S^3$ the Wilson loop preserves $1/16$ of the original supersymmetries.

For special curves, when there are extra relations between the coordinates
and their derivatives,
there will be more solutions and the Wilson loops will preserve more
supersymmetry. We will demonstrate this in some special cases below.

To explicitly find the two combinations of $\bar Q$ and
$\bar S$ which leave the Wilson loop invariant, notice that in singling out
three of the scalars the $R$-symmetry group $SU(4)$ is broken
down to $SU(2)_A \times SU(2)_B$, where $SU(2)_A$ corresponds
to rotations of $\Phi^1,\Phi^2,\Phi^3$ while $SU(2)_B$ rotates
$\Phi^4,\Phi^5,\Phi^6$. Then we recognize that the operators
appearing in (\ref{newconstraints}) are just the generators of
$SU(2)_R$, the anti-chiral part of the Lorentz group, and the generators
of $SU(2)_A$, and the above equations simply state
that $\epsilon^-_1$ is a singlet of the diagonal sum of $SU(2)_R$
and $SU(2)_A$, while it is a doublet of $SU(2)_B$. More explicitly,
we can always choose a basis in which $\rho^i$ act
as Pauli matrices on the $SU(2)_A$ indices,
such that the equations above become
\begin{equation}
(\tau^R_k + \tau^A_k) \epsilon^-_1 = 0, \qquad k=1,2,3.
\label{spinsum}
\end{equation}
If we split the $SU(4)$ index in $\epsilon^-_1$ as
\begin{equation}
\epsilon_A^{1,\,\dot{\alpha}}=\epsilon_a^{1,\,\dot{\alpha} \dot a},
\end{equation}
where $\dot a$ and $a$ are respectively $SU(2)_A$ and $SU(2)_B$ indices,
then the solution to (\ref{spinsum}) can be written as
\begin{equation}
\epsilon_{1\,a} = \varepsilon_{\dot{\alpha} \dot a}
\epsilon_a^{1, \, \dot{\alpha} \dot a}.
\end{equation}
Using any of the equations in (\ref{constraints}) we can determine
$\epsilon_0$
\begin{equation}
\label{eps0 eps1}
\epsilon^-_0 = \tau^R_3\rho^3 \epsilon^-_1 =
\tau^R_3 \tau^A_3 \epsilon^-_1 = -\epsilon^-_1,
\end{equation}
where in the last equality we used (\ref{spinsum}). Our conclusion is then
that the Wilson loops we introduced preserve the two supercharges
\begin{equation}
\bar\cQ^a= \varepsilon^{\dot{\alpha} \dot a}
\left ( \bar{Q}^a_{\dot{\alpha} \dot a}
- \bar{S}^a_{\dot{\alpha} \dot a} \right).
\label{supercharges}
\end{equation}
Besides these fermionic symmetries, our Wilson loop operators obviously preserve the bosonic symmetry $SU(2)_B$. Using the commutation relations of the superconformal algebra given in (\ref{bigalgebra2}), it is easy to verify that the above supercharges, together with the
$SU(2)_B$ generators $T_{ab}$, form the following superalgebra
\begin{equation}
\begin{aligned}
&\left \{ \bar\cQ^a\,,\bar\cQ^b \right \} = 2 T^{ab}\,, \\
&\left [ T^{ab}\,,\bar\cQ^c \right] = \varepsilon^{ca} \bar\cQ^b-\frac{1}{2}
\varepsilon^{ba} \bar\cQ^c\,, \\
&\left [ T^{ab}\,, T^{cd} \right] = \varepsilon^{ca} T^{bd} + \varepsilon^{db} T^{ac}\,. \\
\end{aligned}
\label{Osp12}
\end{equation}
This is an $OSp(1|2)$ subalgebra of the superconformal group.

\subsection{Topological twisting}
\label{Topological twisting}
As mentioned from the onset, this construction is related to a topological
twisting of $\cN=4$ SYM. The twisting consists of replacing
$SU(2)_R$ with the diagonal sum of $SU(2)_R$ and $SU(2)_A$,
which we can denote as $SU(2)_{R'}$, so that the twisted Lorentz
group is $SU(2)_L \times SU(2)_{R'}$.

This twisting
was first considered in \cite{Yamron:1988qc} and further studied in
\cite{Vafa:1994tf} (it is their case {\it ii)}).
After the twisting the supercharges decompose under
$SU(2)_L \times SU(2)_{R'} \times SU(2)_B$ as
\begin{equation}
(\bf{2},\bf{1},\bf{2},\bf{2}) + (\bf{1},\bf{2},\bf{2},\bf{2})
\rightarrow
(\bf{2},\bf{2},\bf{2}) + (\bf{1},\bf{3},\bf{2}) +
(\bf{1},\bf{1},\bf{2}).
\label{twisting}
\end{equation}
From the above it is clear that the two supercharges ${\bar \cQ}^a$
are in the $(\bf{1},\bf{1},\bf{2})$, and therefore they become scalars
after the twisting. As usual, one would then like to regard them as BRST
charges, and the Wilson loops will be observables in their cohomology.

What is new in our case is that those would-be BRST charges are not made
only out of the Poincar\'e supersymmetries $Q$, but include also the
super-conformal ones $S$. Consequently those $\bar\cQ^a$ do not anti-commute,
but rather they close on the $SU(2)_B$ generators (\ref{Osp12}).
This is not a major obstacle, in the resulting topological theory one
would have to consider invariance under $\bar\cQ^a$ up to $SU(2)_B$
rotations, which is what is done in the framework of equivariant
cohomology.

We will not pursue this direction further here.

\section{Examples}
\label{example-sec}

We will now present some examples of Wilson loop operators with
enhanced supersymmetry which are special cases of our general
construction. Among several new interesting operators, we will be
also able to recover some previously known examples, like the well
studied $1/2$ BPS circular  Wilson loop
\cite{Erickson:2000af,Drukker:2000rr} and the $1/4$ BPS circle of
\cite{Drukker:2006ga}, and even a subclass (those living in a
$\bR^3$ subspace) of the $Q$-invariant Wilson loops
\cite{Zarembo:2002an} will arise in a particular ``flat limit''.
To illustrate the richness of the construction, we will determine
in detail the explicit supersymmetries and various supergroups
preserved by the different examples. The relevant notations and
conventions are given in Appendix~\ref{algebra_conv}, and some
technical details of the calculations are collected in Appendix
\ref{supergroups}. For a comprehensive reference on superalgebras
see for example \cite {Frappat:1996pb}.

\subsection{Great circle}
\label{great_circle}

We can first show that the well known $1/2$ BPS circular Wilson
loop is included in our construction as a special example, this is
simply a great circle on the $S^3$. In fact, it is
easy to see that by our construction a maximal circle will couple
to a single scalar. For example, for a circle in the $(1,2)$ plane
\begin{equation}
x^{\mu}=(\cos t,\sin t,0,0) \label{circle_12}
\end{equation}
the pull-back on the loop of the left-invariant one forms (\ref{one-forms})
appearing in (\ref{susy-loop}) is
\begin{equation}
\sigma_1^R=\sigma_2^R=0\,, \qquad
\frac{1}{2}\,\sigma_3^R=dt\,,
\label{great_sigmas}
\end{equation}
so that the corresponding Wilson loop will couple only to
$\Phi^3$. As a consequence, vanishing of the supersymmetry
variation leads to the single constraint
\begin{equation}
\rho^3\gamma^5 \epsilon_0 = i \gamma_{12} \epsilon_1\,,
\label{half-constraints}
\end{equation}
and therefore the loop preserves $16$ ($8$ chiral and $8$
anti-chiral) combinations of $Q$ and $S$ and is indeed a $1/2$ BPS
operator. Using (\ref{half-constraints}) we may write down the
sixteen supercharges as
\begin{equation}
\cQ^A = i \gamma_{12} Q^A + \left(\rho^3 S\right)^A\,, \qquad
\bar\cQ_A = i \gamma_{12} \bar{Q}_A-\left(\rho^3\bar{S}\right)_A ,
\label{great_supercharges}
\end{equation}
where $A=1, \ldots , 4$ and for simplicity we have omitted Lorentz
indices. Furthermore, it is not difficult to show that the $1/2$
BPS circle also preserves the bosonic group $SL(2,\mathbb{R})
\times SU(2) \times SO(5)$. Here, the $SO(5) \subset SO(6)$
simply follows from the fact that the loop couples to a single
scalar. The remaining symmetries $SL(2,\mathbb{R}) \times SU(2)$
correspond to the subgroup of the conformal group $SO(5,1)$ which
leaves the loop (\ref{circle_12}) invariant. It is not difficult
to see that the $SU(2)$ factor is generated by
\begin{equation}
L_1 \equiv \frac{1}{2} \left ( P_3 - K_3 \right ),\qquad
L_2 \equiv \frac{1}{2} \left ( P_4 - K_4\right ),  \qquad
L_3\equiv J_{34}\,,
\label{SU2}
\end{equation}
where $P_{\mu}$ are translations, $K_{\mu}$ are special
conformal transformations and $J_{\mu\nu}$ are Lorentz generators
which can be realized geometrically as
\begin{equation}
P_{\mu} = -i \partial_{\mu}\,,\qquad
K_{\mu} = -i (x^2\partial_{\mu}-2 x_{\mu} x^{\nu} \partial_{\nu})\,,
\qquad
J_{\mu\nu} = i ( x_{\mu} \partial_{\nu} - x_{\nu}\partial_{\mu})\,.
\end{equation}
Finally, the $SL(2,\mathbb{R})$
symmetry is the M\"oebius group in the $(1,2)$ plane generated by
\begin{equation}
I_1 \equiv \frac{1}{2} \left ( P_1 + K_1 \right ),\qquad
I_2 \equiv \frac{1}{2} \left ( P_2 + K_2\right ), \qquad
I_3\equiv J_{12}\,.
\label{mobius}
\end{equation}
All these bosonic symmetries, together with the above
supercharges, form the supergroup $OSp(4^{\star}|4)$ (for an
explicit calculation of this superalgebra, see for example
\cite{Bianchi:2002gz}). Notice that this is the same supergroup
preserved by the $1/2$ BPS straight line (although the explicit
realization in terms of generators of $PSU(2,2|4)$ is different).
This is of course expected since a straight line and a circle are
related by a conformal transformation (an inversion).

A $1/2$ BPS straight line, being of the class invariant under $Q$,
has trivial expectation value. On the other hand the $1/2$ BPS
circle is non-trivial. In perturbation theory, using the Feynman gauge,
the combined gauge-scalar
propagator between two points along a loop is a non-zero constant, so that
the problem of summing all non-interacting graphs (ladder
diagrams) is captured by the Hermitian Gaussian matrix model
\cite{Erickson:2000af,Drukker:2000rr}
\begin{equation}
\langle W \rangle =
    \frac{1}{\mathcal{Z}}\int
    {\cal D} M\, \frac{1}{N}
    \mbox{Tr} \, e^M\, \exp\left(-\frac{2N}{\lambda}\mbox{Tr}
    M^2\right)\,,
\label{mat_model}
\end{equation}
where $M$ is an $N\times N$ Hermitian matrix and
$\lambda=g^2_{YM}N$ is the 't Hooft coupling. It was checked in
\cite{Erickson:2000af} that interacting graphs do not contribute
to order $\lambda^2$, leading to the conjecture that they may
never do so. A more general argument explaining the appearance of
the matrix model was given in \cite{Drukker:2000rr}, using the
above mentioned fact that the circular loop is related to the
straight line by a conformal transformation. This would naively
imply that both Wilson loops are trivial, however the conformal
transformation is singular, and the difference between the two
operators is localized at the singular point, leading then to a
matrix model. Notice however that this argument does not imply
that the matrix model has to be Gaussian, and it is still an open
problem to prove that (\ref{mat_model}) fully captures the VEV of
the $1/2$ BPS circle. Nonetheless, this conjecture has so far
passed an extensive series of non-trivial tests. For example, the
large $\lambda$, $N$ limit of ($\ref{mat_model}$) can be matched
against the classical action of a string world-sheet in $AdS$, and
certain $1/N$ corrections were also correctly reproduced by
D-branes corresponding to Wilson loops in large representations of
the gauge group
\cite{Drukker:2005kx,Yamaguchi:2006tq,Gomis:2006sb}. A
new possible point of view on the matrix model will be discussed
in Section~\ref{S2-section}, where we will argue that all loops
inside a great $S^2 \subset S^3$ (including in particular the
$1/2$ BPS circle) seem to be related to the analogous observables
in the perturbative sector of two-dimensional Yang-Mills, which
can indeed be exactly solved in terms of the same Gaussian matrix
model.

\subsection{Hopf fibers}
\label{fibers_subsection}

A new interesting system contained in
our general construction can be obtained by using the description
of $S^3$ as an Hopf fibration, namely as a $S^1$ bundle over
$S^2$. Explicitly, one can write the $S^3$ metric as
\begin{equation}
ds^2 = \frac{1}{4} \left( d\theta^2 + \sin\theta^2 d\phi^2 +
(d\psi+\cos\theta \,d\phi)^2 \right), \label{Hopf_metric}
\end{equation}
where the range of the Euler angles is $0 \le \theta \le \pi$, $0
\le \phi \le 2\pi$ and $0 \le \psi \le 4\pi$. The $S^1$ fiber is
parameterized by $\psi$, while the base $S^2$ by $(\theta,\phi)$.
These coordinates are related to the cartesian $x^{\mu}$ by
\begin{equation}
\begin{aligned}
&x^1=-\sin\frac{\theta}{2} \sin\frac{\psi-\phi}{2}\,, \qquad
x^2=\sin\frac{\theta}{2} \cos\frac{\psi-\phi}{2}\,, \\
&x^3=\cos\frac{\theta}{2} \sin\frac{\psi+\phi}{2}\,, \qquad
 \ \ x^4=\cos\frac{\theta}{2} \cos\frac{\psi+\phi}{2}\,.
\end{aligned}
\label{euler-angles}
\end{equation}
Consider now a Wilson loop along a generic fiber. This loop will
sit at constant $(\theta,\phi)$, while $\psi$ varies along the curve.
The fibers are non-intersecting great circles of the $S^3$,
so they will each couple to a single scalar, but the interesting fact is that
all the circles in the same fibration will couple to the same scalar,
in this case $\Phi^3$.
An easy way to check this is to write the left-invariant one forms
(\ref{one-forms}) in terms of the Euler angles
\begin{equation}
\begin{aligned}
\sigma^R_1&=-\sin\psi\,d\theta + \cos\psi \sin\theta\,d\phi\,, \\
\sigma^R_2&=\cos\psi\,d\theta + \sin\psi \sin\theta\,d\phi\,, \\
\sigma^R_3&=d\psi+\cos\theta\,d\phi\,. \label{euler-sigmas}
\end{aligned}
\end{equation}
If $\theta$ and $\phi$ are constant and $\psi(t)=2t$ (with $0 \le
t \le 2\pi$), it follows that along the loop
$\sigma^R_1=\sigma^R_2=0$ and $\frac{1}{2}\sigma^R_3 =dt$, as in
(\ref{great_sigmas}). An equivalent way to express this fact is
that a fiber only follows the vector field $\xi^R_3 =
\partial_\psi$ dual to $\sigma^R_3$. Since it is a great circle,
a single loop like this is $1/2$ BPS and without loss of generality we
can take it, as before, to sit in the $(1,\,2)$ plane ({\it i.e.} $\theta=\pi$).

The new feature we want to consider is when there is more than a single
fiber, with the other one at $(\theta,\,\phi)$. If they are not coincident
then the second one will break some of the symmetry of the single
circle. As we shall show, it will project down to the anti-chiral supercharges
and reduce the bosonic symmetries to $U(1)\times SO(5)$.

But before we get there, it is instructive to see how the symmetries of
the single great-circle act on the other fiber. The three-sphere is
mapped to itself by an $SO(4,1)$ subgroup of the
conformal group generated by the rotations $J_{\mu\nu}$ and by
$\frac{1}{2}(P_{\mu}+K_{\mu})$. We have seen in the previous
subsection (\ref{mobius}) that an $SL(2,\mathbb{R})$ subgroup of
this group, obtained by restricting to $\mu,\nu=1,2$, leaves a circle in the
$(1,2)$ plane invariant. So while it will not
move the first fiber at $\theta=\pi$, this $SL(2,\mathbb{R})$
will act non-trivially on the other fiber.%
\footnote{We thank Lance Dixon for suggesting this.}

To see this explicitly, we write the action of the generators
(\ref{mobius}) in terms of the Euler angles as
\begin{equation}
\begin{aligned}
I_1&=i\,\frac{\sin(\psi-\phi)/2}{\sin\theta/2}
\left(\sin\theta\,\partial_\theta-\cot\frac{\psi-\phi}{2}
(\partial_\phi-\partial_\psi)\right)\,,\\
I_2&=-i\,\frac{\cos(\psi-\phi)/2}{\sin\theta/2}
\left(\sin\theta\,\partial_\theta+\tan\frac{\psi-\phi}{2}
(\partial_\phi-\partial_\psi)\right)\,,\\
I_3&=-i(\partial_\phi-\partial_\psi)\,.
\end{aligned}
\end{equation}
Since all the loops are invariant under $\psi$, we can ignore all the
$\partial_\psi$, and then the three generators act as conformal
transformations on the base.

These symmetries allow us to map any point on the base (excluding
$\theta=\pi$) to any other. Therefore, when considering two fibers
we can take the second one at $\theta=0$, which means that it lies in
the $(3,\,4)$ plane.

With this it is easy to check the supersymmetries preserved by the two
fibers. The first circle imposes the constraint (\ref{half-constraints})
\begin{equation}
\rho^3\gamma^5 \epsilon_0 = i \gamma_{12} \epsilon_1\,,
\end{equation}
and analogously the new one (keeping note of the orientation) will impose
\begin{equation}
\rho^3\gamma^5 \epsilon_0 = -i \gamma_{34} \epsilon_1\,,
\end{equation}
In particular we see that
$\gamma_{12}\epsilon_1=-\gamma_{34}\epsilon_1$,
so $\epsilon^1$ is a negative eigenstate of
$\gamma^5=-\gamma^1\gamma^2\gamma^3\gamma^4$, {\it i.e.} it is
anti-chiral, so the loops preserve half the supersymmetries of a single
circle, or are $1/4$ BPS. By the symmetry argument above this is true for
any other fiber (or more than two fibers), which can also be verified
directly, by a somewhat tedious calculation.

The corresponding supercharges
preserved by the system will be essentially the same as the ones
associated to the $1/2$ BPS maximal circle
(\ref{great_supercharges}), except that we only select the
negative chirality
\begin{equation}
\bar\cQ_A = i \gamma_{12} \bar{Q}_A -\left(\rho^3\bar{S}\right)_A
\,. \label{fibers_supercharges}
\end{equation}
As for the bosonic symmetries, notice that of the
$SL(2,\mathbb{R}) \times SU(2) \times SO(5)$ symmetry of the
single fiber, the only remaining symmetry on the space-time side
that remains is rotations of the $\psi$ angle
\begin{equation}
J^R_3 = \frac{1}{2} \left(J_{12}-J_{34}\right) \,.
\end{equation}
Besides this, we have of course the
$SO(5)$ symmetry following from the fact that the fibers only
couple to one scalar. These bosonic symmetries form together with
the fermionic generators (\ref{fibers_supercharges}) the
supergroup $OSp(2|4)$, whose even part is indeed $SO(2) \times
Sp(4) \simeq U(1) \times SO(5)$. This supergroup can be seen as
the subgroup of $OSp(4^*|4)$ obtained by dropping the positive
chirality charges in (\ref{great_supercharges}). From the point of
view of the algebra, it is also natural to understand why the
symmetries involving $P_{\mu}$ and $K_{\mu}$ are lost for the Hopf
fibers system, as those symmetries arise from commutators of
charges in (\ref{great_supercharges}) with opposite chirality.

The symmetry argument above allowed us in the case of two circles
to move them relative to each-other. In perturbation theory one
finds an even stronger statement, the combined gauge-scalar propagator
between {\em any} two points on {\em any} two fibers is the same
constant as for the single circle.

Consider for example the
propagator between a point $x^{\mu}(t;\theta_0,\phi_0)$ on one
fiber and a point $y^{\mu}(s;\theta_1,\phi_1)$ on a second fiber.
Since both circles only couple to $\Phi^3$, the propagator is
\begin{equation}
\left \langle \Big{(} i\,\dot x^{\mu} A_{\mu}^a(x) + \Phi_3^a(x)
\Big{)} \Big{(} i\,\dot y^{\mu} A_{\mu}^b(y) + \Phi_3^b(y)
    \Big{)} \right \rangle =  \frac{g_{YM}^2}{4\pi^2}
\frac{1-\dot x \cdot \dot y}{(x-y)^2}\,\delta^{ab} =
\frac{g_{YM}^2}{8\pi^2}\,\delta^{ab}\,,
\end{equation}
as can be checked using the explicit parametrization
(\ref{euler-angles}). Thus this system of non-intersecting circles on
$S^3$ is reminiscent of the BPS system
of parallel straight lines in flat space. In that case the lines
do not interact between each other (the propagators vanish) and
the observable is trivial. Here we find that the fibers do
interact, however the ``interaction strength'' is just a constant
independent of the relative distance.

Since the propagator is a constant, all ladder diagrams
contributing to the correlator of several Hopf fibers can be
exactly summed up using the same Gaussian matrix model describing
the $1/2$ BPS circle, but with a different insertion compared to
(\ref{mat_model}). Concretely, for a system made of $k$ fibers,
the ladder diagrams contribution will be equal to
\begin{equation}
\langle W_k \rangle_{\mbox{\scriptsize ladders}} = \left\langle
\left( \frac{1}{N} \mbox{Tr} \, e^M \right)^k
\right\rangle_{\mbox{\scriptsize m.m.}}\,, \label{fibers_matrix}
\end{equation}
where the expectation value on the right hand side is taken in the
Gaussian matrix model as in (\ref{mat_model}). Of course it would
be an interesting non-trivial calculation to also evaluate the
contribution (if any) of diagrams with internal vertices. At large
$N$ the correlator in (\ref{fibers_matrix}) will be the same as
$k$ non-interacting circles and will be reproduced at strong
coupling by $k$ disconnected string surfaces in $AdS$. An
interesting problem, which we will not further pursue here, would
be to study the possible contribution of the connected string
configuration in $AdS$.

\subsection{Great $S^2$}
\label{S2-subsection}

An infinite subfamily of operators which turns out to be very
interesting is obtained by restricting the loop to lie on a great
$S^2$ inside $S^3$. For concreteness, we may define this
two-sphere by the condition $x^4=0$. From the definition of the
invariant one forms one can see that on this maximal $S^2$ the
left and right forms are no longer independent, rather
\begin{equation}
\sigma^L_i=-\sigma^R_i=-2 \varepsilon_{ijk}x^j\,d x^k\,,
\label{S2-oneforms}
\end{equation}
which can also be written as a cross-product.
Then it is not difficult to realize that (\ref{susy-variation})
has more solutions. Using that the left forms are related to the
action of the Lorentz generators on positive chirality spinors
\begin{equation}
dx^{\mu} x^{\nu} \gamma_{\mu\nu} \epsilon^+ =
\frac{i}{2}\, \sigma_i^L\tau_i^L \epsilon^+\,,
\end{equation}
the relation $\sigma^L_i=-\sigma^R_i$ implies that
(\ref{susy-variation}) is solved not only by the antichiral
spinors satisfying
\begin{equation}
\tau_i^R \epsilon_1^- = \rho_i \epsilon_0^-\,,
\end{equation}
but also by positive chirality spinors obeying
\begin{equation}
\tau_i^L\epsilon_1^+=-\rho_i\epsilon^+_0\,.
\end{equation}
Combining the two chiralities, this can be also written as
\begin{equation}
i\gamma_{jk}\epsilon_1 =\varepsilon_{ijk}\rho_i \gamma^5\epsilon_0\,.
\label{S2constraints}
\end{equation}
So, contrary to the general $S^3$ case in (\ref{constraints}), we
see that now the constraints are not chiral and hence the
supersymmetries are doubled. The generic Wilson loop on $S^2$ will
therefore give a $1/8$ BPS operator. One can solve the constraints
in the same way as described in Section~\ref{susy-subsection},
but we will now get two
copies of the solution, one for each chirality. The four
supercharges may be written explicitly as
\begin{equation}
\cQ ^a= (i\tau_2)^{\alpha}_{\ \dot a} \left ( Q^{\dot a
a}_{\alpha} + S^{\dot a a}_{\alpha} \right ), \qquad \bar\cQ^a=
\varepsilon^{\dot{\alpha} \dot a} \left ( \bar{Q}^a_{\dot{\alpha}
\dot a} - \bar{S}^a_{\dot{\alpha} \dot a} \right ).
\label{S2-super}
\end{equation}
The bosonic symmetry is also enlarged compared to the generic
curve on $S^3$. In fact, besides invariance under the group
$SU(2)_B \subset SO(6)$ which rotates $\Phi^4,\Phi^5,\Phi^6$,
there is an extra $U(1)$ symmetry generated by
\begin{equation}
\frac{1}{2} \left ( P_4 - K_4 \right )\,, \label{S2-U1}
\end{equation}
which follows from the fact that the loops satisfy $x^4=0$. The
presence of this extra symmetry may be also understood from the
algebra of the supercharges. In fact, one can see that
anticommuting charges of opposite chirality precisely produces the
$U(1)$ generator (\ref{S2-U1}). In Appendix~\ref{S2-supergroup} we
give a detailed derivation of the algebra generated by these
symmetries and prove that it is a $SU(1|2)$ superalgebra. The even
part of this superalgebra is $U(1) \times SU(2)_B$ and the four
fermionic generators transforming as $\bf{2}^+$ + $\bf{2}^-$ under
the even symmetries can be obtained by defining appropriate linear
combinations of the supercharges (\ref{S2-super}).

A generic smooth curve on $S^2$ exhibits a curious property, whose
precise significance would be interesting to explore in more
depth: The gauge coupling for that curve is given, using vector
notation in $\bR^3$, by $\dot{\vec{x}}$ while from
(\ref{S2-oneforms}) the scalar coupling is the cross-product
$\vec{x}\times \dot{\vec{x}}$. If we take $|\dot{\vec{x}}|=1$,
then $\vec{x}\times \dot{\vec{x}}$ is also a vector on $S^2$ and
we can consider Wilson loops along that path in space-time. The
corresponding scalar coupling will be
\begin{equation}
\left(\vec{x}\times \dot{\vec{x}}\right)\times\left(\vec{x}\times
\ddot{\vec{x}}\right)=-\vec{x}\left(\dot{\vec{x}}\cdot
\vec{x}\times \ddot{\vec{x}}\right)\propto \vec{x}\, .
\label{duality}
\end{equation}
The proportionality constant $\vec{x}\times \ddot{\vec{x}}$ is
non-zero if the curve is nowhere a geodesic ({\it i.e.} it is
never part of a great circle). We see then that for any smooth,
nowhere geodesic curve on $S^2$ there is a dual curve with gauge
and scalar coupling interchanged.%
\footnote{It is possible to extend this to curves with sections that are
geodesic, in the dual loops they will manifest themselves as cusps
(and vice-versa).}
In Section~\ref{JS2-subsection} we comment on the extension
of this map to the dual $AdS_5\times S^5$. Only on the boundary
is it a map between $AdS_5$ and $S^5$, otherwise it mixes the
coordinates in a somewhat more complicated way (see
(\ref{acsreduced}) and the discussion after it).

In the following subsections we discuss some examples of special
loops inside $S^2$ preserving some extra supersymmetries. The case
of the general loops belonging to this class is presented in great
detail in Section~\ref{S2-section}, where we provide evidence that they are
related to Wilson loops in two-dimensional Yang-Mills theory.

\subsubsection{Latitude}
\label{latitude-subsection}

\begin{figure}
\begin{center}
\includegraphics[width=100mm]{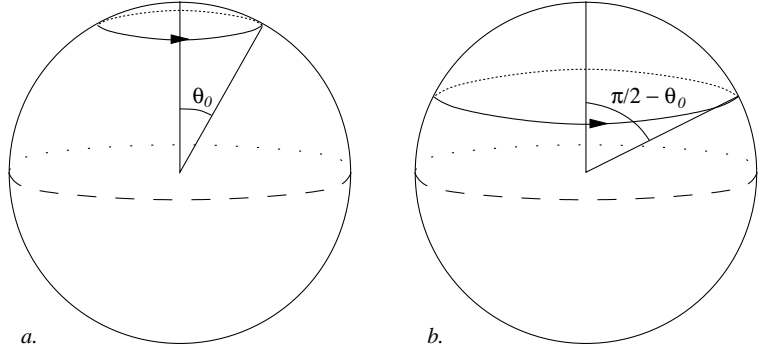}
\parbox{13cm}{
\caption{
Quarter-BPS Wilson loop along a latitude. In a. we show
the Wilson loop
along a latitude at angle $\theta_0$ on an $S^2 \subset \bR^4$.
b. depicts the scalar couplings which follow a dual
latitude on $S^2 \subset S^5$. Notice
that if we took b. to be the path of the loop in space,
then a. would describe the associated scalar
couplings. This is an explicit example of the duality between
scalar and gauge field couplings discussed in the text.
\label{latitude-fig}}}
\end{center}
\end{figure}

Taking the loop to be at the equator of the $S^2$ will clearly
give the $1/2$ BPS circle described in Section~\ref{great_circle}.
More generally we can take the loop to be a non-maximal circle,
{\it i.e.} a latitude of the $S^2$. Concretely, we can parameterize the
loop as
\begin{equation}
x^\mu = (\sin\theta_0 \cos t\, , \sin\theta_0 \sin t\, ,
\cos\theta_0,\,0) \,.
\label{quarter-circle}
\end{equation}
Computing the scalar couplings for this curve according to
(\ref{S2-oneforms})
\begin{equation}
\frac{1}{2}\,\sigma_i^R
= \varepsilon_{ijk} x^j\, dx^k = \sin\theta_0
(-\cos\theta_0 \cos t\, , -\cos\theta_0 \sin t\, , \sin\theta_0)\,dt\,,
\label{scalars_latitude}
\end{equation}
one can see that they also describe a latitude on the $S^2 \subset
S^5$ associated to $\Phi^1$, $\Phi^2$, $\Phi^3$, but the circle
sits at $\pi/2 - \theta_0$, see figure \ref{latitude-fig}. In
particular, when the loop is a maximal circle, $\theta_0 = \pi/2$,
the curve in scalar space reduces to a point (the north pole) and
one falls back to the $1/2$ BPS circle described in
Section~\ref{great_circle}.
This family of loops is essentially the same as the operators considered in
\cite{Drukker:2006ga}: The operator we describe here and the one
in \cite{Drukker:2006ga} are simply related by a conformal
transformation (a dilatation and a translation along $x^3$) which
moves the circle from the equator to a parallel.\footnote{Also
compared to \cite{Drukker:2006ga} $\theta_0$ is replaced here by
$\pi/2-\theta_0$.}

As can be seen from (\ref{scalars_latitude}), such an operator
couples to three scalars, but it can be shown that the
supersymmetry equations will give only two independent
constraints. Indeed, one can see that the supersymmetry variation
vanishes at every point along the loop provided that the following
two conditions are satisfied
\begin{equation}
\begin{gathered}
\cos\theta_0 \big{(}\gamma_{12} + \rho_{12}\big{)}
\epsilon_1 = 0, \\
\rho^3\gamma^5 \epsilon_0 = \big{[} i \gamma_{12} + \gamma_3
\rho^2\gamma^5 \cos\theta_0 (\gamma_{23} + \rho_{23}) \big{]}
\epsilon_1 \,. \label{quarter-constraints}
\end{gathered}
\end{equation}
If $\cos\theta_0 \ne 0$, one has two independent constraints and
the loop preserves $1/4$ of the supersymmetries. In the special
case $\cos\theta_0 = 0$ the first constraint disappears and one
recovers the $1/2$ BPS maximal circle condition
(\ref{half-constraints}).

One may solve the constraints (\ref{quarter-constraints}) as
described in Section~\ref{susy-subsection}
by viewing $\gamma_i$ and $\rho_i$ as Pauli
matrices acting on Lorentz and $SU(2)_A$ indices respectively. In
particular, the first line in (\ref{quarter-constraints}) may be
written as
\begin{equation}
(-i\gamma_{12} + \tau_3^A) \epsilon_1 = 0.
\label{first-constraint}
\end{equation}
For a generic loop we had three such equations (for the
anti-chiral spinor), which meant that the only solution had to be
a singlet of the diagonal $SU(2)_R+SU(2)_A$ group. Here we find
only one such equation for each of the chiralities, such that a
$U(1)$ charge ($\tau_3^\text{total}$) has to vanish. So in
addition to the singlet, this constraint allows one of the states
of the triplet. Explicitly, we can write the two solutions of
(\ref{first-constraint}) as
\begin{equation}
\begin{aligned}
\epsilon_{1,\,a}^{(1)}& = \epsilon_{1,\,\dot 1\,
a}^2-\epsilon_{1,\,\dot 2\,a}^{1}
=(i\tau_2)^{\dot a}_{\ \alpha} \,\epsilon_{1,\, \dot a\,a }^\alpha\\
\epsilon_{1,\,a}^{(2)} &= \epsilon_{1,\,\dot 2\,a}^{1} +
\epsilon_{1,\,\dot 1\,a}^{2} =(\tau_1)^{\dot a}_{\ \alpha} \,
\epsilon_{1,\,\dot a a }^\alpha \,,
\label{ep1-solutions}
\end{aligned}
\end{equation}
and similarly for the other chirality. The $\epsilon_0$ spinors
can be obtained by solving the second line of the constraints. For
the singlet spinor $\epsilon_1^{(1)}$, the term proportional to
$\cos\theta_0$ does not contribute and the solution is the same as
the one for the great $S^2$ loops given in equation (\ref{S2-super}),
that is
\begin{equation}
\cQ_{(1)}^a= (i\tau_2)^{\alpha}_{\ \dot a} \left ( Q^{\dot a
a}_{\alpha} + S^{\dot a a}_{\alpha}\right )\,,\qquad \qquad
\bar\cQ_{(1)}^a= \varepsilon^{\dot \alpha \dot a} \left (
\bar{Q}^a_{\dot \alpha \dot a} - \bar{S}^a_{\dot \alpha \dot
a}\right )\,. \label{latitude-sup1}
\end{equation}
As for the solutions corresponding to $\epsilon_{1,(2)}$, because
of the $\gamma_3$ in the term proportional to $\cos\theta_0$, the
second constraint in (\ref{quarter-constraints}) will relate
$\epsilon_0$ of a given chirality to a combination of
$\epsilon_1$'s of both chiralities. Explicitly one can write the
resulting conserved supercharges as
\begin{equation}
\begin{aligned}
&\cQ_{(2)}^a= \frac{1}{\sin\theta_0}(\tau_3 \varepsilon)^{\dot\alpha
\dot a} \left ( \bar Q^a_{\dot\alpha \dot a} - \bar S^a_{\dot\alpha \dot
a}\right ) +\cot\theta_0 \, (i\tau_2)^{\alpha}_{\ \dot a}
\left(Q_{\alpha}^{\dot a a}- S_{\alpha}^{\dot a a} \right)\,, \\
&\cQ^{\prime\,a}_{(2)}=
\frac{1}{\sin\theta_0}(\tau_1)^{\alpha}_{\ \dot a} \left (
Q^{\dot a a}_{\alpha}+S^{\dot a a}_{\alpha}\right ) + \cot\theta_0
\, \varepsilon^{\dot \alpha \dot a} \left(\bar{Q}_{\dot \alpha
\dot a}^a +\bar{S}_{\dot \alpha \dot a}^a\right) \,.
\label{latitude-sup2}
\end{aligned}
\end{equation}
The bosonic symmetries preserved by this loop turn out to be
$SU(2) \times U(1) \times SU(2)_B$. Besides the obvious $SU(2)_B$
symmetry, the other $SU(2)$ is essentially equivalent to the
$SU(2)$ preserved by the maximal circle (\ref{SU2}), except that
one should conjugate those generators by a dilatation and a
translation along $x^3$ which will move the circle from the
equator to a latitude. The resulting generators are similar to
(\ref{SU2}), but they are $\theta_0$ dependent and now involve
also the dilatation generator $D$. The explicit expressions are
given in Appendix~\ref{latitude-supergroup}, where we present the
detailed calculation of the superalgebra associated to this Wilson
loop. The remaining $U(1)$ symmetry mixes Lorentz and $R$-symmetry
and is given by the combination $J_{12} + J_{12}^A$, where
$J_{12}^A$ is the generator of $SU(2)_A$ rotating $\Phi_1$ and
$\Phi_2$. This follows from the fact that the loop coordinates
$x^{i}$ and the scalar couplings (\ref{one-forms}) satisfy the equation
$x^2\sigma_1^R - x^1\sigma_2^R= 0$. In \ref{latitude-supergroup} we
show that the eight supercharges and these bosonic generators can
be organized to form a $SU(2|2)$ superalgebra.

This example is particularly interesting because it turns out that
in perturbation theory the combined gauge-scalar propagator is
also constant, and it is equal to the one for 1/2 BPS circle with
the simple rescaling $g^2_{YM} \rightarrow g^2_{YM}
\sin^2\theta_0$ \cite{Drukker:2006ga}. This led to the conjecture
that this $1/4$ BPS Wilson loop is also captured by the matrix
model (\ref{mat_model}) with a rescaling of the coupling constant.
The $AdS$ string solution dual to this operator is explicitly
known, as reviewed in Appendix~\ref{latitude-sol-appendix}, and its
classical action perfectly agrees with the strong coupling limit
of the matrix model result. An explicit D3 solution describing the
Wilson loop in a large symmetric representation was also found in
\cite{Drukker:2006zk}, where it was shown again agreement with the
matrix model, including all $1/N$ corrections at large $\lambda$.
More details on these results and the implications for the
conjectured relation of the $S^2$ loops to 2d Yang-Mills are
discussed in Section~\ref{S2-section}.

\subsubsection{Two longitudes}
\label{longitudes-subsection}
\begin{figure}
\begin{center}
\includegraphics[width=100mm]{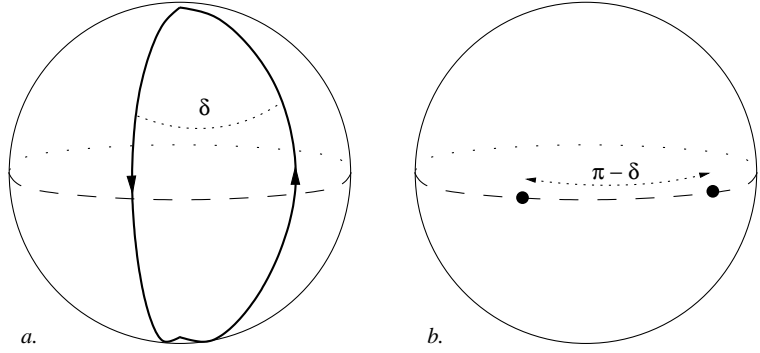}
\parbox{13cm}{
\caption{
Quarter-BPS Wilson loop made of two longitudes. In a. we
show the loop on $S^2 \subset \bR^4$ obtained by taking two half circles,
or longitudes, with opening angle $\delta$. The corresponding scalar
couplings in b. turn out to be two points on the equator of
$S^2 \subset S^5$ separated by an angle $\pi-\delta$.
\label{longi-fig}}}
\end{center}
\end{figure}

A further example of a family of $1/4$ BPS Wilson loops that are also
a special case of loops on a great $S^2$ can be obtained as follows.
Consider a loop made of two arcs of length $\pi$ connected at an
arbitrary angle $\delta$, {\it i.e.} two longitudes on the two-sphere.
We can parameterize the loop in the following way
\begin{equation}
\begin{array}{ll}
x^{\mu}=(\sin t,\, 0,\, \cos t,\, 0)\,,\qquad
& 0 \le t \le \pi\,, \\
x^{\mu}=( -\cos \delta \sin t,\, -\sin \delta \sin t,\,\cos
t,\,0)\,,\qquad &\pi \le t \le 2 \pi\,.
\end{array}
\label{arcs}
\end{equation}
The corresponding Wilson loop operator will couple to $\Phi^2$
along the first arc and to $- \Phi^2 \cos \delta + \Phi^1 \sin
\delta$ along the second one, see figure \ref{longi-fig}. Notice
that such an operator is related by a stereographic projection to a
Wilson loop of the type invariant under $Q$ \cite{Zarembo:2002an}
given by two semi-infinite rays on the plane with an
opening angle $\delta$. Using this observation we were able to
construct the explicit dual string solution for this Wilson loop,
which is presented in Appendix \ref{longitudes-sol-appendix}.

It is straightforward to study the supersymmetry variation of this
operator. Each arc, being (half) a maximal circle, is $1/2$ BPS
and will produce a single constraint
\begin{equation}
\begin{array}{ll}
\text{First arc:}
&\rho^2\gamma^5 \epsilon_0 = i \gamma_{31} \epsilon_1\,,   \\
\text{Second arc:}\quad &(\rho^2\gamma^5 \cos\delta-\rho^1\gamma^5
\sin \delta) \epsilon_0 = i (\gamma_{31} \cos \delta  -\gamma_{23}
\sin \delta) \epsilon_1\,.
\end{array}
\end{equation}
Combining the two equations, we see that the system has to
satisfy, as long as $\sin \delta \ne 0$,
\begin{equation}
\rho^2\gamma^5 \epsilon_0 = i \gamma_{31} \epsilon_1\,,   \qquad
\rho^1\gamma^5 \epsilon_0 = i \gamma_{23} \epsilon_1 \, .
\label{longi-constraints}
\end{equation}
These constraints are of course consistent and therefore the loop
will preserve $1/4$ of the supersymmetries. When $\sin \delta =
0$, the second equation in (\ref{longi-constraints}) disappears
and the loop becomes $1/2$ BPS (in the case $\delta = \pi$, it is
just the maximal circle discussed above, while in the case $\delta
= 0$, the loop is made of two coincident half circles with
opposite orientations). No further supersymmetries will be broken when
one adds more circles or half-circles that all intersect at the
north and south poles.

To solve the above constraints, we can proceed as usual by first
eliminating $\epsilon_0$. This gives the equation
\begin{equation}
(-i\gamma_{12} + \tau_3^A) \epsilon_1 = 0\,,
\end{equation}
which is the same equation encountered for the latitude discussed
in the previous subsection. The two solutions for positive
chirality are given in (\ref{ep1-solutions}) and similarly one can
get the negative chirality ones. From the equation $\rho^2\gamma^5
\epsilon_0 = i \gamma_{31} \epsilon_1$ one can then get the two
solutions for $\epsilon_0$ as
\begin{equation}
\epsilon_{0,(1)} = \gamma^5\epsilon_{1,(1)}\,, \qquad
\epsilon_{0,(2)} = -\gamma^5\epsilon_{1, (2)}\,.
\end{equation}
Thus the eight supercharges which annihilate the Wilson loop made
of two longitudes are
\begin{equation}
\begin{aligned}
&\cQ_{(1)}^a= (i\tau_2)^{\alpha}_{\ \dot a} \left ( Q^{\dot a
a}_{\alpha} + S^{\dot a a}_{\alpha}\right ),\qquad \cQ_{(2)}^a=
(\tau_1)^{\alpha}_{\ \dot a}
\left ( Q^{\dot a a}_{\alpha} - S^{ \dot a a}_{\alpha}\right ),  \\
&\bar\cQ_{(1)}^a= \varepsilon^{\dot \alpha \dot a} \left (
\bar{Q}^a_{\dot \alpha \dot a} - \bar{S}^a_{\dot \alpha \dot
a}\right ),\qquad \quad \, \bar\cQ_{(2)}^a= (\tau_3
\varepsilon)^{\dot \alpha \dot a} \left ( \bar{Q}^a_{\dot \alpha
\dot a} + \bar{S}^a_{\dot \alpha \dot a}\right ).
\end{aligned}
\label{longi-SUSY}
\end{equation}
The loop also preserves the bosonic symmetry group $U(1) \times
U(1) \times SO(4)$. The $SO(4) \subset SO(6)$ factor simply
comes from the fact the this loop only couples to $\Phi_1$ and
$\Phi_2$ so that we are free to rotate
$\Phi^3,\Phi^4,\Phi^5,\Phi^6$. To understand the $U(1)^2$
symmetry, one can look at what are the compatible symmetries of
two circles in the $(1,3)$ and $(2,3)$ planes. Recalling our
discussion of the great circle, one can see that there are two
shared symmetry generators, namely $\frac{1}{2}(P_4-K_4)$ and
$\frac{1}{2}(P_3+K_3)$. These two generators commute and give a
$U(1)^2$ symmetry.%
\footnote{Throughout we studied the symmetries only at the level
of the algebra, so we are not distinguishing between $U(1)$ and
$\bR$.} These bosonic symmetries, together with the eight
supercharges (\ref{longi-SUSY}), form the direct product
superalgebra $SU(1|2) \times SU(1|2)$, as we show in Appendix
\ref{longitudes-supergroup}.

\subsection{Hopf base}
\label{hopf-section}
Consider a curve parameterized by the Euler
angles $\theta$ and $\phi$, which form the base of the
Hopf fibration (\ref{Hopf_metric}). A family of loops with
enhanced supersymmetry can be obtained if along the fibers we
choose
\begin{equation}
\psi(t)=-\int_0^t dt'\,\dot\phi(t')\cos\theta(t')\,,
\label{psi-eqn}
\end{equation}
which guarantees that the pull-back of $\sigma_3^R$ along the loop
vanishes, see
(\ref{euler-sigmas}), so the operator will only couple to $\Phi^1$
and $\Phi^2$. A generic curve of this form will break all the
chiral supersymmetries, and for the anti-chiral ones will
introduce the constraints
\begin{equation}
\rho^2\epsilon_0^- = \tau_2^R \epsilon_1^-\,,   \qquad
\rho^1\epsilon_0^- = \tau_1^R \epsilon_1^- \, .
\end{equation}
This is the anti-chiral part of equation
(\ref{longi-constraints}), and consequently the loop will preserve
the anti-chiral supersymmetries in (\ref{longi-SUSY})
\begin{equation}
\bar\cQ_{(1)}^a= \varepsilon^{\dot \alpha \dot a} \left (
\bar{Q}^a_{\dot \alpha \dot a} - \bar{S}^a_{\dot \alpha \dot
a}\right ),\qquad \bar\cQ_{(2)}^a= (\tau_3 \varepsilon)^{\dot
\alpha \dot a}\ \left ( \bar{Q}^a_{\dot \alpha \dot a} +
\bar{S}^a_{\dot \alpha \dot a}\right )\,. \label{base-susy}
\end{equation}
Therefore such operators are $1/8$ BPS.

The example of the two longitudes is a special case of these loops
where the entire loop is contained within an $S^2$, so in addition
to the four anti-chiral supercharges (\ref{base-susy}), it also
preserves four chiral supercharges. To relate them explicitly,
note that among the Euler angles only $\theta$ varies along the
two arcs of (\ref{arcs}) while $\phi$ and $\psi$ are kept fixed
with $\psi+\phi=\pi$, $\psi+\phi=3\pi$ or $\psi+\phi=5\pi$.

The equation for $\psi$ (\ref{psi-eqn}) leads to an integral condition,
namely that the loop is closed. It can actually be restated in a nice way
as a condition on the area bound by the loop on the base
\begin{equation}
\int d\phi\,d\theta\sin\theta
=\int_0^{2\pi}dt\,\dot\phi(t)\,(1-\cos\theta(t))
=\phi(2\pi)+\psi(2\pi)\,.
\end{equation}
Since $\psi$ has period $4\pi$ and so does $\psi+\phi$, we deduce
from this equation that the area bound by the curve should be
quantized in units on $4\pi$.

The bosonic symmetry preserved by such a loop is just the $SO(4)$
rotating $\Phi^3$, $\Phi^4$, $\Phi^5$ and $\Phi^6$. The
superalgebra will be the same as the one of the Wilson loop made
of two longitudes, but restricted to the antichiral sector.
Defining linear combinations as in (\ref{Osp12sq-comb}), one
obtains the same algebra given in (\ref{Osp12square}), the only
difference being the we should use the negative chirality. It is
easy to see that this is an $OSp(1|2) \times OSp(1|2)$
superalgebra. Notice that a diagonal subgroup of this algebra is
just the $OSp(1|2)$ preserved by all our loops.

\subsubsection{Latitude on the base}
\label{Hopf-latitude}
As mentioned before, the longitudes discussion in
Section~\ref{longitudes-subsection} are also special examples of
loops on the Hopf base.

Beyond this example we found one simple family of loops in this
class to which we have explicit string solutions. They are given by taking a
latitude curve on the Hopf base
\begin{equation}
\phi=kt\,,\qquad \theta=\theta_0\,,\qquad 0\leq t\leq2\pi\,,
\end{equation}
where in general we have allowed a multiply wrapped latitude with
winding $k$. From equation (\ref{psi-eqn}) it follows that $\psi$
is also linear in $t$
\begin{equation}
\psi=-kt\cos\theta_0\,.
\end{equation}
The periodicity of $\psi$ implies that $k\cos\theta_0$ should be
an integer such that the area above the loop on the base is a
multiple of $4\pi$.

Let us take $k=k_1+k_2$ and $k\cos\theta_0=k_1-k_2$. Then in terms
of the Cartesian coordinates (\ref{euler-angles}) this curve is
\begin{equation}
x^1=\sqrt{\frac{k_2}{k}}\sin k_1t\,,\quad
x^2=\sqrt{\frac{k_2}{k}}\cos k_1t\,,\quad
x^3=\sqrt{\frac{k_1}{k}}\sin k_2t\,,\quad
x^4=\sqrt{\frac{k_1}{k}}\cos k_2t\,.
\end{equation}
This is a motion on a torus inside $S^3$ where the curve wraps the
two cycles $k_1$ and $k_2$ times. In general (see
Section~\ref{torus-subsection} and Appendix~\ref{torus-appendix})
one could take any torus inside $S^3$, but the extra conditions for loops
on the Hopf base require the ratio of the lengths of the cycles to be
$\sqrt{k_2/k_1}$. If $k_1=k_2$ this is a (multiply wrapped) circle.

The scalar couplings for these loops turn out to be quite simple,
\begin{equation}
\frac{1}{2}\,\sigma_1^R=\sqrt{k_1k_2}\,\cos(k_2-k_1)t\,dt\,,\qquad
\frac{1}{2}\,\sigma_2^R=\sqrt{k_1k_2}\,\sin(k_2-k_1)t\,dt\,,
\label{scalars_hopf_lat}
\end{equation}
so we just have a periodic motion, as in the case of the latitude on the
great $S^2$ in Section~\ref{longitudes-subsection} (and taking
the limit when the curve approaches the north-pole).

Since the path of this loop in $\bR^4$ is periodic, the dual string solution
describing it can be found by using the techniques of
\cite{Drukker:2005cu}. The detailed calculation is presented in
Appendix~\ref{torus-appendix}, where the action of the surface in
$AdS_5\times S^5$ describing a generic toroidal loop is computed.
For the application to the latitude discussed in this section, we
can use all the expressions from the general case of
\ref{torus-appendix} with the replacement
\begin{equation}
\sin\frac{\theta_0}{2}=\sqrt{\frac{k_2}{k}}\,,\qquad
\cos\frac{\theta_0}{2}=\sqrt{\frac{k_1}{k}}\,.
\end{equation}
Going over the calculation one sees that many of the expressions
simplify and the final result for the action
(\ref{torus-total-action}), where without loss of generality we have chosen $k_1\leq k_2$, is
\begin{equation}
\cS=-\left(2k_1-\sqrt{k_1 k_2}\right)\sqrt{\lambda}\,.
\end{equation}
It would be very interesting to see if the expectation value of the
loop could possibly be
computed exactly in gauge theory and compared at strong coupling
with this string calculation.

\subsection{More toroidal loops}
\label{torus-subsection}
As mentioned in the last subsection, the tools used for calculating
the loops associated with latitudes on the Hopf base can immediately
be applied to general doubly-periodic loops on {\em any} torus in
$S^3$.

We take the curve to be of the form
\begin{equation}
\begin{gathered}
x^1=\sin\frac{\theta}{2}\sin k_1t\,,\qquad
x^2=\sin\frac{\theta}{2}\cos k_1t\,,\\
x^3=\cos\frac{\theta}{2}\sin k_2t\,,\qquad
x^4=\cos\frac{\theta}{2}\cos k_2t\,.
\end{gathered}
\end{equation}
The scalar couplings for these loops are also simple,
\begin{equation}
\begin{aligned}
\frac{1}{2}\,\sigma_1^R
&=\frac{k_1+k_2}{2}\,\sin\theta\cos(k_2-k_1)t\,dt\,,\\
\frac{1}{2}\,\sigma_2^R
&=\frac{k_1+k_2}{2}\,\sin\theta\sin(k_2-k_1)t\,dt\,,\\
\frac{1}{2}\,\sigma_3^R &=\left(k_2 \cos^2{\theta\over 2}-k_1 \sin^2{\theta\over 2}\right)\,dt\,.
\end{aligned}
\end{equation}
Those expressions are similar to the ones for the latitude on $S^2$
in Section~\ref{latitude-subsection}. The string solution dual to these loops is presented in Appendix~\ref{torus-appendix}.

Let us just comment that these loops are a natural generalization of
the latitudes on the Hopf base, in the same way that the 1/4 BPS
latitude generalized the $Q$-invariant loops of
\cite{Zarembo:2002an}. Here too, compared with
(\ref{scalars_hopf_lat}) there is an extra constant coupling to the
third scalar $\Phi^3$.

It is tempting to guess that these loops arise by considering other
$S^2$ spaces inside $S^3$, where the equation for $\psi$
(\ref{psi-eqn}) is modified by the constant $\mu$ to
\begin{equation}
\dot\psi=-\mu\cos\theta\,\dot\phi\,,
\end{equation}
Such a construction would in turn lead to these general toroidal
loops with
\begin{equation}
\sin\frac{\theta}{2}
=\sqrt{\frac{k_2(1+\mu)-k_1(1-\mu)} {2k\mu}}\,.
\end{equation}

While it is clear that those loops, like all the others we constructed,
preserve 2 supercharges, we have not substantiated
whether they preserve some extra supersymmetries. If so, it would
be interesting to identify the general curve with those
supersymmetries, since those curves might give interpolating
families between the Hopf base and the great $S^2$. As an
indication that this might work, note that for
$k_2(1-\mu)-k_1(1+\mu)=0$ this is again the great circle and when
$k_2=0$, we end up with the latitude on the maximal $S^2$ of
Section~\ref{latitude-subsection}.

\subsection{Infinitesimal loops}
\label{Infinitesimal loops}
We conclude our list of examples by showing that in a particular
flat limit we can recover from our construction a subclass of the
loops of \cite{Zarembo:2002an}. If a loop is concentrated entirely
near one point, say $x^4=1$, one will not see the curvature of the
sphere anymore. More precisely, we can take a limit in which we
send the radius of $S^3$ to infinity while keeping the size of the
loop fixed, so that we end up with a curve on flat $\mathbb{R}^3$.
In this limit the left and right forms will then become exact
differentials
\begin{equation}
\sigma_i^{R,L}\sim 2\,dx_i\,,\qquad i=1,2,3\,,
\end{equation}
so the Wilson loop (\ref{susy-loop}) will reduce to
\begin{equation}
W=\frac{1}{N}\Tr\,\cP\exp \oint dx^i \left( i A_i +\Phi^i\right)\,.
\end{equation}
This is indeed a subclass of the $Q$-invariant loops constructed by
Zarembo in \cite{Zarembo:2002an} where the curve is restricted to
be on $\mathbb{R}^3$.
Studying the supersymmetry variation of such operator one can see
that generically it will only preserve two combinations of
Poincar\'e supersymmetries defined by the constraints
\begin{equation}
\left(\gamma^i-i\rho^i\gamma^5\right)\epsilon_0=0\,,\qquad
i=1,2,3\,.
\end{equation}
If the curve is restricted further to lie only in a 2-plane or a
line near $x^4=1$, the supersymmetry will be further enhanced. For
certain shapes, like a straight line or a circle on the plane,
also combinations of superconformal supersymmetries may be
preserved.

This should explain why in this case the expectation value of
these loops is trivial. The planar loops come from infinitesimal
ones on $S^3$, so it is quite natural that their expectation
values is unity. This might also explain why the construction of
the D3-brane solution dual to the Wilson loop in this limit was
singular \cite{Drukker:2006zk}.


\section{Wilson loops as pseudoholomorphic surfaces}
\label{J-section}

After going over the construction of the supersymmetric Wilson
loops and presenting many examples, expanding on
\cite{Drukker:2007dw}, in this part of the paper we will present
completely new results on the general string solutions dual to
those Wilson loops. Their underlying geometry will turn out to be
surprisingly simple and associated to the existence of an almost
complex structure, which we will call ${\cal J}$, on the subspace
of $AdS_5\times S^5$ in which the string solutions dual to the
loops live. As we shall show, the string surfaces satisfy the
``pseudo-holomorphic equations'' associated to this almost complex
structure which are a simple generalization of the usual
Cauchy-Riemann equations one encounters in complex geometry. An
analogous picture for the class of $Q$-invariant Wilson loops was
proposed in \cite{Dymarsky:2006ve}. As already mentioned in the
field theory discussion, see Section~\ref{Infinitesimal loops},
these latter loops are trivial in the sense that their expectation
value is expected to be identically one. On the other hand we know
that the expectation value of the loops constructed in this paper
is non-trivial. We will show that the loop expectation value
receives a nice geometrical interpretation in terms of the
integral on the string world-sheet of the fundamental two-form
associated to ${\cal J}$.

For the reasons just mentioned it will be useful to begin this
section by reviewing the concept of a pseudo-holomorphic
surface.\footnote{For a comprehensive discussion see \cite{dusa}.}
Let $\Sigma$ be a two-dimensional surface with complex
structure\footnote{An almost complex structure on a
two-dimensional surface is always integrable \cite{dusa}.}
$\text{\j}^{\alpha}_{\ \beta}$, ($\alpha,\beta=1,2$), embedded in
a space $M$ with almost complex structure ${\cal J}^M_{ \ N}$.
This surface is said to be pseudo-holomorphic if it satisfies
\be\label{pseudo} V_{\alpha}^M\equiv\partial_\alpha X^M-\kappa
\,{\cal J}^M_{ \ N}\,\text{\j}^{\
\beta}_{\,\alpha}\,\partial_\beta X^N=0\,. \ee The possible
choices $\kappa=\pm1$ correspond to (pseudo)holomorphic and
anti-holomorphic embeddings. In our discussion we will assume
$\kappa=1$. These equations are a natural generalization of the
Cauchy-Riemann equations on the complex plane, to which they
reduce when we identify $\Sigma$ and $M$ with $\mathbb{R}^2$ and
use the standard complex structure
\be\text{\j}={\cal J}=\left(%
\begin{array}{cr}
  0 & -1 \\
  1 & 0 \\
\end{array}%
\right). \ee The solutions of the pseudo-holomorphic equations
(\ref{pseudo}) are surfaces calibrated by ${\cal J}$. Indeed if we
introduce the positive definite quantity \be {\cal
P}=\frac{1}{4}\int_\Sigma \sqrt{g}\,g^{\alpha\beta}
G_{MN}V_{\alpha}^M V_{\beta}^N \ee and expand ${\cal P}$ we obtain
\be {\cal P}=A(\Sigma)-\int_\Sigma {\cal J}\geq 0\,
\label{area-bound} \ee where $A(\Sigma)$ is the area of the
surface $\Sigma$ and ${\cal J}$ denotes the pull-back of the
fundamental two-form \be {\cal J}=\half\, {\cal
J}_{MN}\,dX^M\wedge dX^N. \ee For a pseudo-holomorphic surface
${\cal P}=0$, and one concludes that \be A(\Sigma)=\int_\Sigma
{\cal J}. \label{area} \ee Note that if
 $\cJ$ is closed, its integral is the same for all surfaces in
the same (relative) homology class and then the bound in
(\ref{area-bound}) applies to them all. Therefore a string surface
calibrated by a closed two-form is necessarily a minimal surfaces
in its homology class.

In our context the ambient space will be a subspace of
$AdS_5\times S^5$ and $\Sigma$ will be the string world-sheet on
which the complex structure can be expressed in terms of the
world-sheet metric $g_{\alpha\beta}$ and the flat epsilon symbol
$\varepsilon^{\alpha\delta}$ (see (\ref{epsconvention})) as
\be
\text{\j}^\alpha_{\ \beta}=
\frac{1}{\sqrt{g}}\varepsilon^{\alpha\delta}g_{\delta\beta}\,.
\ee

The $AdS$ dual description of the $Q$-invariant loops was found in
\cite{Dymarsky:2006ve}. The loops are constructed by associating
to every tangent vector in $\bR^4$ one of the scalars, in a way
related to the topological twisting of an $SO(4)$ subgroup of the
$R$-symmetry group and the Euclidean Lorentz group.

When thinking of a D3 in flat ten dimensional space this leads to
a natural association of the four coordinates parallel to the brane
and four of the transverse directions. Taking the near-horizon limit
of the metric after accounting for the brane's back-reaction leads to
$AdS_5\times S^5$ in the Poincar\'e patch with coordinates
$\left(x^\mu,\,y^m,\,u^i\right)$ with $\mu,m=1,2,3,4$ and $i=1,2$ and
metric
\be
ds^2=\left(y^2+u^2\right)dx^\mu dx^\mu+\frac{1}{
y^2+u^2}\left(dy^m dy^m+du^i du^i\right),
\ee
the corresponding string solutions live in the $u^i={\rm const.}$ subspace.

It is now natural to relate the coordinates $x^\mu$ and $y^m$ with
$\mu=m$ with the closed 2-form \be {J}=\half{J}_{MN}\,dX^M\wedge
dX^N\equiv\delta_{\mu m}dx^\mu\wedge dy^m\,, \label{Jzar} \ee as
it is invariant under the twisted group. It is easy to see that
${J}^M_{ \ N}$ squares to minus the identity and therefore it
defines an almost complex structure on the relevant subspace of
$AdS_5\times S^5$.  The string solutions dual to these loops turn
out to be pseudo-holomorphic surfaces with respect to this almost
complex structure and satisfy
\begin{equation}
(y^2+u^2)\partial^\alpha x^{\mu}-
\text{\j}^\alpha_{\ \beta}\partial^\beta y^m\delta^\mu_m=0\,.
\label{pseudodym}
 \end{equation}
Since the two-form $J$ is closed, they are minimal calibrated
surfaces with (divergent) world-sheet area given by (\ref{area}).
Using the closure of the calibration two-form ${J}$ it is
immediate to re-express the integral of ${J}$ as a contour
integral on the world-sheet boundary obtaining \be A(\Sigma)=
\frac{1}{\epsilon}\int dt|\dot{x}|\,, \ee where the formally
divergent integral has been regularized by computing it at
$z=\epsilon$. The classical action $S_{cl}(\Sigma)$ is the finite
part of the world-sheet area and therefore vanishes, implying
that the Wilson loops have trivial expectation value \be \langle W
\rangle=e^{-\sqrt{\lambda}\, S_{cl}(\Sigma)/2\pi}=1. \ee Despite
the existence of this beautiful structure, the only explicit
solutions known are the straight line and the $1/4$ BPS circle,
which is the limit of the latitude when $\theta_0\to0$ (see
Section~\ref{latitude-subsection}). In
Appendix~\ref{longitudes-sol-appendix} we construct another explicit
solution for a loop in this class. This loop is made of two rays
in the plane at arbitrary opening angle and is related to the
longitudes example of Section~\ref{longitudes-subsection} by a
stereographic projection (Figure~\ref{cusp-fig}).

In the rest of this section  we will see that it is possible to
extend these ideas to the class of supersymmetric Wilson loops
presented in Section~\ref{Introduction}. Those loops follow an
arbitrary path on $S^3$ and couple to three scalars,
parameterizing an $S^2$. Therefore they will be described by a
string ending along a path in an $S^3\times S^2$ on the boundary
of $AdS_5\times S^5$.

For a generic curve on $\bR^4$ or $S^4$ the string may extend into
all of $AdS_5$, but when it is restricted to $\bR^3$ or $S^3$, it
will
remain inside an $AdS_4$ subspace. Likewise we assume%
\footnote{For a curve coupling to two scalars and wrapping
$S^1\subset S^5$ the solution will have to extend into $S^2\subset
S^5$, for topological reasons. This is indeed the case for the
circular $Q$-invariant loop \cite{Zarembo:2002an} and our
assumption is that a similar phenomenon does not occur with
boundary data in $S^3\times S^2$.} that the string will remain
inside the $S^2\subset S^5$, so the full solution will reside
inside an $AdS_4\times S^2$ subspace which we label by ${\cal X}$.
This assumption will be later justified by proving that the
solutions to the pseudo-holomorphic equation in this subspace are
extrema of the action.

The metric we employ is ($\mu=1,\cdots,4$, $i=1,\,2,\,3$) \be
 ds^2=\frac{1}{z^2}\,dx^\mu dx^\mu+z^2 dy^i
dy^i\,,\qquad z^2\equiv\frac{1}{y^iy^i}\,, \label{X-metric} \ee
subject to the constraint \be x^2+z^2=1,\qquad x^2\equiv x^\mu
x^\mu. \ee We will see that the string solutions dual to the loops
are pseudo-holomorphic with respect to an almost complex structure
${\cal J}$ on ${\cal X}$ which we construct next. The fundamental
two-form associated to $\cJ$ will turn out to be not closed
suggesting the interpretation of our loops as  ``generalized
calibrated submanifolds''. We will also argue that the non-closure
of $\cJ$ seems to be related to the fact that the loops have
non-trivial expectation values.

\subsection{Almost complex structure on $AdS_4\times S^2$}
\label{acs section}

We want to motivate the construction of the almost complex
structure relevant to the $AdS$ description of the generic loops
on $S^3$ by taking the supersymmetry conditions derived in field
theory as our starting point, see (\ref{susy-fieldtheory}). They
can be summarized as
\be
\gamma_{\mu\nu}\,\epsilon_1^-
=-{i}\sigma_{\mu\nu}^i\tilde\rho_i\,\epsilon_0^-\,,
\label{susy1}
\ee
or equivalently as
\be
\gamma_{\nu}\tilde\rho_{i}\,\epsilon_1^-
=-{i}\sigma_{\mu\nu}^i\gamma_\mu\,\epsilon_0^-\,,
\label{susy2}
\ee
where $(\gamma_{\mu},\tilde\rho_i)$
denote seven of the 10-dimensional (flat) anti-commuting gamma matrices%
\footnote{To make them anti-commute they are related to the
field-theory gamma matrices in (\ref{deltaW}) by
$\tilde\rho_i=\rho_i\gamma^5$.}
and $\sigma^i_{\mu\nu}$ denote the components of the
left-invariant one-forms on $S^3$ (\ref{one-forms-decompose}).
We can also express the algebra of the $SU(2)_A$ rotating the three scalars
(and $y^i$) as%
\footnote{The extra minus sign is due to $\gamma^5$.}
\be
\tilde\rho_{ij}\,\epsilon_0^-=-
i\,\varepsilon_{ijk}\tilde\rho_{k}\,\epsilon_0^-\,.
\label{susy3}
\ee

The almost complex structure ${\cal J}$ in the dual string side is
ultimately expected to encode all these conditions. We can rewrite
these relations in terms of curved-space gamma matrices%
\footnote{The indices $M,\,N$ include all seven directions, but to
avoid ambiguities we will never substitute their values for them,
only for $\mu,\,\nu$ and $i,\,j$.}
$\Gamma_M=(\Gamma_\mu,\Gamma_i)=(z^{-1}\gamma_\mu,z\,\tilde\rho_i)$
(remembering (\ref{eps0 eps1}) that $\epsilon_0^-=-\epsilon_1^-$)
as \be
\begin{aligned}
z\,\Gamma_{M\,\mu}\,\epsilon_0^- &= -i\,{\cal J}^N_{ \
M;\,\mu}\,\Gamma_N\,\epsilon_0^-
\label{cond}\\
z\,\Gamma_{M\,i}\,\epsilon_0^- &= i\,{\cal J}^N_{ \
M;\,i}\,\Gamma_N\,\epsilon_0^-\,,
\end{aligned}
\ee
with
\begin{equation}
\label{cond2} {\cal J}^\mu_{ \
\nu;\,i}={z^2}\,\sigma^i_{\mu\nu},\qquad {\cal J}^\nu_{ \
i;\,\mu}=-z^4{\cal J}^i_{ \ \nu;\,\mu}
=z^2\,\sigma^i_{\nu\mu},\qquad {\cal J}^i_{ \
j\,;\,k}=-z^2\,\varepsilon_{ijk}\,,
\end{equation}
and with all the other components of ${\cal J}^N_{ \ M;P}$
vanishing. We can interpret (\ref{cond}) as a multiplication table
for the curved gamma matrices acting on $\epsilon_0$: The product
of two gamma matrices is re-expressed in terms of another gamma
matrix ${\cal J}^N_{ \ M;\,P}\Gamma_N$. In fact, this
multiplication table up to factors of $z$ is basically the
octonion multiplication table, which can be regarded as a higher
dimensional generalization of the usual cross-product in
$\mathbb{R}^3$. We present it in Appendix~\ref{S6app} and review
how it can be used to define an almost complex structure on the
round 6-sphere. In analogy to (\ref{notation}) it is then natural
to introduce the following matrix \be {\cal J}^M_{ \ N}={\cal
J}^M_{ \ N;\,P}\,X^P, \label{notnot} \ee where $M$ and $N$ denote
row and column indices respectively. From (\ref{cond2}) and
(\ref{notnot}) we can read the various components of \be
{\cal J}=\left(%
\begin{array}{cc}
  {\cal J}^\mu_{\  \nu} &  {\cal J}^\mu_{ \ j} \\
   {\cal J}^i_{\  \nu} &  {\cal J}^i_{ \ j} \\
\end{array}
\right),
\ee
 to be
\begin{equation}
{\cal J}^\mu_{ \ \nu}={z^2}\,\sigma^i_{\mu\nu}\,y^i,\qquad {\cal
J}^\nu_{ \ i}=-z^4{\cal J}^i_{ \ \nu}
=z^2\,\sigma^i_{\nu\mu}\,x^\mu,\qquad {\cal J}^i_{ \
j}=-z^2\varepsilon_{ijk}\,y^k\,. \label{calJ-components}
\end{equation}
Explicitly
\be
\label{acsx4}
{\cal J}=\left(%
\begin{array}{ccc}
 z^2 \left(%
\begin{array}{cccc}
  0 & y_3 & -y_2 & -y_1 \\
  -y_3 & 0 & y_1 & -y_2\\
  y_2 & -y_1 & 0 & -y_3\\
y_1& y_2& y_3&0\\
\end{array}%
\right) &
 & z^2\left(%
\begin{array}{ccc}
  -x_4 & -x_3 & x_2 \\
  x_3 & -x_4 & -x_1 \\
  -x_2 & x_1 & -x_4 \\
x_1& x_2 &x_3
\end{array}%
\right)\\
 z^{-2}\left(%
\begin{array}{cccc}
  x_4 & -x_3 & x_2& -x_1 \\
  x_3 & x_4 & -x_1 & -x_2\\
  -x_2 & x_1 & x_4 & -x_3 \\
\end{array}%
\right)  &
 & z^2 \left(%
\begin{array}{ccc}
  0 & -y_3 & y_2 \\
  y_3 & 0 & -y_1 \\
  -y_2 & y_1 & 0 \\
\end{array}%
\right) \\
\end{array}%
\right)\,. \ee

To show that ${\cal J}$ defines an almost complex structure on
${\cal X}=AdS_4\times S^2$, note that a generic tangent vector
$p^M=\left(p_1,p_2,p_3,p_4,q_1,q_2,q_3\right)$ in $T{\cal X}$
satisfies the condition \be x^\mu p^\mu-z^4 y^i q^i=0\,,
\label{derive-tangent} \ee which comes from differentiating the
constraint $x^2+z^2=1$. Then it is easy to see that ${\cal J}^N_{
\ M}{p}^M$ is still a tangent vector so that ${\cal J}$ is a well
defined map on the tangent space $T{\cal X}$.  Furthermore if we
consider the action of ${\cal J}^2$ we obtain an expression very
similar to what one gets for $S^2$ (see (\ref{S2-J^2}) in
Appendix~\ref{S6app}) and with the aid of (\ref{derive-tangent})
one finds that \be {\cal{J}}^2({p})=-p\,. \ee Therefore ${\cal J}$
defines an almost complex structure on ${\cal X}=AdS_4\times S^2$.

As in the case of the almost complex structure for the strings
dual to the $Q$-invariant loops (\ref{Jzar}), our almost complex
structure ${\cal J}$ reflects the topological twisting associated
to our loops. As discussed in Section~\ref{Topological twisting},
this twisting reduces the product of the groups $SU(2)_R$ and
$SU(2)_A$ to their diagonal subgroup $SU(2)_{R'}$ which is then
regarded as part of the Lorentz group. This can be seen directly
from our construction as ${\cal J}^\mu_{ \ \nu}$ is given by the
contraction of the components of the one-forms $\sigma_i^R$ with
the $y^i$ coordinates on which the $SU(2)_A$ group acts. Similar
remarks can be made for the ${\cal J}^i_{ \ \nu}$ sub-block. At a
more formal level the twisting manifests itself through the
condition \be \left({\cal J}^{\mu\nu}_{ \ \ \,
;\,i}\Gamma_{\mu\nu} -{\cal J}^{jk}_{ \ \ \,
;\,i}\Gamma_{jk}\right)\epsilon_0^-=0\,, \ee which simply
expresses the invariance of $\epsilon_0$ under the twisted
$SU(2)_{R'}$ action \be \left(\sigma^i_{\mu\nu}\gamma_{\mu\nu}
+\varepsilon_{ijk}\tilde\rho_{jk}\right)\epsilon_0^-=0\,. \ee

Since this almost complex structure captures those properties of our
Wilson loops, we expect the string solutions describing the Wilson loops
in $AdS_5\times S^5$ to be compatible with it, {\it i.e.} that the
world-sheet is pseudo-holomorphic with
respect to ${\cal J}$. We do not have a proof of this, but in the
remainder of this section we will study such pseudo-holomorphic
surfaces and show that their properties match with the expected
behavior of the string duals.

In order to write the pseudo-holomorphic equations associated to
${\cal J}$ we introduce the vector
$X^M=\left(x_1,x_2,x_3,x_4,y_1,y_2,y_3\right)$ in ${\cal X}$ and
the equations are \be {\cal J}^M_{ \ N}\partial_\alpha
X^N-\sqrt{g}\,\varepsilon_{\alpha\beta}\partial^\beta X^M=0.
\label{basic} \ee For brevity in the following we will refer to
the pseudo-holomorphic equations (\ref{basic}) as the ${\cal
J}$-equations. As we will show, surfaces satisfying those
equations are supersymmetric and are classical solutions of the
string action.

It is possible to repackage three of the ${\cal J}$-equations in
form notation as
\be
\star_2 dy^i=
\frac{1}{2z^2} \,\sigma^i+ \frac{z^2}{ 2}\,\eta^i\,,\qquad
i=1,2,3\,.
\label{eqnew}
\ee
On the left-hand side we used the Hodge dual with respect to the
world-sheet metric and on the right-hand side we used the
pull-backs to the world-sheet of the one-forms (we use the same
notations for the forms and their pull-backs)
\be
\begin{aligned}
&\sigma^1= 2  (x_2\, d x_3-x_3\, dx_2+x_4\, dx_1-x_1\, d x_4)\,, \cr
&\sigma^2 =   2 (x_3\, d x_1-x_1\, d x_3+x_4\, d x_2-x_2\,d x_4)\,,\cr
&\sigma^3 =  2 (x_1\, d x_2-x_2\, d x_1+x_4\, d x_3-x_3\, d x_4)\,,
\label{sigmapull}
\end{aligned}
\ee
which are defined in the same way as the right-forms on $S^3$
(\ref{one-forms}) but we extend the definition to arbitrary radius.
The other forms are the pull-backs of the $SU(2)_A$ currents
\be
\begin{aligned}
{\eta}^1&= 2  (y_2\,d y_3-y_3\,d y_2)\,, \\
{\eta}^2& = 2 (y_3\,dy_1-y_1\,d y_3)\,, \\
{\eta}^3 &= 2 (y_1\,dy_2-y_2\, dy_1)\,.
\end{aligned}
\label{rhopull}
\ee

We will try to show that the $\cJ$ equations are satisfied by the
strings dual to the supersymmetric loops on $S^3$. As a first
support for this claim consider the asymptotic form of the surface
near the boundary of $AdS_5$. As we approach the boundary, taking
$z$ to zero, $x^\mu$ as well as $y^i/y$ approach constants, given
by the boundary conditions. In the conformal gauge we denote the
two world-sheet directions as $n$ and $t$, normal and tangent to
the boundary respectively. It can be shown in general
\cite{Drukker:1999zq} that $|\partial_n z|=|\partial_t x|$. In our
case we can take (\ref{eqnew}) which in the $z\to0$ limit reduces
to \be
\partial_n y^i\simeq
\frac{1}{2z^2} \,\sigma^i_{\mu\nu}x^\mu\partial_t x^\nu\,.
\label{near-boundary} \ee Given that $y^i$ scale as $z^{-1}$ we
get \be z\,y^i\simeq \frac{\sigma^i_{\mu\nu}x^\mu\partial_t x^\nu}
{|\partial_t x|}\,. \ee The left-hand side represents the boundary
conditions on the $S^2$, which exactly match the scalar couplings
of the Wilson loop (\ref{susy-loop}) captured by the right-hand
side.

Another way to see this is by looking at (\ref{acsx4}), where
in the $z\to0$ limit, as we approach the $AdS_5$,
the lower-left sub-matrix ${\cal J}^i_{\ \nu}$
dominates. The entries in this sub-block are the components of the
forms $\sigma_i^R$ which define the coupling of the scalars
$\Phi^i$ to the Wilson loop operator in the field theory.
Therefore we can view ${\cal J}$ as the natural bulk extension of
those couplings.

Lowering the indices of the almost complex structure we obtain an
antisymmetric tensor ${\cal J}_{MN}$. We can therefore introduce
the following fundamental two-form\footnote{For symbol economy we will use the same symbol $\cJ$
to denote both the almost
complex structure and the associated fundamental two-form. It will always be clear from the context
what $\cJ$ refers to.} \be {\cal J}=\half\,{\cal
J}_{MN}\,d X^M\wedge d X^N =\frac{1}{ 4}\,
y^i\left(d\sigma^i-z^4d\eta^i \right) -\half\, \sigma^i\wedge
dy^i. \label{fundform} \ee where the one-form $\sigma^i$ and
$\eta^i$ were defined in (\ref{sigmapull}) and (\ref{rhopull}).
Later in Section~\ref{calib section} we will discuss our string as
surfaces calibrated by ${\cal J}$. For now we limit ourselves to
observe that this is not a standard calibration as  ${\cal J}$ is not closed \be \label{dJ} d{\cal
J}=-\frac{1}{4}\,dy^i\wedge d\sigma^i+z^4\, dy_1\wedge dy_2\wedge
dy_3. \ee Written out explicitly $d{\cal J}$ reads\footnote{For
brevity in what follows we omit the $\wedge$ symbol and use the
notation $dx_{\mu\nu}=dx_\mu\wedge dx_\nu$ and
$dy_{123}=dy_1\wedge dy_2\wedge dy_3$.}
 \be -dy_1 dx_{23}-dy_1
dx_{41}-dy_2 dx_{31}-dy_2 dx_{42}-dy_3 dx_{12}-dy_3 dx_{43}+z^4
dy_{123}\,, \ee which is remarkably similar to the expression of
associative three form preserved by the exceptional group $G_2$,
see (\ref{associative}).

The non-closure of $\cJ$ for a calibrated string is unusual and
raises the issue of whether the solutions of the
$\cJ$-equations are automatically solutions of the $\sigma$-model. 
To prove that this is indeed the case, we consider the equations
of motion for the $\sigma$-model in $AdS_5\times S^2$ (the
equations of motion for the extra three coordinates in $S^5$ are
automatically satisfied by setting them to constants)
\be
 \nabla_{\alpha}\left(G_{MN}\partial^\alpha X^N\right)
=\partial_{\alpha}\left(G_{MN}\partial^\alpha X^N\right)
-\half\,\partial_M G_{PN}\partial_\alpha X^P\partial^\alpha X^N=0
\ee with metric $G_{MN}$ as in (\ref{X-metric}) and
$\nabla_{\alpha}$ denoting the pull-back of the covariant
derivative with respect to $G_{MN}$.
 We now show that the equations of motion for the $x^\mu$ and $y^i$
coordinates are satisfied once we assume that the string lives in
the $AdS_4\times S^2$ subspace and is a solution of the $\cJ$
equations. Using the ${\cal J}$-equations  we can write the
equations of motion for $x^\mu$ and $y^i$ as \be
\label{variationX} \epsilon^{\alpha\beta}\partial_\alpha
X^P\partial_{\beta}X^N \left({\partial_P {\cal J}_{MN}}-\half\,
{\partial_M G_{QP} }\, {\cal J}^Q_{\ N}\right)=0\,. \ee When
$M=\mu$ the second term in (\ref{variationX}) does not contribute
and it is very easy to see that this condition is indeed
satisfied. For $M=i$, on the other hand, the left hand side of
(\ref{variationX}) becomes after switching to form notation \be
\frac{1}{2} \left(d\sigma^i-z^4
d\eta^i\right)\left(\delta^{ik}-z^2 y^i y^k\right)\,. \ee This
expression vanishes since, by using the ${\cal J}$ equations and
the orthogonality condition $x^\mu dx^\mu-z^4 y^i dy^i=0$, one can
show after some algebra that \be d\sigma^i-z^4 d\eta^i=z^4 y^i\,
G_{MN}\partial_\alpha X^M\partial^\alpha X^N\,d^2\sigma\,. \ee

\subsection{Supersymmetry}

A good check that the solutions of the ${\cal J}$-equations
describe our Wilson loops comes from studying the supersymmetries
preserved by those strings. In this subsection we will prove that
strings satisfying those equations are indeed supersymmetric and
are invariant under precisely the same supercharges which
annihilate the dual operator on the field theory side.

The $\kappa$-symmetry condition for a fundamental string is \be
\label{kappa} \left(\sqrt{g}\,
\varepsilon^{\alpha\beta}\partial_\alpha X^M\partial_\beta
X^N\Gamma_{MN}-i\,G_{MN}\partial_\alpha X^M\partial^\alpha
X^N\right)\epsilon_{AdS}=0\,, \ee where $\epsilon_{AdS}$ is the
$AdS_5\times S^5$ Killing spinor. The most convenient form for the
Killing spinor is \cite{Claus:1998yw} \be
\epsilon_{AdS}=\frac{1}{\sqrt{z}}\left(\epsilon_0+z
\left(x^\mu\Gamma_\mu-y^i\Gamma_i\right)\epsilon_1\right)\,, \ee
where $\epsilon_0$ and $\epsilon_1$ are constant 16 component
Majorana-Weyl spinors. In fact they are the exact analogues of the
spinors representing the Poincar\'e and conformal supersymmetries
in the dual $\cN=4$ theory (\ref{conformal spinor}), as can be
seen by going to the $AdS$ boundary where $\epsilon_{AdS}$ reduces
to \be \epsilon_{AdS}\mathop{\sim}_{z\to0} \frac{1}{\sqrt{z}}
\left(\epsilon_0+ x^\mu\gamma_\mu\, \epsilon_1\right). \ee

To prove (\ref{kappa}) we first use the ${\cal J}$-equations and
rewrite the term multiplying $\epsilon_{AdS}$ as \be
\partial_\alpha X^M\partial^\alpha X^N
\left({\cJ}^P_{\ \ N}\Gamma_M\Gamma_P
-iG_{MN}\right)=
\partial^\alpha X^P\Gamma_P\,
\partial_\alpha X^M\left({\cal J}^{N}_{\ \ M}\Gamma_N
-i\,\Gamma_M\right)\,.
\ee
It will therefore be enough to prove
\be
\partial_\alpha X^M
\left({\cal J}^{N}_{\ \ M}\Gamma_N-i\,\Gamma_M\right)
\epsilon_{AdS}=0\,. \label{gen} \ee This equation should be
satisfied by the same supersymmetry parameters as in the
gauge-theory calculation in Section~\ref{susy-subsection}. They
were all collected in (\ref{cond}) in terms of the components of
$\cJ$. Using first that $\epsilon_0=-\epsilon_1$, the left-hand
side of (\ref{gen}) becomes (switching to form notation) \be
\begin{aligned}
&i\,dX^M
\left(-iX^P{\cal J}^N_{\ \ M;\,P}\Gamma_N\,\epsilon_0+z \left(x^\mu\Gamma_{M\mu}-y^i\Gamma_{M\,i}\right)\epsilon_0\right)\\
&\hskip1cm
-i\, dX^M\left(\Gamma_M\,\epsilon_0-i z X^P {\cal
J}^N_{\ \ M;\,P}\Gamma_N\left(x^\mu\Gamma_\mu-y^i\Gamma_i\right)\epsilon_0\right).
\end{aligned}
\ee
The terms in the first line vanish once we impose on $\epsilon_0$ and
$\epsilon_1$ the conditions in (\ref{cond}).
Using that $x^2+z^2=1$ and that $x^\mu dx^\mu-z^4 y^i dy^i=0$
allows to prove that also the terms in the second line vanish.

Beyond allowing us to prove $\kappa$-symmetry, equation (\ref{gen})
is quite interesting in its own right. First multiplying it by%
\footnote{$\partial_{\bar{z}}\equiv
\partial_\sigma-i\partial_{\tau}$,
$\partial_{{z}}\equiv\partial_\sigma+i\partial_{\tau}$.}
$\partial_{\bar{z}}X^N\Gamma_N$ gives
\be
\partial_{\bar{z}}X^M\partial_{\bar{z}}X^M\,\epsilon_{AdS}\,,
\ee
which holds because of the Virasoro constraint. Multiplying by
$\partial_{z}X^N\Gamma_N$ leads to
\be
-i\partial_{z}X^M\partial_{\bar{z}}X^N
\left(\Gamma_{MN}+G_{MN}\right)\epsilon_{AdS}=0\,,
\ee
which is the $\kappa$ symmetry condition rewritten in the
$z,\bar{z}$ basis. We also observe that, by using the
pseudo-holomorphic equations, one can recast the condition
(\ref{gen}) simply as
\be
\partial_{\bar{z}}X^M\Gamma_M\epsilon_{AdS}
=\Gamma_{\bar{z}}\,\epsilon_{AdS}=0\,,
\label{rew}
\ee
 where $\Gamma_{\bar{z}}$ is the pull-back to the
world-sheet of the gamma matrices.

\subsection{Wilson loops and generalized calibrations}
\label{calib section}

In this section we will discuss the string dual to our Wilson
loops from the point of view of calibrated submanifolds. More
precisely we will argue that the natural geometrical description
of the corresponding string solutions is in the context of
``generalized calibrations''
\cite{Gutowski:1999iu,Gutowski2,Gutowski:1999dr}.\footnote{See
also \cite{Gauntlett:2003di} for a general discussion on
calibrations.} The main result is that the classical action of the
strings (and hence the expectation value of the loops) is given by
the integral on the world-sheet of the fundamental two-form
 ${\cal J}$. This is because, as discussed in the introduction of
 Section~\ref{J-section},
 the world-sheet area of a pseudo-holomorphic surface $\Sigma$
 can be computed by
integrating the pull-back of the fundamental two-form ${\cal J}$
(\ref{fundform}), \be A(\Sigma)=\int_\Sigma {\cal J}\, .
\label{are} \ee This equation suggests that our loops can be
viewed as two-dimensional calibrated submanifolds with the
two-form ${\cal J}$ as calibration. As already observed this is
not a standard calibration though as the fundamental two-form
${\cal J}$ is not closed, see (\ref{dJ}).

Without worrying about this issue for now, note that it is
possible to rewrite the two-form ${\cal J}$ as a sum of two
contributions \be \label{splitJ}{\cal J}={\cal J}_0+d\Omega \ee
with \be {\cal J}_0=-\frac{1}{4}\,y^i\left(d\sigma^i +z^4
d\eta^i\right),\qquad \Omega=\half\, y^i \sigma^i. \label{j0} \ee
Using Stokes theorem the world-sheet area is then \be
A(\Sigma)=\int_\Sigma {\cal J}_0+\int_{\partial \Sigma}\Omega\,.
\label{AJ0Omega} \ee This expression is generically divergent and
requires regularization. It can be seen by studying the
asymptotics near the boundary $z\sim0$ (see the discussion around
(\ref{near-boundary})) that the contribution of $\cJ_0$ is finite.

The integral of $\Omega$ is therefore divergent, but this is exactly the
divergence that needs to be subtracted from the area. To see that we again
use the manipulations as in (\ref{near-boundary}) to rewrite it as
\be
\int_{\partial \Sigma}\Omega=\half \int_{\partial \Sigma}dt\,
y^i{\sigma}^i_{\mu\nu}x^\mu\partial_\eta x^\nu
=-\half \int_{\partial \Sigma}dt\,\sqrt{g}\,
z^2\partial^n{y^i}\,.
\label{diver}
\ee
Here $dt$ is the line element tangent to the boundary and $\partial^n$
the normal derivative. The last expression is an integral over the momentum
$P_{y^i}$ conjugate to the coordinates $y^i$, which in turn can be related
to $P_z$, the momentum conjugate to $z$. Therefore we can rewrite
\be
\int_{\partial \Sigma}\Omega
=-\int_{\partial \Sigma}dt\,y^i P_{y^i}
=\int_{\partial \Sigma}dt\,z\, P_z\,.
\ee
The rigorous procedure to get a finite answer for the Wilson loops is
by a Legendre transform over the radial coordinate $z$
\cite{Drukker:1999zq}. It will
therefore precisely cancel the entire contribution of $\Omega$.

The $AdS$/CFT prediction for the expectation value of the Wilson
loop in the strong coupling regime is then
\begin{equation}
\label{prediction}
\exp{\left(-\frac{\sqrt{\lambda}}{2\pi}\int_{\Sigma}{\cal
J}_0\right)}.
\end{equation}
We can go further and derive a simpler expression for ${\cal
J}_0$. Applying the $d$ operator on equation (\ref{eqnew})
yields\be \half \left(d\sigma^i+z^4 d\eta^i\right)+\half\, dz^4
\eta^i-d(z^2\star_2 dy^i)=0\,. \ee Taking the inner product of
this equation with ${y}^i$ we derive the following relation for
${\cal J}_0$ \be {\cal J}_0=-\half\, y^i\cdot d(z^2\star_2 d y^i).
\ee By writing $y^i=\theta^i/z$ with $\theta^i\theta^i=1$, $\cJ_0$
can be proven to be equal to \be -\half
\sqrt{g}\,\left(\theta^i\cdot \nabla^2
\theta^i-\frac{\nabla^2z}{z}\right)d^2\sigma \ee where $\nabla^2$
is the world-sheet Laplacian. The regularized area can therefore
be written in a rather simple form as\be\label{j00}
\int_{\Sigma}{\cal J}_0=\half\int_{\Sigma}d^2\sigma\sqrt{g}
\left(\partial_\alpha \theta^i\,\partial^\alpha\theta^i
+\frac{\nabla^2z}{z} \right)\,, \ee or equivalently \be
\label{J0final} \half\int_{\Sigma}d^2\sigma\sqrt{g}\,
\left(\partial_\alpha\theta^i\,\partial^\alpha\theta^i
+\frac{1}{z^2}\,\partial_\alpha z\,\partial^\alpha z+\nabla^2 \log
z\right)\,. \ee The last term can also be rewritten as a boundary
term \be \half\int_{\Sigma}d^2\sigma\sqrt{g}\,{\nabla^2\log z}
=\half\int_{\partial\Sigma}dr\frac{\partial_\sigma z}{z}\,. \ee
Unfortunately we are not able to re-express also the first two
terms in (\ref{J0final}) as integrals on the contour of the Wilson
loops at the boundary. This is unfortunate, as it would have
allowed to compute the expectation value of the Wilson loop
without the need of an explicit string solution. We leave this
issue to future investigations.

Before we end this subsection we turn back to the issue of the
non-closure of $\cJ$. As already observed a surface calibrated
with respect to a closed form is a minimal surface in its homology
class. Such a statement will not apply in our case and we should
instead study our string solutions within the framework of
generalized calibrations. Those are defined in complete analogy to
calibrations, only without demanding closure of the form
\cite{Gutowski:1999iu,Gutowski2,Gutowski:1999dr}. Given a $k$-form
$\psi$ which is not closed, a generalized calibrated submanifold
is a $k$-dimensional submanifold which is a minimum of the
(energy) functional \be E(M)=\rm{Vol}(M)-\int_M \psi.
\label{psigen} \ee Since we do not require closure of $\psi$, a
minimum of $E(M)$ is not necessarily a minimal-volume manifold.

Generalized calibrations appear very naturally in the discussion of
D-branes in curved backgrounds.
Their actions typically include a Wess-Zumino term in addition to
the Dirac-Born-Infeld term and therefore cannot be seen as
volume-minimizing submanifolds. In these cases the non-closure of
$\psi$ can be due to torsion or to the presence of
background or worldvolume fluxes. Equation (\ref{psigen}) can be
thought as a BPS condition for these branes.

The above discussion points to a connection between $\cJ$ being a
generalized calibration and our loops having a non-trivial
expectation value (in contrast to the $Q$-invariant loops). This
interpretation is suggested by (\ref{splitJ})-(\ref{prediction}),
where we see that while the exact piece reduces to a divergent
boundary contribution canceled by a counter-term, the non closed
piece $\cJ_0$ gives a finite non-trivial expectation value. In
comparing equations (\ref{AJ0Omega}) and (\ref{psigen}) it is also
tempting to consider $\int d\Omega$ as the analogue of the area
functional $\text{Vol}(M)$ and $\cJ_0$ as the analogue of $\psi$.
It would be interesting to see if there is some realization of
$\cJ_0$ in terms of a pull-back of a flux to the world-sheet.

Another interesting feature of our loops is the existence of
unstable solutions. It was found in \cite{Drukker:2006ga} and
reviewed in Appendix~\ref{latitude-sol-appendix} that there are
two classical string solutions describing the latitude loop, one
is a minimum and the other not. This should be quite general since
our scalar couplings define a curve on $S^2$ and therefore the
string can wrap the north or the south pole (or in principle also
wrap the sphere multiple times).  This phenomenon might be related
to the non-closure of $\cJ$.

\subsection{Loops on $S^2$ and strings on $AdS_3\times S^2$}
\label{JS2-subsection}
We now present an application of the general formalism so far
discussed to the subclass of supersymmetric Wilson loops on $S^2$
which were constructed in Section~\ref{S2-subsection} and will
be studied further in Section~\ref{S2-section}.
Recall that in the field theory, after setting $x_4=0$, the couplings
to the scalars $\Phi^i$ can be written in vector notations
as (\ref{S2-oneforms})
\be
\label{cross}
\frac{1}{2}\,\vec\sigma^R=\vec x\times d\vec x\,,
\ee
An interesting way to think of (\ref{cross}) is as
\be
\label{cross2}
\frac{1}{2}
\sigma^R_i=J^i_{\ j}d x^j\,,
\ee
where $J$ is the almost complex structure of unit
2-sphere (\ref{acss2}). This almost complex
structure appears then very naturally in the definition of
these Wilson loops.

The dual string solutions in the bulk live in the subspace
$AdS_3\times S^2\subset{\cal X}$ gotten by restricting to $x_4=0$.
This clearly implies that on the world-sheet also $\partial_\alpha
x^4=0$, and one of the pseudo-holomorphic equations (\ref{basic})
becomes \be
y^i\partial_{\alpha}x^i+x^i\partial_{\alpha}y^i=0,\qquad
i=1,2,3\,. \ee This can be easily integrated to a constant
\be\label{twist} x_1\, y_1+x_2\, y_2+x_3\, y_3=C\,. \ee Hence the
strings are restricted to live inside a four-dimensional subspace
of $AdS_3\times S^2$ given by this constraint.

The remaining equations in (\ref{basic}) can be repackaged in
terms of the following almost complex structure
\be\label{acsreduced}
{\cal J}=\left(%
\begin{array}{cc}
  z^2\left(%
\begin{array}{ccc}
  0 & y_3 & -y_2 \\
  -y_3 & 0 & y_1 \\
  y_2 & -y_1 & 0 \\
\end{array}%
\right) & z^2\left(%
\begin{array}{ccc}
  0 & -x_3 & x_2 \\
  x_3 & 0 & -x_1 \\
  -x_2 & x_1 & 0 \\
\end{array}%
\right) \\
  z^{-2}\left(%
\begin{array}{ccc}
  0 & -x_3 & x_2 \\
  x_3 & 0 & -x_1 \\
  -x_2 & x_1 & 0 \\
\end{array}%
\right) & z^2\left(%
\begin{array}{ccc}
  0 & -y_3 & y_2 \\
  y_3 & 0 & -y_1 \\
  -y_2 & y_1 & 0 \\
\end{array}%
\right) \\
\end{array}%
\right) \ee  which should be thought as defined on the
four-dimensional subspace of $AdS_3\times S^2$ given by
(\ref{twist}).

Note that all the sub-blocks of the almost complex structure
(\ref{acsreduced}) are proportional to the almost complex
structure of $S^2$ (\ref{acss2}). Therefore this construction
naturally extends the map from the gauge couplings to the scalars
(\ref{cross2}), (\ref{duality}) to the bulk of $AdS_3\times S^2$.

For some examples in this $S^2$ sub-sector the explicit string solutions
have been written down explicitly and are collected in
Appendix~\ref{solutions-appendix}.
These solutions are dual to the latitude and two
longitudes Wilson loops discussed in Section~\ref{latitude-subsection}
and Section~\ref{longitudes-subsection}. Using them we can explicitly
test the validity of the ${\cal J}$-equations. Translating from
polar and spherical coordinates, the solution (\ref{lat-solution}) is
\begin{equation}
\label{latpoinc}
\begin{gathered}
x_1=\frac{\tanh{\sigma_0}\cos{\tau}}{\cosh{\sigma}},\qquad
x_2=\frac{\tanh{\sigma_0}\sin{\tau}}{\cosh{\sigma}},\qquad
x_3=\frac{1}{\cosh\sigma_0}\,,\qquad
z=\tanh{\sigma_0}\tanh{\sigma}\,,
\\
y_{1}=-\frac{\cos{\tau}}
{z\cosh(\sigma_0\pm\sigma)}\,,\qquad
y_{2}=-\frac{\sin{\tau}}
{z\cosh(\sigma_0\pm\sigma)}\,,\qquad
y_{3}=\frac{\tanh(\sigma_0\pm\sigma)}{z}\,.
\end{gathered}
\end{equation}
where the $\pm$ sign depends on whether the string wraps over the
north or the south poles.

It is immediate to check that this solution
satisfies $x^2+z^2=1$ and that
$x_1y_1+x_2 y_2+x_3y_3$ is a constant (\ref{twist}).
It is also not difficult to check that it satisfies the
${\cal J}$-equations.

Before going to the two-longitudes solution we recall (see
Section~\ref{longitudes-subsection} and
Appendix~\ref{longitudes-sol-appendix}) that it is related by a
stereographic projection to the cusp solution on the plane. This
solution has vanishing regularized action and is therefore
expected to be solution of the pseudo-holomorphic equation
associated to (\ref{Jzar}) as we now verify. For convenience we
write the metric of the relevant subspace of $AdS_5\times S^5$ as
\be \frac{1}{y^2} \left(dx_1^2+dx_2^2\right)+y^2
\left(dy_1^2+dy_2^2\right) \ee so that the pseudo-holomorphicity
condition becomes \bea \label{long1}
\partial_\alpha x^\mu-
y^2\,\sqrt{g}\,
\epsilon_{\alpha\beta}\partial^\beta y^m \delta^\mu_m=0,\qquad
\mu=1,2\,,\qquad m=1,2.
\eea
In these coordinates the cusp solution found in
Appendix~\ref{longitudes-sol-appendix} reads%
\footnote{This solution describes only half the world-sheet, the other half
is a mirror image of it and all the ensuing statements apply to it too.}
\bea
x_1&=&r\cos{\phi}(v),\qquad
x_2=r\sin{\phi}(v),\\
y_1&=&\frac{\cos{\varphi(v)}}{rv}\,,\qquad
y_2=\frac{\sin{\varphi(v)}}{rv}\,,
\label{sign}
\eea
where $r$ and $v$ are world-sheet coordinates (not in the conformal gauge)
and
\bea
\phi&=&\arcsin\frac{v}{p}- \frac{1}{\sqrt{1+p^2}}
\arcsin\sqrt{\frac{1+1/p^2}{1+1/v^2}}\,,\\
\varphi&=&\frac{1}{\sqrt{1+p^2}}
\arcsin\sqrt{\frac{1+1/p^2}{1+1/v^2}}\,.
\label{phivarphi}
\eea
Calculating the induced world-sheet metric, one finds
\bea
g_{rr}&=&\frac{1+v^2}{r^2 v^2},\qquad
\,g_{rv}=\frac{1}{rv},\qquad
g_{vv}=\frac{p^2(1+v^2)-v^4}{v^2(p^2-v^2)(1+v^2)},\\
\sqrt{g}&=&\frac{p}{r v^2\sqrt{p^2-v^2}} \,.
\label{ws-metric}
\eea
With these expression
one can check that the supersymmetric cusp solution indeed satisfies
(\ref{long1}).

Now we are ready to move over to the two-longitudes solution,
which is related to the cusp solution by a coordinate change (a conformal
transformation on the boundary).
In Appendix~\ref{longitudes-sol-appendix} it is written
in global coordinates and mapping them to the Poincar\'e patch we have
\begin{equation}
\label{longpoinc}
\begin{gathered}
x_1=\frac{2r}{1+r^2+r^2v^2}\cos{\phi}\,,
\qquad x_2=\frac{2r}{1+r^2+r^2v^2}\sin{\phi}\,,\qquad
x_3=\frac{r^2+r^2v^2-1}{{1+r^2+r^2v^2}}\,,\\
y_1=\frac{\sin{\varphi}}{z}\,,\qquad
y_2=\frac{\cos{\varphi}}{z}\,,\qquad y_3=0\,,\qquad z=\frac{2r
v}{1+r^2+r^2v^2}\,,
\end{gathered}
\end{equation}
with the same $\phi(v)$ and $\varphi(v)$ as before (\ref{phivarphi}).

As for the latitude solution, for this solution too it is clear that
$x^2+z^2=1$ and that
$x_1y_1+x_2 y_2+x_3y_3$ is a constant (\ref{twist}).
Using the same expressions for the world-sheet metric (\ref{ws-metric})
we can also check that it satisfies the
${\cal J}$-equations.

As discussed in Section~\ref{calib section}, the string solutions dual
to the Wilson loops can be interpreted as (generalized)
calibrations. As such their world-sheet area can be
computed by the integral of the pull-back of ${\cal J}$ to the
world-sheet. Using (\ref{latpoinc}) and (\ref{longpoinc}) it is
easy to verify explicitly this fact for the latitude and two
longitudes loops, for which we obtain respectively
\be
\int\cJ=
\int
d\sigma\, d\tau\left(\frac{1}{\sinh^2\sigma}+
\frac{1}{\cosh^2(\sigma+\sigma_0)}\right),
\ee
and
\be
\int\cJ=
\int dr\, dv\,\frac{p}{r v^2\sqrt{p^2-v^2}}\,.
\ee
These results are in agreement with
the expected (un-regularized) world-sheet area for these solutions.
To obtain the regularized area we need to subtract the boundary
term contribution from $\int{\cal J}$. The correct regularized
area is then obtained from integrating ${\cal J}_0$ (\ref{j0}),
which yields for the latitude and two longitudes respectively
\bea
\int\cJ_0
&=&\int d\tau\int_0^\infty d\sigma
\left(-\frac{1}{\cosh^2\sigma}
+\frac{1}{\cosh^2(\sigma+\sigma_0)}\right)=-2\pi\sin\theta_0\,,\\
\int\cJ_0
&=&2\int_0^p dv \int_0^\infty dr
\frac{-4 pr}{\sqrt{p^2-v^2}(1+r^2(1+v^2))^2}=
-2\frac{\pi p}{\sqrt{1+p^2}}\,.
\eea
The factor $2$ in the second line comes from accounting of the two
branches of the two-longitudes solution. These results are in
agreement with those obtained by different methods in Appendix
\ref{solutions-appendix}.


\section{Loops on a great $S^2$ and 2d Yang-Mills theory}
\label{S2-section}

In the present section we focus on loops defined on the great $S^2$
presented above in Section~\ref{S2-subsection}. We will provide
some evidence, expanding on the discussion in \cite{Drukker:2007yx},
that these loops are actually
equivalent to the usual, non-supersymmetric Wilson loops of
Yang-Mills theory on a 2-sphere in the Wu-Mandelstam-Leibbrandt
(WML) prescription
\cite{Wu:1977hi,Mandelstam:1982cb,Leibbrandt:1983pj}.

We shall start by analyzing the structure of the combined
``gauge + scalar'' propagator in Feynman gauge on the sphere and
we shall prove that it effectively reduces to the propagator of pure
2d Yang-Mills theory in the generalized  Feynman gauge with gauge
parameter $\xi=-1$ and with the WML prescription to regularize the
poles. The equivalence of the propagators in the two theories leads to
the agreement between the leading terms in the perturbative calculation.
In some examples, where there is a conjectured matrix-model reduction
of the perturbative expansion this agreement extends to the full series.
Furthermore in all the examples where we have explicit solutions to the
string equations describing those loops in $AdS$, the result of that
calculation agrees with the strong coupling expansion of the two dimensional
theory.

We should mention, however, that we have not been able to
substantiate this correspondence beyond the leading order calculation
and those examples, in particular we have not been able to compute
interacting graphs for generic loops.  It is then conceivable that the
two dimensional theory
describing those loops might be more complicated, with the same kinetic term
as YM, but with different (potentially also non-local)
interactions.

If this correspondence holds, it would be one of those miracles
of $\cN=4$ SYM, where there seems
to be a ``consistent truncation'' to the sphere and we
can simply ignore all the fields away from it. The other remarkable fact of this correspondence is that YM in 2d is invariant under area preserving diffeomorphisms. So a
subsector of the superconformal theory is invariant under all transformations
which change angles but keep areas constant. One interesting direction to investigate would be then to
find out if those properties manifest themselves in a deeper way in the entire
theory beyond this subsector.

\subsection{Perturbative expansion}

Consider a loop (\ref{susy-loop}) restricted to a unit $S^2$
(defined by $x_4=0$), where the scalar coupling reduces to
$\sigma^R_i=2 \varepsilon_{ijk}x^j\,d x^k$.  Expanding the
exponent to second order in the fields and computing the expectation
value will then give the following contractions of the gauge fields
and the scalars
\begin{equation}
\vev{W}\simeq 1-\frac{1}{2N}\Tr {\cal P}\int dx^i\,dy^j
\left[\vev{A_i(x)\,A_j(y)}- \varepsilon_{i k l }\varepsilon_{j m n
}\, x^k y^m \vev{\Phi^l(x)\,\Phi^n(y)}\right]\,.
\end{equation}
In the Feynman gauge, where the propagators are
\begin{equation}
\vev{A_i^a(x)\,A_j^b(y)}=\frac{g_{4d}^2}{4\pi^2}
\frac{\delta^{ab}\, g_{ij}}{(x-y)^2}\, , \qquad
\vev{\Phi^{aI}(x)\,\Phi^{bJ}(y)}=\frac{g_{4d}^2}{4\pi^2}
\frac{\delta^{ab}\, \delta^{IJ}}{(x-y)^2}\,,
\end{equation}
and using that $\varepsilon_{ikl}\varepsilon_{jml}
=\delta_{ij}\delta_{km}-\delta_{im}\delta_{jk}$,
we find (choosing a definite ordering of the loop parameters)
\begin{equation}
\vev{W}\simeq 1-\frac{g_{4d}^2 N}{8\pi^2}\oint_{s\ge t} ds \,
dt\,  \dot x^i(s)\,\dot y^j(t)\,
\left(\frac{1}{2}g_{ij} -\frac{(x-y)_i
(x-y)_j}{(x-y)^2}\right)\,. \label{eff-prop}
\end{equation}
Here we have also used that $x^2=y^2=1$ (and consequently
$\dot x^i\,  x_i=\dot y^i\, y_i=0$), and we have normalized
the $SU(N)$ generators as $\Tr(T^aT^b)=\delta^{ab}/2$. The
super-Yang-Mills coupling constant $g_{YM}$ has been relabeled $g_{4d}$
to distinguish it from the two-dimensional coupling $g_{2d}$ that will
appear in the following.

Notice that the combined ``gauge + scalar'' propagator in the expression
above is not generically a constant, as was the case for the 1/2 BPS
circle,  a fact which led to the identification of that operator with the
zero-dimensional Gaussian matrix model of
\cite{Erickson:2000af,Drukker:2000rr}.
But still, instead of having mass-dimension 2,
as expected in a four-dimensional theory it is dimensionless.
This is the first indication
that this effective propagator may serve as a vector propagator in two
dimensions.

\subsubsection{Near-flat loops}
\label{near-flat-subsection}
As a first step toward making contact with the propagator of
Yang-Mills theory on a 2-sphere, we start with the easier case of
small loops near the north pole of the $S^2$,  $x_3\simeq1$.
These loops live on an almost flat surface and, as discussed in
Section~\ref{Infinitesimal loops}, in the infinitesimal limit,
one recovers the construction of
\cite{Zarembo:2002an}. We may approximate
\begin{equation}
x_i=
\left(x_1,\,x_2,\,\sqrt{1-x_1^2-x_2^2}\right)
\simeq\left(x_1,\,x_2,\,1-\frac{x_1^2+x_2^2}{2}\right)\,.
\end{equation}
For the derivatives with respect to the loop parameter one has
\begin{equation}
\dot x_i
\simeq
\left(\dot x_1,\,\dot x_2,\,-x_1\dot x_1-x_2\dot x_2\right)\, ,
\end{equation}
while the distance is unmodified to leading order
\begin{equation}
(x-y)^i(x-y)_i \simeq (x-y)^r(x-y)_r\,,
\end{equation}
where now Latin indices from the end of the alphabet ($r,\,s,\ldots$)
run only over the directions 1 and 2.

Since it is always contracted with the tangent vectors, we may simplify
the propagator appearing in (\ref{eff-prop}) to
\begin{equation}
\Delta_{ij}^{ab}(x-y)=\frac{g_{4d}^2\delta^{ab}}{4\pi^2}\left(
\frac{1}{2}g_{ij}+\frac{y_i\, x_j}{(x-y)^2}\right)\,.
\end{equation}
Looking at $\dot x^i\dot y^j$ contracted with this expression
one obtains to quadratic order (we omit the overall coefficient with the coupling constant)
\begin{equation}
\begin{aligned}
\dot x^i\dot y^j\Delta_{ij}
& \simeq
\dot x^r\dot y^s\left(\frac{1}{2}\delta_{rs}+
\frac{y_r \,x_s}{(x-y)^2}\right)
-\frac{\dot x^r(y^s\dot y^s)y^r}{(x-y)^2}
-\frac{(x^r\dot x^r)\dot y^s x^s}{(x-y)^2}
+\frac{(x^r\dot x^r)(y^s\dot y^s)}{(x-y)^2}\\
&=\dot x^r\dot y^s\left(\frac{1}{2}\delta_{rs}-
\frac{(x-y)_r(x-y)_s}{(x-y)^2}\right)\,.
\label{2dprop0}
\end{aligned}
\end{equation}
While this last expression looks very similar to the propagator in  (\ref{eff-prop}), it is
completely different. Here everything is written in terms of 2d vectors
and one cannot drop the $\dot x^r\, x_r$ and $\dot y^r\, y_s$ terms, since they are no longer
zero for a general 2d curve.

We want now to analyze
\begin{equation}
\Delta^{ab}_{rs}(x-y)\equiv \frac{g^2_{4d}\delta^{ab}}{4\pi^2}
\left(\frac{1}{2}\delta_{rs}-
\frac{(x-y)_r(x-y)_s}{(x-y)^2}\right)\label{2dprop}
\end{equation}
in more detail. A simple proof that it can really be interpreted as a
propagator consists in checking that it is annihilated by an appropriate
two-dimensional kinetic operator. It is easy to verify that
${\cal D}^{rs}=-\delta^{rs}\partial^2+2\partial^r\partial^s$
does indeed the job. This is a Laplacian in generalized Feynman gauge
with gauge parameter $\xi=-1$.
The full gauge-fixed Euclidean action in this gauge reads
\bea
L=\frac{1}{g_{2d}^2}\left[\frac{1}{4}
\left(F_{rs}^a\right)^2
-\frac{1}{2}\left(\partial_r A^{a,r}\right)^2
+\partial_r b^a\left(D^r c\right)^a\right]\, ,
\label{actionxi}
\eea
where
\be
F^a_{rs}=\partial_{[r}A^a_{s]}+f^{abc}A^b_r\,A^c_s\, , \qquad \left(D_r c\right)^a=\partial_r c^a+f^{abc}A^b_r\, c^c\, .
\ee

It is instructive to present also an alternative proof, based on the use of Maxwell's equations
\begin{equation}
\partial_r F^{rs}=0\,.
\end{equation}
Here $F^{rs}$ is an abelian field strength which in two dimensions has only one component, $F_{12}$, and Maxwell's equations imply that it is a constant.

If equation (\ref{2dprop}) is a legitimate propagator, then the two-dimensional gauge field can be expressed as (here we suppress the color indices)
\begin{equation}
A_r (x)=\int dy \, \Delta_{rs}(x-y)\, J^s(y)\, ,
\end{equation}
where the current $J^s(y)$ can be taken to be  localized on the loop  so that
\begin{equation}
A_r(x)=\oint ds\, \Delta_{rs}(x-y)\, \dot y^s(s)\, .
\end{equation}
Differentiating this expression one finds the corresponding field strength
\begin{equation}
F_{rs}(x)=\partial_{[r}A_{s]}(x)=-\frac{g_{4d}^2}{4\pi^2}\oint ds\,
\frac{\dot y_{r}(x-y)_s-\dot y_{s}(x-y)_r}{(x-y)^2}\,.
\end{equation}
Using the complex variable $z=x_1-y_1+i(x_2-y_2)$, this becomes
\begin{equation}
F_{12}(x)=i\frac{g_{4d}^2}{4\pi^2}\oint \frac{dz}{z}\,,\label{F12z}
\end{equation}
which is $-g_{4d}^2/2\pi$ if the source surrounds $x$ and vanishes otherwise. Then
$F_{12}$ is constant in patches and (\ref{2dprop}) is indeed a propagator.

Before moving on to the $S^2$ case we pause for a moment to notice that the expectation value of a loop will be a function of the loop's area. Consider for example a small circle  with radius $r$ sitting at the north pole. The propagator (\ref{2dprop}) does not depend on the radius of the circle but the tangent vectors $\dot x^r \dot y^s$ do, so that the final result will scale as $r^2$. More precisely
\begin{equation}
\oint ds\, dt\, \dot x^r(s) \dot y^s(t)\Delta_{rs}(x-y)=-\frac{1}{2}g_{4d}^2 r^2=-\frac{g_{4d}^2}{2\pi} {\cal A}_1\, ,
\label{area1}
\end{equation}
where ${\cal A}_1$ is the area of the loop. This result can be generalized to a loop of arbitrary shape $\cC$ by using (\ref{F12z})
\begin{equation}
\oint_\cC ds\,dt\,  \dot x^{r}(s)\dot y^s(t)\Delta_{rs}\left(x-y\right)=\oint_\cC ds \, \dot x^{r}(s)A_{r} (x)=\int_{\Sigma_1}F_{12}=-\frac{g_{4d}^2}{2\pi}\cA_1\,,
\end{equation}
where $\Sigma_1$ is the surface enclosed by the loop.

\subsubsection{Generic loops on $S^2$}

We now consider generic loops extending over the whole sphere. To see
that the expression in (\ref{eff-prop}) is a vector propagator on $S^2$
we change coordinates and parameterize the sphere in terms of complex
coordinates $z$ and $\bar z$ as
\begin{equation}
x_i = \frac{1}{1+z\bar z}\left(z+\bar z,\, -i(z-\bar
z),\,1-z\bar z\right)\,. \label{complex-coords}
\end{equation}
In these coordinates, the $S^2$ metric takes the standard
Fubini-Study form
\begin{equation}
ds^2 = \frac{4 \,dz\, d\bar z}{(1+ z \bar z)^2}\,. \label{metric}
\end{equation}

From the near-flat case we expect the correct gauge choice to be the generalized Feynman
gauge with gauge
parameter $\xi=-1$.  The Yang-Mills term in the action (\ref{actionxi})
becomes for the theory on the sphere
\begin{equation}
L=\frac{\sqrt{g}}{g_{2d}^2} \left[\frac{1}{4}(F_{ij}^a)^2
-\frac{1}{2}(\nabla^iA^a_i)^2\right] = -\frac{\sqrt{g}}{g_{2d}^2}
(g^{z \bar z})^2\left[ (\nabla_zA_{\bar z}^a)^2+(\nabla_{\bar
z}A_z^a)^2\right]\, ,
\end{equation}
where in the last equality we have ignored interaction terms, and
the covariant derivatives are taken with respect to the metric
(\ref{metric}). A simple calculation shows that the propagators
\begin{equation}
\begin{aligned}
\Delta^{ab}_{zz}(z,w)&
=\delta^{ab}\frac{g_{2d}^2}{\pi}\frac{1}{(1+ z \bar z)}\frac{1}{(1+w \bar w)}\frac{\bar z-\bar w}{z-w}\, ,\\
\Delta^{ab}_{\bar z\bar z}(z,w)& =\delta^{ab}\frac{g_{2d}^2}{\pi}
\frac{1}{(1+ z \bar z)}\frac{1}{(1+w \bar w)} \frac{z-w}{\bar
z-\bar w}\, ,
\label{props}
\end{aligned}
\end{equation}
satisfy
\begin{equation}
\frac{2}{g_{2d}^2}(g^{z \bar z})^2 \nabla^2_{\bar z}
\Delta^{ab}_{zz}(z,w) =\delta^{ab}
\frac{1}{\sqrt{g}}\delta^2(z-w)\, ,
\end{equation}
and similarly for $\Delta_{\bar z\bar z}$.
By doing the change of variables to the complex coordinates
(\ref{complex-coords}), one can then see that the effective
propagator in (\ref{eff-prop}) agrees with the 2d vector
propagators (\ref{props}) when the 2d and 4d couplings are related
by
\begin{equation}
\label{g2 g4}
g_{2d}^2 = -\frac{g_{4d}^2}{4\pi}\,.
\end{equation}
Notice that $g_{2d}^2$ has 2 dimensions of mass, as becomes obvious after reinserting the appropriate powers of the radius of the $S^2$ in the formula above.

The alternative argument based on the Maxwell's equations can also be repeated in this instance.
Given a source along the curve $y$ and using the effective propagator on $S^2$,
the gauge field at $x$ is
\begin{equation}
A_i=\frac{g_{2d}^2}{\pi}\int dy^j
\left(\frac{1}{2}\delta_{ij}
-\frac{(x-y)_i (x-y)_j}{(x-y)^2}\right)\,,
\end{equation}
and the resulting field-strength, gotten by differentiation and
projection in the directions tangent to the sphere, is
\begin{equation}
F_{ij}=-\frac{g_{2d}^2}{\pi}\int ds\,
\frac{-\dot y_i y_j+\dot y_j y_i}{(x-y)^2}\,.
\end{equation}
The associated dual scalar $\tilde F = \frac{1}{2} \epsilon_{ijk} F_{ij} x_k$ reads
\begin{equation}
\tilde F=-\frac{g_{2d}^2}{\pi}\int ds\,
\frac{\varepsilon_{ijk}\,\dot y^i y^j x^k}{(x-y)^2}\,.
\end{equation}
To evaluate $\tilde{F}$ explicitly we define $\theta(s)$ to be the
angle between the points $x$ and $y$. Then the numerator is
proportional to the one-form normal to $d\theta$, which we label
by $d\phi$. This gives
\begin{equation}
\tilde F=\frac{g_{2d}^2}{\pi}\int d\phi\,
\frac{\sin^2\theta}{2(1-\cos\theta)} =\frac{g_{2d}^2}{\pi}\int
d\phi\,\cos^2\frac{\theta}{2}
=\frac{g_{2d}^2}{2\pi}\int_{\Sigma_2} d\theta\,d\phi\sin\theta
=2 g_{2d}^2\frac{\cA_2}{\cA}\,,
\end{equation}
where $\cA_2$ is the area of the part of the sphere enclosed by
the loop and not including $x$ and $\cA$ the total area. Clearly
this is a constant unless $x$ crosses the loop. Then it is
simple to evaluate the Wilson loop at the quadratic order using
Stokes' theorem for the $x$ integral in (\ref{eff-prop}). We get
\begin{equation}
\vev{W}=1-\frac{N}{4}\int_{\Sigma_1}\tilde F
+O(g_{2d}^4)
=1-g_{2d}^2N\,\frac{\cA_1\cA_2}{2\cA}+O(g_{2d}^4)\,,
\label{g2-result}
\end{equation}
and the result is the product of the areas of the two parts of the sphere
separated by the loop and it clearly does not depend on the order
of the $y$ and $x$ integrals.

\begin{figure}
\begin{center}
\includegraphics[width=50mm]{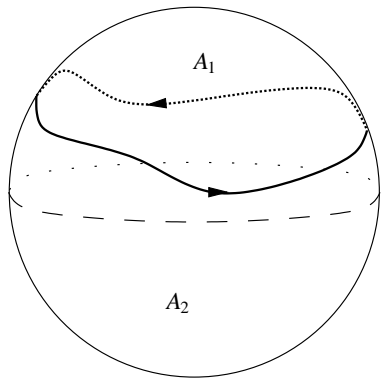}
\parbox{13cm}{
\caption{ An arbitrary curve on $S^2$ divides it into two
surfaces, one with area $\cA_1$ and the other with area $\cA_2$.
In all the calculations that we did the expectation value of the
Wilson loop turns out to be a function only of the product of
those two areas. \label{s2-fig}}}
\end{center}
\end{figure}

We were unfortunately not able to calculate higher-order graphs
for loops of arbitrary shape, neither in four dimensions nor
explicitly in two. Note that as opposed to the light-cone gauge,
the preferred gauge choice in two dimensions, in our generalized
Feynman gauge there are interaction vertices and the ghosts do not
decouple, so the calculation is non-trivial. As an example of this
complexity, we report  in Appendix~\ref{WML-appendix} the
computation of the interacting graphs at order $\lambda^2$ in the
$\xi=-1$ gauge, in the hope that this could be matched at some
intermediate stage with a similar calculation in four dimensions.
We were not able to find such a matching for a general curve, but
were able to carry it through in the case of a circular loops. We
find that also in this gauge, as expected from gauge invariance,
the interacting graphs cancel, but this cancellation is achieved in
a very non-trivial way.

It would of course be extremely useful  to better understand the
relation between 4d and 2d interactions, for example it would be
nice to study 4d gauge choices such that the combined gauge and scalar
propagators reduce to the light-cone gauge propagator in 2d, where
computations are trivial. Then one could hope that in such a
4d gauge it would be possible to show that by integrating the interacting
vertices over the directions transverse to the sphere, they cancel,
as they do in the corresponding gauge in 2d.

In any case, two-dimensional Yang-Mills is a soluble theory \cite{Migdal:1975zg,
Rusakov:1990rs}, so we can use known results (derived by other
methods) and compare them to some results in four dimensions,
including some strong coupling results from the $AdS$ dual of
$\cN=4$ SYM, which we will do in the next subsection.

The above perturbative calculation (\ref{g2-result}) of the Wilson loop in two dimensions is
very similar to the one performed by Staudacher and Krauth in
\cite{Staudacher:1997kn} on $\bR^2$ in light-cone gauge. The important part in their calculation  is not the choice of gauge, but the
choice of regularization prescription of a pole in the derivation
of the configuration-space propagator. The one they used, which
can be applied also in Euclidean signature, was
proposed by Wu, Mandelstam, and Leibbrandt
(WML) \cite{Wu:1977hi,Mandelstam:1982cb,Leibbrandt:1983pj}.

Going back for a moment to the near-flat case and changing coordinates
from $x_1, x_2$ to  $x_\pm=x_1\mp i x_2$, it is easy to see that our
(\ref{2dprop}) has the exact same structure of the WML  propagator
on the plane as in \cite{Staudacher:1997kn}, up to a factor of 2
\begin{equation}
\left< A_+(x) A_+(y)\right>\propto \frac{x_+-y_+}{x_- - y_-}\, .
\end{equation}
In our gauge, with $\xi=-1$, there is a propagator also for
$A_-$ (but no mixed term).  In the light-cone gauge one sets $A_-=0$
and the $A_+$ propagator is double ours.
The same applies for the sphere, where one may take $A_{\bar z}=0$
as the light-cone gauge and
then, using the same prescription, the propagator for $A_z$ would
be double the one in (\ref{props}).

Staudacher and Krauth were able to sum up all the
ladders and find that the Wilson loop is given by
\begin{equation}
\vev{W}=\frac{1}{N}L_{N-1}^1\left(g_{2d}^2\cA_1\right)
\exp\left[-\frac{g_{2d}^2\cA_1}{2}\right]\,,
\label{2d-result}
\end{equation}
where $L_{N-1}^1$ is a Laguerre polynomial and $\cA_1$ is the area
enclosed by the loop. This is equal to the expectation value of a
Wilson loop in the Gaussian Hermitian matrix model
(\ref{mat_model}), after a rescaling of the coupling
constant\footnote{This result is valid for $U(N)$ gauge group. The
exact formula for $SU(N)$ can be easily deduced from this one, see
\cite{Staudacher:1997kn}.}. This expression has an obvious
generalization to $S^2$ with the simple replacement $\cA_1\to
\cA_1\cA_2/\cA$, where the combination of the areas is the same as
appeared in (\ref{g2-result}).

The reader may be puzzled by those formulas, since they do not agree
with the exact solution of YM in two dimensions 
\cite{Douglas:1993iia,Gross:1994mr}. This confusion was
resolved by Bassetto and Griguolo \cite{Bassetto:1998sr}, who showed
that (\ref{2d-result}) may be extracted from
the exact result by restricting to the zero instanton sector following the
expansion of \cite{Witten:1992xu} (see also  \cite{Gross:1994mr}). 
It was therefore concluded that
the perturbative calculation of \cite{Staudacher:1997kn}, using the
light-cone gauge and the WML prescription
for performing the momentum integrals does not capture non-perturbative
effects.

The two dimensional propagator we found is thus not in the same gauge,
but it also is defined by the WML prescription. Since we expect
the result not to depend on gauge, we conclude that the result of
the perturbative 2-dimensional YM sum that our four-dimensional
Wilson loops seem to point to is given by
\begin{equation}
\vev{W}
=\frac{1}{N}L_{N-1}^1\left(-g_{4d}^2\,\frac{\cA_1\cA_2}{\cA^2}\right)
\exp\left[\frac{g_{4d}^2}{2}\,\frac{\cA_1\cA_2}{\cA^2}\right]\,.
\label{4d-result}
\end{equation}
The expansion of this
expression to order $g_{4d}^2$ agrees with the aforementioned result
(\ref{g2-result}). In the next subsection we will provide further
evidence that this expression correctly captures the
Wilson loops in four dimensions.

Note that in relating our observables in 4-dimensions and those in 2d, see
(\ref{g2 g4}), the real 4-dimensional coupling is, interestingly,
matched with an imaginary one in 2-dimensional. This could be
associated to the fact that the supersymmetric loops in Euclidean
$\cN=4$ SYM (\ref{Wilson-loop}) have an imaginary scalar coupling and
are non-unitary observables. In many cases their expectation values are
greater than 1 (which is manifested in the dual $AdS$ by negative
action) and this seems to be represented in the 2-dimensional model by
this change in sign of the square of the coupling.

\subsection{Examples and strong coupling checks}

Beyond the agreement at leading order in perturbation theory, which
led us to propose that Wilson loops on $S^2$ may be described by
2-dimensional YM, in this section we test this hypothesis further.
We compare the result of some perturbative and some strong coupling
calculations of specific operators in four dimensions with the exact
(perturbative) result in two dimensions (\ref{4d-result}).

To compare with results from $AdS$ we will need the asymptotic
behavior of (\ref{4d-result}) at large $N$ and large $g_{4d}^2 N$.
In this limit it reduces to
\begin{equation}
\vev{W}
\simeq\frac{\cA}{\sqrt{g_{4d}^2N\cA_1\cA_2}}
I_1\left(\frac{2\sqrt{g_{4d}^2N \cA_1\cA_2}}{\cA}\right)
\simeq\exp\left(\frac{2\sqrt{g_{4d}^2N \cA_1\cA_2}}{\cA}\right)\,,
\label{bessel}
\end{equation}
with $I_1$ a modified Bessel function of the first kind.

\subsubsection{Latitude}

Let us start by considering the circle at latitude $\theta_0$ discussed
above in Section~\ref{latitude-subsection}. This loop was studied in
\cite{Drukker:2006ga}, where it was shown that its combined
gauge+scalar propagator  is the same as the propagator of the 1/2 BPS circle modulo a rescaling of the coupling constant, $g_{4d}^2\rightarrow g_{4d}^2\sin^2\theta_0$. Assuming the vanishing of interacting graphs at all orders in perturbation theory (as is usually also assumed for the 1/2 BPS circle \cite{Erickson:2000af,Drukker:2000rr}), one can then resum all the ladders with a matrix model computation and show that the expectation value of the latitude is equal to (\ref{4d-result}) after the replacement
$\cA_1\cA_2/\cA^2\to\frac{1}{4}\sin^2\theta_0$. For a latitude
the areas of the patches bound by the curve are
\begin{equation}
\cA_1=2\pi (1-\cos\theta_0)\,,\qquad
\cA_2=2\pi (1+\cos\theta_0)\,,
\end{equation}
so indeed $\cA_1\cA_2=\frac{1}{4}\cA^2\sin^2\theta_0$, as claimed.

One can test this all-order result also from a string computation in $AdS_5\times S^5$ \cite{Drukker:2006ga}, from which one finds that the classical action of the string is ${\cal S}=-\sqrt{g^2_{4d}N}\sin\theta_0$, consistently with the strong coupling limit of the matrix model result%
\footnote{In string theory one also finds a second, unstable surface with ${\cal S}=+\sqrt{g^2_{4d}N}\sin\theta_0$, which matches another saddle point of the matrix model.}
(\ref{bessel}).
Finally, a further check can be obtained for loops in high dimensional
symmetric representations of the gauge group \cite{Drukker:2006zk}:
The loop is calculated in this case using a D3-brane rather than a fundamental string and, again, the resulting action agrees with the matrix model result, including all $1/N$ corrections at large $g^2_{4d}N$.

\subsubsection{Two longitudes}
\label{long-string-appendix}

The second example we consider are the two longitudes discussed
in Section~\ref{longitudes-subsection}. In this case it is not
obvious {\it a priori} that there exists an all-order matrix model
computation, since the rungs connecting the two different arcs are
not constant.

For the two longitudes separated by an angle $\delta$ the areas of
the two patches are given by
\begin{equation}
\cA_1=2\delta\,,\qquad
\cA_2=2(2\pi-\delta)\,.
\end{equation}
And those factors then come into the one-loop expression
(\ref{g2-result})
\begin{equation}
\frac{g_{4d}^2 N}{8 \pi^2}\delta(2\pi-\delta)\,.\label{g2fin}
\end{equation}
This can also be verified by a direct integration of the
combined propagator along the loop.

This clearly agrees with the weak coupling expansion of (\ref{4d-result}),
as is true for all our supersymmetric loops on a great $S^2$,
but for the latitude loops we can also test this expression at strong
coupling, since we have explicit string solutions in $AdS_5\times S^5$.
Those are described in detail in
Appendix~\ref{longitudes-sol-appendix}, where it is found by a
stereographic projection to a cusp in the plane and then
calculated by generalizing \cite{Drukker:1999zq}.
The result for the classical action (\ref{longi-action}) reads
\begin{equation}
\cS=-\frac{\sqrt{g_{4d}^2N\delta(2\pi-\delta)}}{\pi}\,.
\end{equation}
Recalling that the expectation value of the Wilson loop is the exponent of minus the
classical action, we exactly recover equation (\ref{bessel}).

We see then that also in this case the perturbative and the strong coupling
results are related to the 1/2 BPS circle by a simple rescaling of the
coupling constant,
$g_{4d}^2\rightarrow g^2_{4d}\delta(2\pi-\delta)$.
This suggests that the expectation value of this loop may also be
captured by a matrix model, although the propagators are not
constant in this case.


\section{Discussion}

In this paper we have studied a family of supersymmetric Wilson loops
in $\cN=4$ SYM which were proposed in \cite{Drukker:2007dw}.
The construction assumes the loops are restricted to an $S^3$ submanifold
of space-time (or Euclidean space) and then for a curve of {\em
arbitrary} shape we give a prescription for the scalar couplings that
guarantees that the resulting loop is globally supersymmetric.
This idea is inspired by the supersymmetric loops which have
trivial expectation values \cite{Zarembo:2002an}, but our loops are
more interesting observables.

We proposed several different angles to study those loops. First we
analyzed their general properties, like the supersymmetry they preserve.
We studied the dual string surfaces in $AdS_5\times S^5$, and
concentrating on loops on $S^2$ we pointed out a possible connection
to YM theory in two dimensions. We also mentioned briefly the
connection to topologically twisted YM.

In the general analysis we
focused on certain subclasses of loops which have enlarged supersymmetry
and studied them in detail. One example is $1/2$ BPS---a great circle, a
few cases were $1/4$ BPS: The latitude line on $S^2$, two half-circles,
or the longitudes on $S^2$, and the ``parallel circles'' or Hopf fibers
on $S^3$. A general loop on $S^2$ preserves $1/8$ of the supersymmetries,
as do loops built on the base of the Hopf fibration.
Some special cases of $1/16$ BPS loops are the infinitesimal ones,
which reside in a limit where one recovers the
``trivial'' loops of \cite{Zarembo:2002an}. Another example that is $1/16$
BPS and where we found the string solutions are general toroidal loops.

This analysis shows the richness of these operators we have
constructed. One can focus on  subsectors with fewer
operators and more supersymmetry, which may simplify some
calculations, or one can go to the more general cases which are far
less restrictive but also more complicated. From an algebraic point of
view we found a myriad of different subalgebras of $PSU(2,2|4)$
preserved by the different subsectors: $OSp(1|2)$, $SU(1|2)$,
$OSp(1|2)^2$, $SU(1|2)^2$, $OSp(2|4)$, $SU(2|2)$ and $OSp(4^\star|4)$.
We have included an extensive analysis of those symmetries in
Section~\ref{example-sec} to facilitate future study of those
subsectors.

Our next angle was that of the dual string theory on $AdS_5\times
S^5$, where the Wilson loops (in the fundamental representation) are
described by fundamental strings and in our case are restricted
to live within an $AdS_4\times S^2$ subspace. For some of the
specific examples enumerated in Section~\ref{example-sec}
we have explicit solutions of the string equations of motion. We
gathered them all in Appendix~\ref{solutions-appendix}. While some of
those solutions were known before, most of them (the ``longitudes'',
the ``latitudes on the Hopf base'' and the ``toroidal loops'') are
new.

But beyond the explicit solutions in those special examples we
found some general properties satisfied by the strings describing
those loops (following similar ideas in \cite{Dymarsky:2006ve}).
First we found an almost complex structure on the $AdS_4\times
S^2$ subspace where the string solution lives. Its structure is
inspired by the supersymmetry properties of the loops and is a
generalization of the almost complex structure on $S^6$ (see
Appendix~\ref{S6app}). We then showed that a string that is
pseudo-holomorphic with respect to this almost complex structure
has the correct boundary conditions, preserves the right
supersymmetries and satisfies the $\sigma$-model equations of
motion. In the specific examples where we had explicit solutions
the strings are indeed pseudo-holomorphic and we are inclined to
believe that this condition will be satisfied in general, though
we do not have an existence proof.

Another approach at studying those loops was to find an analogous
theory with the same operators. This was inspired by the fact that
the circle seems to be captured by a 0-dimensional matrix model
\cite{Erickson:2000af,Drukker:2000rr}. We presented some
evidence that when the loops are restricted to a great $S^2$ and
preserve four supercharges they may be described by a perturbative
calculation in 2-dimensional bosonic YM on $S^2$. As with the
$AdS$ calculation mentioned in the previous paragraph, we do not
have a proof of this equivalence, but all the explicit checks that
we could make worked.

The checks include the ladder diagrams for {\em all} the loops on $S^2$
(in a certain gauge, see Appendix~\ref{WML-appendix}),
explicit string theory results for the ``latitude'' and ``longitudes'' examples
as well as an agreement with the 0-dimensional matrix model. A peculiar
fact is that the Wilson loops do not agree with the full result of YM in
2 dimensions, but rather to a perturbative sector excluding instanton
contributions \cite{Bassetto:1998sr} (the instantons of 2-dimensional
YM are abelian monopoles). This feature of the agreement might appear somewhat unnatural. On
one side in fact there is a perfectly defined set of operators of $\cN =4$ SYM, 
while on the other side the zero-instanton
sector of two-dimensional YM  is not clearly defined. This is because the instanton numbers in this theory are not topological quantities (the instantons are unstable and can unwind in the $U(N)$ space).\footnote{We thank David Gross for raising this issue.}
It would be then extremely interesting to
understand whether the full 2-dimensional result, including
instanton corrections is also related to such Wilson loops in some way.

A remarkable fact about this purported correspondence is that
2-dimensional YM is invariant under area-preserving diffeomorphisms.
So by restricting to a sphere of fixed radius and adding the scalar
couplings we found operators in $\cN=4$ theory whose expectation
value depends on certain areas on the sphere. We find this quite a
surprising result in a conformal theory.

One last approach to study our loops is through a topologically twisted
version of $\cN=4$ SYM. We presented the relevant twist, where three
of the six scalars become a triplet under the twisted Lorentz group and
the other three are singlets. The novel feature about our loops is that
they are not invariant under the usual supersymmetry generators, but
rather under a linear combination with the super-conformal ones. This
means that those operators are observables in the twisted theory where
the BRST charges are made out of those linear combinations. We have
not constructed this theory in any detail but we think it would be
interesting to do so. We did use this twisting to motivate the
string-theory construction in Section~\ref{J-section} and we also
expect it to be useful in trying to prove that those Wilson loops may
be calculated in terms of a lower-dimensional theory, like 2-dimensional
YM, or in proving invariance under area-preserving diffeomorphisms.

Beyond the operators studied in this paper (and the ones in
\cite{Zarembo:2002an}) we find it quite likely that there are other
supersymmetric Wilson loops. These non-local operators, as well as
surface operators (for example
\cite{Constable:2002xt,Gukov:2006jk,Buchbinder:2007ar})
and domain walls \cite{DeWolfe:2001pq} are much less studied than
local operators but they have very interesting properties.

While this is quite an extensive report on supersymmetric Wilson loops on
$S^3$ where we presented many new results, it is also satisfying to see
how many interesting questions were left unanswered. This is an indication
to us that we have touched on an interesting subsector of $\cN=4$ SYM
which is very rich, yet one where exact results are feasible.


\subsection*{Acknowledgments}

We are happy to thank C. Beasley, G. Bonelli, J. Gauntlett, L. Griguolo, S. Hartnoll, S. Itzhaki, S.
Kim, T. Okuda, P. Olesen, Y. Oz, J. Plefka, M. Staudacher, 
N. Suryanarayana, A. Tanzini, A. Tomasiello, and M. Wolf for
interesting discussions. N.D. would like to thank the University
of Barcelona, the Galileo Galilei Institute and Perimeter
Institute for their hospitality in the course of this work and the
INFN for partial financial support. The work of S.G. is supported
in part by the Center for the Fundamental Laws of Nature at
Harvard University and by NSF grants PHY-0244821 and DMS-0244464.
D.T. acknowledges the kind hospitality of the Michigan Center for
Theoretical Physics at the final stages of this project. D.T. is
supported in part by the Department of Energy under Contract
DE-FG02-91ER40618 and in part by the NSF grant 
PHY05-51164.

\newpage


\appendix


\section{Superconformal algebra}
\label{algebra_conv}
In this appendix we collect our conventions
for the $\cN=4$ superconformal algebra $PSU(2,2|4)$, following
\cite{Beisert:2004ry}. We denote by $J^{\alpha}_{\ \beta}$, $\bar
J^{\dot\alpha}_{\ \dot\beta}$ the generators of the $SU(2)_L
\times SU(2)_R$ Lorentz group, and by  $R^A_{\ B}$ the 15
generators of the $R$-symmetry group $SU(4)$. The remaining bosonic
generators are the translations $P_{\alpha\dot\alpha}$, the
special conformal transformations $K^{\alpha\dot\alpha}$ and the
dilatations $D$. Finally the 32 fermionic generators are the
Poincar\'e supersymmetries $Q^A_{\alpha}$, $\bar Q_{\dot\alpha A}$ and
the superconformal supersymmetries $S^{\alpha}_A$, $\bar
S^{\dot\alpha A}$.

The commutators of any generator with $J^{\alpha}_{\ \beta}$, $\bar
J^{\dot\alpha}_{\ \dot\beta}$ and $R^A_{\ B}$ are canonically
dictated by the index structure, while commutators with the
dilatation operator $D$ are given by $\big{[}D\,,\cal G \big{]} =
\mbox{dim}(\cal G)\,\cal G$, where $\mbox{dim}(\cal G)$ is the
dimension of the generator $\cal G$.

The remaining non-trivial commutators are
\begin{equation}
\begin{aligned}
&\big{\{} Q^A_{\alpha}\,, \bar Q_{\dot\alpha B} \big{\}}=
\delta^A_B P_{\alpha\dot\alpha}\,, \qquad \quad \big{\{}
S^{\alpha}_A \,,\bar S^{\dot\alpha B} \big{\}}
= \delta^B_A K^{\alpha\dot\alpha}\,, \\
&\big{[}K^{\alpha\dot\alpha}\,, Q^A_{\beta} \big{]}=
\delta^{\alpha}_{\beta}\bar S^{\dot\alpha A}\,, \qquad \quad ~
\big{[}K^{\alpha\dot\alpha}\,, \bar Q_{\dot\beta A} \big{]} =
\delta^{\dot\alpha}_{\dot\beta} S^{\alpha}_A\,, \\
&\big{[}P_{\alpha\dot\alpha}\,, S^{\beta}_{A} \big{]}= -
\delta^{\beta}_{\alpha}\bar Q_{\dot\alpha A}\,, \qquad \quad
\!\!\!~ \big{[}P_{\alpha\dot\alpha}\,, \bar S^{\dot\beta A}
\big{]} =
-\delta^{\dot\beta}_{\dot\alpha} Q_{\alpha}^A\,, \\
&\big{\{} Q^A_{\alpha} \,, S^{\beta}_{B} \big{\}} = \delta^A_B
J^{\beta}_{\,\,\alpha} +
\delta^{\beta}_{\alpha} R^A_{\ B} + \frac{1}{2} \delta^A_B \delta^{\beta}_{\alpha} D\,, \\
&\big{\{} \bar Q_{\dot\alpha A} \,, \bar S^{\dot\beta B} \big{\}}
= \delta^B_A \bar J^{\dot\beta}_{\,\,\dot\alpha} -
\delta^{\dot\beta}_{\dot\alpha} R^B_{\ A} + \frac{1}{2} \delta^B_A \delta^{\dot\beta}_{\dot\alpha} D\,, \\
&\big{[} K^{\alpha\dot\alpha}\,,
P_{\beta\dot\beta}\big{]}=\delta^{\dot\alpha}_{\dot\beta}
J^{\alpha}_{\ \beta} + \delta^{\alpha}_{\beta} \bar
J^{\dot\alpha}_{\ \dot\beta} + \delta^{\alpha}_{\beta}
\delta^{\dot\alpha}_{\dot\beta} D\,.
\end{aligned}
\label{bigalgebra1}
\end{equation}

For the analysis of the supersymmetries preserved by the various
Wilson loop operators discussed in the paper, it is natural to
consider the breaking of the $R$-symmetry group $SU(4) \rightarrow
SU(2)_A \times SU(2)_B$. Explicitly, we can split the $\bf 4$ and
$\bf \bar 4$ indices of $SU(4)$ as
\begin{equation}
{\cal G}^A \rightarrow {\cal G}^{\dot{a} a} \qquad \quad {\cal
G}_A \rightarrow {\cal G}_{\dot{a} a}\,,
\end{equation}
where $\dot a$ and $a$ are respectively $SU(2)_A$ and $SU(2)_B$
fundamental indices.

All $SU(2)$ indices can be raised/lowered by using the appropriate
epsilon tensor, for which we adopt the conventions
\begin{equation}
\label{epsconvention}
\begin{aligned}
&\varepsilon^{rs} = \left(\begin{array}{cc}
0&1\\
-1&0 \end{array} \right) \qquad \quad \varepsilon_{rs} =
\left(\begin{array}{cc}
0&-1\\
1&0 \end{array}  \right) \cr &{\cal G}^{r} = \varepsilon^{rs}
{\cal G}_s\,, \qquad \qquad \quad {\cal G}_r = \varepsilon_{rs}
{\cal G}^s\,,
\end{aligned}
\end{equation}
where the indices $r,s$ belong to either $SU(2)_L$, $SU(2)_R$,
$SU(2)_A$, or $SU(2)_B$.

The $R$-symmetry generators decompose under $SU(4) \rightarrow
SU(2)_A \times SU(2)_B$ as $\bf{15} \rightarrow (\bf{3},1) +
(1,\bf{3}) + (\bf{3},\bf{3})$. This can be explicitly written as
\begin{equation}
R^A_{\ B} \rightarrow R^{\dot a a}_{\ \ \dot b b} =
\frac{1}{2}\delta^a_b \dot T^{\dot a}_{\ \dot b} + \frac{1}{2}
\delta^{\dot a}_{\dot b} T^{a}_{\ b} + \frac{1}{2} M^{\dot a a}_{\
\ \dot b b} \label{SU4-break}
\end{equation}
where $\dot T^{\dot a}_{\ \dot b}$ and $T^{a}_{\  b}$ are
respectively the $SU(2)_A$ and $SU(2)_B$ generators, and the 9
generators in the $(\bf{3},\bf{3})$ are given by $M^{\dot a a}_{\
\ \dot b b}$, which is traceless in each pair of indices
\begin{equation}
\delta^{\dot b}_{\dot a} M^{\dot a a}_{\ \ \dot b b}=
\delta^{b}_{a} M^{\dot a a}_{\ \ \dot b b}=0\,.
\end{equation}
Inserting the decomposition (\ref{SU4-break}) in the $SU(4)$
algebra
\begin{equation}
\big{[} R^A_{\ B}\,, R^C_{\ D} \big{]} = \delta^A_D R^C_{\
B}-\delta^C_B R^A_{\ D}\,,
\label{SU4}
\end{equation}
and projecting onto singlets of $SU(2)_A$ and of $SU(2)_B$,
one can verify that $\dot T^{\dot a}_{\ \dot b}$ and $T^{a}_{\
b}$ satisfy $SU(2)$ commutation relations with standard
normalization
\begin{equation}
\big{[} \dot T^{\dot a}_{\ \dot b}\,, \dot T^{\dot c}_{\ \dot d}
\big{]} = \delta^{\dot a}_{\dot d} \dot T^{\dot c}_{\ \dot
b}-\delta^{\dot c}_{\dot b} \dot T^{\dot a}_{\ \dot d}\,, \qquad
\big{[} T^{a}_{\ b}\,, T^{ c}_{\  d} \big{]} = \delta^{ a}_{ d}
T^{ c}_{\  b}-\delta^{ c}_{ b} T^{ a}_{\  d}\,.
\end{equation}
One can also check that $\dot T^{\dot a}_{\ \dot b}$ and $T^a_{\
b}$ act on the supercharges according to canonical $SU(2)$
commutation rules. For example starting from
\begin{equation}
\big{[} R^A_{\ B}\,,Q^C_{\alpha} \big{]} = -\delta^C_B
Q^A_{\alpha} + \frac{1}{4} \delta^A_B Q^C_{\alpha}\,,
\end{equation}
the above decomposition (\ref{SU4-break}) can be seen to imply
\begin{equation}
\big{[} \dot T^{\dot a}_{\ \dot b}\,, Q^{\dot c c}_{\alpha}
\big{]} = -\delta_{\dot b}^{\dot c} Q^{\dot a c}_{\alpha} +
\frac{1}{2} \delta^{\dot a}_{ \dot b} Q^{\dot c
c}_{\alpha}\,,\qquad \big{[}  T^{ a}_{\  b}\,, Q^{\dot c
c}_{\alpha} \big{]} = -\delta_{ b}^{ c} Q^{\dot c a}_{\alpha} +
\frac{1}{2} \delta^{ a}_{  b} Q^{\dot c c}_{\alpha}\,,
\label{SU2-supcharges}
\end{equation}
and similarly for the other supercharges.

Commutators involving the $M^{\dot a a}_{\ \ \dot b b}$ may be
written more conveniently in the basis defined by
\begin{equation}
M^{\dot a a}_{\ \ \dot b b} = (\tau_{\dot{m}})^{\dot a}_{\ \dot b}
(\tau_m)^a_{\ b}\, M_{\dot{m}m}\,, \qquad
\dot T^{\dot a}_{\ \dot b} = (\tau_{\dot{m}})^{\dot a}_{\ \dot b}\,
\dot T_{\dot{m}}\,, \qquad
T^a_{\ b} = (\tau_m)^a_{\ b}\, T_m\,,
\label{pauli-basis}
\end{equation}
where $\dot{m}$, $m$ are indices in the $\bf 3$ of $SU(2)_A$ and
$SU(2)_B$ respectively, and $\tau_{\dot{m}}$, $\tau_m$ are Pauli matrices.
Projecting (\ref{SU4}) onto the $({\bf 3},\, {\bf 3})$ representation
of $SU(2)_A\times SU(2)_B$ under the decomposition
(\ref{SU4-break}), one can obtain the following commutation
relations
\begin{equation}
\begin{aligned}
&\big{[} \dot T_{\dot{m}} \,,M_{\dot{n}m} \big{]}
= i \varepsilon_{\dot{m}\dot{n}\dot{p}} M_{\dot{p}m}\,,
\qquad
\big{[} T_m\,,M_{\dot{m}n} \big{]} = i \varepsilon_{mnp} M_{\dot{m}p}\,, \\
&\big{[} M_{\dot{m}m}\,,M_{\dot{n}n} \big{]} = i \left (\delta_{mn}
\varepsilon_{\dot{m}\dot{n}\dot{p}} \dot T_{\dot{p}}
+ \delta_{\dot{m}\dot{n}} \varepsilon_{mnp} T_p
\right)\,. \label{M-algebra}
\end{aligned}
\end{equation}
For completeness, we may also list the action of the $M_{\dot{m}m}$ on
the supercharges, which can be written as
\begin{equation}
\begin{aligned}
&\big{[} M_{\dot{m}m}, Q^{\dot a a}_{\alpha} \big{]} = -\frac{1}{2}
(\tau_{\dot{m}})^{\dot a}_{\ \dot b} (\tau_m)^a_{\ b} Q^{\dot b
b}_{\alpha}\,, \qquad \big{[} M_{\dot{m}m}, S^{\dot a a}_{\alpha}
\big{]} = \frac{1}{2} (\tau_{\dot{m}})^{\dot a}_{\ \dot b} (\tau_m)^a_{\
b}
S^{\dot b b}_{\alpha}\,, \\
&\big{[} M_{\dot{m}m}, \bar Q_{\dot \alpha \dot a}^a \big{]} =
-\frac{1}{2} (\tau_{\dot{m}})^{\dot b}_{\ \dot a} (\tau_m)^a_{\ b} \bar
Q_{\dot \alpha \dot b}^b \,, \qquad \!\! \big{[} M_{\dot{m}m}, \bar
S_{\dot \alpha \dot a}^a \big{]} = \frac{1}{2} (\tau_{\dot{m}})^{\dot
b}_{\ \dot a} (\tau_m)^a_{\ b} \bar S_{\dot \alpha \dot b}^b \,.
\end{aligned}
\end{equation}

As they can be useful for explicit calculations of the
superalgebras presented in Appendix~\ref{supergroups}, we finally
list here the remaining non-trivial commutation relations of the
superconformal algebra written in $SU(2)_A \times SU(2)_B$
notation
\begin{equation}
\begin{aligned}
&\big{\{} Q^{\dot a a}_{\alpha}\,, \bar Q^{b}_{\dot\alpha\dot b}
\big{\}}= -\varepsilon^{a b} \delta^{\dot a}_{\dot b}
P_{\alpha\dot\alpha}\,, \qquad  \!\! \big{\{} S^{\dot a
a}_{\alpha} \,,\bar S^{b}_{\dot\alpha \dot b} \big{\}}
= -\varepsilon^{a b} \delta^{\dot a}_{\dot b} K_{\alpha\dot\alpha}\,, \\
&\big{[}K_{\alpha\dot\alpha}\,, Q^{\dot a a}_{\beta} \big{]}=
\varepsilon_{\alpha\beta}\bar S^{\dot a a}_{\dot\alpha}\,, \qquad
\quad ~ \big{[}K_{\alpha\dot\alpha}\,, \bar Q^{a}_{\dot\beta \dot
a} \big{]} =
\varepsilon_{\dot\alpha\dot\beta} S^a_{\alpha \dot a} \,, \\
&\big{[}P_{\alpha\dot\alpha}\,, S^{\dot a a}_{\beta} \big{]}=
\varepsilon_{\alpha\beta} \bar Q^{\dot a a}_{\dot\alpha}\,, \qquad
\quad ~~ \big{[}P_{\alpha\dot\alpha}\,, \bar S^{a}_{\dot\beta \dot
a} \big{]} =
\varepsilon_{\dot\alpha \dot\beta} Q^{a}_{\alpha \dot a} \,, \\
&\big{\{} Q^{\dot a a}_{\alpha} \,, S^{\dot b b}_{\beta} \big{\}}
=
 \varepsilon^{\dot a \dot b} \varepsilon^{ab} J_{\alpha\beta} +
\frac{1}{2}\varepsilon_{\alpha\beta} \left(\varepsilon^{a b}
T^{\dot a\dot b} + \varepsilon^{\dot a \dot b} T^{ab}
- M^{\dot a \dot b\, ab} - \varepsilon^{\dot a \dot b} \varepsilon^{ab}  D \right)\,, \\
&\big{\{} \bar Q^a_{\dot\alpha \dot a} \,, \bar S^b_{\dot\beta
\dot b} \big{\}} = -\varepsilon_{\dot a \dot b} \varepsilon^{ab}
\bar J_{\dot \alpha \dot \beta} + \frac{1}{2}
\varepsilon_{\dot\alpha \dot\beta} \left(\varepsilon^{a b} \dot
T_{\dot a\dot b} - \varepsilon_{\dot a \dot b} T^{ab} +
M^{ab}_{\dot a\dot b} + \varepsilon_{\dot a \dot b}
\varepsilon^{ab}  D \right)\,.
\end{aligned}
\label{bigalgebra2}
\end{equation}

\section{Superalgebra calculations}
\label{supergroups}

In this appendix we collect some of the
explicit calculations of the superalgebras for the different
subsectors of Wilson loop operators presented in Section
\ref{example-sec}.

\subsection{Loops on $S^2$}
\label{S2-supergroup} To determine what is the full superalgebra
preserved by this family of Wilson loop operators, it is first
convenient to rewrite the $U(1)$ generator (\ref{S2-U1}) using
$SU(2)_L \times SU(2)_R$ notation as
\begin{equation}
L \equiv \frac{1}{2} \bI^{\alpha\dot\alpha} \left
(P_{\alpha\dot\alpha}-K_{\alpha\dot\alpha} \right )\,.
\end{equation}
Then using the superconformal algebra (\ref{bigalgebra2}) one can
obtain the following commutation relations
\begin{equation}
\begin{aligned}
&\left \{\cQ^a\,, \cQ^b \right \} = -2 T^{ab}\,, \qquad
\left \{ \bar\cQ^a\,,\bar\cQ^b \right \} = 2 T^{ab}\,, \\
&\left \{ \cQ^a\,, \bar\cQ^b \right \} = -2 \varepsilon^{ab} L \,,\\
&~\left [ L \,, \cQ^a \right ] = \frac{1}{2} \bar\cQ^a\,, \qquad
\quad \left [ L \,, \bar\cQ^a \right ] = \frac{1}{2} \cQ^a\,,
\label{Osp22}
\end{aligned}
\end{equation}
while the commutators of the $SU(2)_B$ generators with the
supercharges and with themselves are canonical, as in
(\ref{Osp12}), and we do not report them here. The algebra
(\ref{Osp22}) is an $OSp(2|2)$ superalgebra (modulo possible
rescalings of the charges to bring it in a standard form).

This superalgebra is isomorphic to $SU(1|2)$ as can be seen by
defining the $L$ eigenstates
\begin{equation}
\cQ^a_{\pm} \equiv
 \frac{1}{2} \left(\cQ^a \pm \bar\cQ^a\right)\,.
\end{equation}
In terms of these charges, the superalgebra above can be written
as
\begin{equation}
\begin{aligned}
&\left \{\cQ^a_+ \,, \cQ^b_+ \right \} = \left \{\cQ^a_- \,, \cQ^b_- \right \} = 0\,,\\
&\left \{\cQ^a_+ \,, \cQ^b_- \right \} = -T^{ab} + \varepsilon^{ab} L \,,\\
&\left [ L\,, \cQ^a_{\pm} \right ] = \pm \frac{1}{2}
\cQ^a_{\pm}\,, \label{SU12}
\end{aligned}
\end{equation}
which is indeed the superalgebra $SU(1|2)$ (again we do not write
the canonical $SU(2)_B$ commutation relations). Notice that from
(\ref{SU12}) we can see that the supercharges $\cQ^a_+$ and
$\cQ^a_-$ do square to zero. However these operators are not
scalar after the twisting (\ref{twisting}), so one may not use
them to define a topological BRST charge in the usual sense.

\subsection{Latitude}
\label{latitude-supergroup} We begin by rewriting the bosonic
generators in $SU(2)_L \times SU(2)_R$ notation in the conventions
given in Appendix~\ref{algebra_conv}. The $SU(2)$ obtained from
(\ref{SU2}) after a translation and dilatation is generated by
\begin{equation}
\begin{aligned}
&L^{(\theta_0)}_1 = \frac{-i}{2\sin\theta_0}
(\tau_3)^{\alpha\dot\alpha} \left(P_{\alpha\dot\alpha} - K_
{\alpha\dot\alpha} \right)
-i \cot\theta_0 \, D \,,\\
&L^{(\theta_0)}_2 = \frac{1}{2}  \bI^{\alpha\dot\alpha}
\left(P_{\alpha\dot\alpha} - K_{\alpha\dot\alpha} \right)\,,\\
&L^{(\theta_0)}_3 = \frac{1}{\sin\theta_0}\left (J_3 + \bar J_3\right)
+ \frac{1}{2} \cot\theta_0\,\bI^{\alpha\dot\alpha}
\left(P_{\alpha\dot\alpha} + K_{\alpha\dot\alpha} \right)\,.
\label{new-SU2}
\end{aligned}
\end{equation}
where $J_3 = \frac{1}{2} (\tau_3)^{\alpha}_{\ \beta} J^{\beta}_{\
\alpha}$ and similarly for $\bar J_3$. The generator of the $U(1)$
symmetry mixing Lorentz and $R$-symmetry can be written as
\begin{equation}
{\cal C} \equiv \frac{1}{\sin\theta_0} \left (\bar J_3 - J_3  +
\dot T_3 \right)\,.
\end{equation}
where $\dot T_3 = \frac{1}{2}(\tau_3)^{\dot a}_{\ \dot b} \dot
T^{\dot a}_{\ \dot b}$, and the normalization by $\sin\theta_0$ is
for later convenience.

We can now check that these bosonic symmetries together with the
eight supercharges in (\ref{latitude-sup1}) and
(\ref{latitude-sup2}) form the superalgebra $SU(2|2)$. To this
purpose, one has to find linear combinations of the above
supercharges which transform as $(\bf{2},\bf{2})+(\bf{2},\bf{2})$
under $SU(2) \times SU(2)_B$. These can be constructed from the
following $L^{(\theta_0)}_3$ eigenstates
\begin{equation}
\cQ_{(1)}^{a,\pm} = \frac{1}{2} \left( \cQ^a_{(1)} \pm
\bar\cQ^a_{(1)} \right)\,,\qquad \quad \cQ_{(2)}^{a,\pm} =
\frac{1}{2} \left( \cQ^a_{(2)} \pm \cQ^{\prime\,a}_{(2)}\right
)\,.
\end{equation}
After some algebra, one finds that the relevant combinations which
give $SU(2)$ doublets are
\begin{equation}
{\cal Q}^a_{\eta} \equiv \frac{1}{\sqrt 2}\left(\begin{array}{c}
\cQ_{(1)}^{a,+} + \cQ_{(2)}^{a,-}  \\
i \cQ_{(1)}^{a,-} - i \cQ_{(2)}^{a,+}
\end{array}\right)\,, \qquad
{\cal S}^a_{\eta} \equiv \frac{1}{\sqrt 2} \left(\begin{array}{c}
i\cQ_{(1)}^{a,+} - i\cQ_{(2)}^{a,-}  \\
\cQ_{(1)}^{a,-} + \cQ_{(2)}^{a,+}
\end{array}\right)\,,
\end{equation}
where $\eta=1,2$ is a fundamental index in the $SU(2)$ in
(\ref{new-SU2}). Defining as usual the generators
\begin{equation}
L^{\eta}_{\ \delta} \equiv (\tau_e)^{\eta}_{\ \delta}
L^{(\theta_0)}_e\,, \qquad e=1,2,3
\end{equation}
the full superalgebra preserved by the latitude Wilson loop can be
finally written as
\begin{equation}
\begin{aligned}
&\big{[}  T^{ a}_{\  b}\,, {\cal Q}^c_{\eta} \big{]} = -\delta^{
b}_{ c} {\cal Q}^{a}_{\eta} + \frac{1}{2} \delta^{ a}_{  b} {\cal
Q}^{c}_{\eta}\,,\qquad \big{[}  T^{ a}_{\  b}\,, {\cal S}^c_{\eta}
\big{]} =
-\delta^{ b}_{ c} {\cal S}^{a}_{\eta} + \frac{1}{2} \delta^{ a}_{  b} {\cal S}^{c}_{\eta}\,, \\
&\big{[}  L^{ \eta}_{\ \delta}\,, {\cal Q}^a_{\gamma} \big{]} =
\delta^{ \eta}_{ \gamma} {\cal Q}^{a}_{\delta} - \frac{1}{2}
\delta^{\eta}_{\delta} {\cal Q}^{a}_{\gamma}\,,\qquad
\big{[}  L^{ \eta}_{\ \delta}\,, {\cal S}^a_{\gamma} \big{]} =
\delta^{ \eta}_{ \gamma} {\cal S}^{a}_{\delta} - \frac{1}{2}
\delta^{\eta}_{\delta}
{\cal S}^{a}_{\gamma}\,, \\
& \left\{ {\cal Q}^a_{\eta}\,,{\cal S}^b_{\delta}\right\} =
\epsilon^{ab} L_{\eta\delta} + \epsilon_{\eta\delta} T^{ab}
-\epsilon^{ab} \epsilon_{\eta\delta}\, {\cal C}\,, \label{SU22}
\end{aligned}
\end{equation}
and all other commutators vanish (except the standard $SU(2)$
algebras for $T^a_{\ b}$ and $L^{\eta}_{\ \delta}$). Notice in
particular that ${\cal C}$ behaves as a central charge of the
algebra. This is the superalgebra $SU(2|2)$, as stated above.

\subsection{Two longitudes}
\label{longitudes-supergroup}

First, to recognize how the $SO(4)$ symmetry rotating
$\Phi^3$, $\Phi^4$, $\Phi^5$ and $\Phi^6$ arises from the algebra of
the fermionic charges (\ref{longi-SUSY}), one can evaluate
commutators of supercharges with the same chirality. This yields
\begin{equation}
\begin{aligned}
&\left \{\cQ_{(1)}^a\,, \cQ_{(1)}^b \right \} = -2 T^{ab}\,,
\qquad \!\!\!\! \left \{\cQ_{(2)}^a\,, \cQ_{(2)}^b \right \}  = -2
T^{ab}\,, \qquad
\!\!\! \! \left \{\cQ_{(1)}^a\,, \cQ_{(2)}^b \right \} = 2 M^{ab}_{\dot 1 \dot 2}\,, \\
&\left \{\bar\cQ_{(1)}^a\,, \bar\cQ_{(1)}^b \right \} = 2
T^{ab}\,,\qquad \left \{\bar\cQ_{(2)}^a\,, \bar\cQ_{(2)}^b \right
\} = 2 T^{ab}\,, \qquad \,\left \{\bar\cQ_{(1)}^a\,,
\bar\cQ_{(2)}^b \right \} = -2 M^{ab}_{\dot 1 \dot 2}\,,
\label{R-sym}
\end{aligned}
\end{equation}
where the $M^{ab}_{\dot a \dot b}$ are the generators in the $(\bf
3,\bf 3)$ of $SU(2)_A \times SU(2)_B$ arising in the decomposition
of $SU(4)$ discussed in Appendix~\ref{algebra_conv}, see
(\ref{SU4-break}). In the basis defined in (\ref{pauli-basis}),
the $R$-symmetry generators in (\ref{R-sym}) may be written as
\begin{equation}
T^{ab} = - (\tau_m \varepsilon)^{ab}\, T_m\,, \qquad M^{ab}_{\dot
1 \dot 2} = - (\tau_m \varepsilon)^{ab}\, M_{\dot{3}m}\,.
\end{equation}
The six generators $T_m,M_{\dot{3}m}$ commute with the $SO(2)$ generated
by $\dot T_3$ (which is the symmetry rotating $\Phi_1$ and
$\Phi_2$), and as expected generate a $SO(4)$ subgroup of $SU(4)$,
as can be seen using the algebra (\ref{M-algebra}). Explicitly,
defining the linear combinations
\begin{equation}
\widehat{\cal T}_m = \frac{1}{2} (T_m + M_{\dot{3}m})\,, \qquad
\widetilde{\cal T}_m = \frac{1}{2} (T_m - M_{\dot{3}m}),
\end{equation}
one finds that
\begin{equation}
\big{[}\widehat{\cal T}_m\,,\widehat{\cal T}_n\big{]} = i
\varepsilon_{mnp} \widehat{\cal T}_p\,,\qquad \big{[}\widetilde{\cal
T}_m\,,\widetilde{\cal T}_n\big{]} = i \varepsilon_{mnp}
\widetilde{\cal T}_p\,, \qquad \big{[}\widehat{\cal
T}_m\,,\widetilde{\cal T}_n\big{]} = 0\,, \label{SO4}
\end{equation}
which is indeed $SU(2) \times SU(2) = SO(4)$. By looking at the
action of the $T_m$ and $M_{\dot{3}m}$ on the supercharges, one can
construct the following orthogonal combinations
\begin{equation}
\begin{aligned}
\widehat{\cQ}^a\equiv\frac{1}{2} \left ( \cQ^a_{(1)} - \cQ^a_{(2)}
\right)\,,\qquad \widetilde{\cQ}^a\equiv\frac{1}{2} \left (
\cQ^a_{(1)} + \cQ^a_{(2)} \right)\,,\qquad \label{Osp12sq-comb}
\end{aligned}
\end{equation}
and analogously for the other chirality. These combinations
satisfy
\begin{equation}
\begin{aligned}
&\left \{ \widehat{\cQ}^a\,, \widehat{\cQ}^b \right \} = 2 (\tau_m
\varepsilon)^{ab}\, \widehat{\cal T} _m\,,\qquad \qquad
\left \{ \widetilde{\cQ}^a\,, \widetilde{\cQ}^b \right \} = 2 (\tau_m \varepsilon)^{ab}\, \widetilde{\cal  T}_m\,, \\
&\,\,\big{[}\widehat{\cal T}_m\,,\widehat{\cQ}^a \big{]} =
-\frac{1}{2} (\tau_m)^a_{\ b} \widehat{\cQ} ^b\,,\qquad \qquad
\,\big{[}\widetilde{\cal T}_m\,,\widetilde{\cQ}^a \big{]} =
-\frac{1}{2} (\tau_m)^a_{\ b} \widetilde{\cQ} ^b\,,
\label{Osp12square}
\end{aligned}
\end{equation}
while all commutators mixing generators in the first and second
column of the above equation vanish. A  similar algebra applies of
course to the negative chirality charges.

The remaining $U(1)\times U(1)$ bosonic symmetry generated by
\begin{equation}
L \equiv \frac{1}{2} \bI^{\alpha\dot\alpha} \left
(P_{\alpha\dot\alpha}-K_{\alpha\dot\alpha} \right )\,,\qquad I
\equiv \frac{1}{2} \tau_3^{\alpha\dot\alpha} \left
(P_{\alpha\dot\alpha}+K_{\alpha\dot\alpha} \right )\,,
\end{equation}
arises from commutators of supercharges of opposite chirality. By
explicitly evaluating the relevant commutators, it is easy to see
that $L$ acts on any supercharge in (\ref{longi-SUSY}) by changing
its chirality, as in (\ref{Osp22}), while acting with $I$ changes
chirality together with flipping a charge of type ``$(1)$'' into a
charge of type ``$(2)$''. One can then see that defining the linear
combinations
\begin{equation}
\widehat{L} = \frac{1}{2} \left( L - I \right)\, \qquad
\widetilde{L} = \frac{1}{2} \left( L + I \right)\,
\end{equation}
together with the $\widehat L$ and $\widetilde L$ eigenstates
\begin{equation}
\widehat{\cQ}^a_{\pm} \equiv \frac{1}{2} \left(\widehat{\cQ}^a \pm
\widehat{\bar\cQ}{}^a \right)\,\qquad  \quad
\widetilde{\cQ}^a_{\pm} \equiv \frac{1}{2} \left(\widetilde{\cQ}^a
\pm \widetilde{\bar\cQ}{}^a \right)\,
\end{equation}
allows one to write the full algebra in the direct product form
\begin{equation}
\begin{aligned}
&\left \{ \widehat{\cQ}^a_+ \,, \widehat{\cQ}^b_{+} \right \} =
\left \{ \widehat{\cQ}^a_- \,, \widehat{\cQ}^b_{-} \right \}=0\,,
\qquad \qquad \quad \left \{ \widetilde{\cQ}^a_+ \,,
\widetilde{\cQ}^b_{+} \right \} =
\left \{ \widetilde{\cQ}^a_- \,, \widetilde{\cQ}^b_{-} \right \}=0\,, \\
&\left \{ \widehat{\cQ}^a_+ \,, \widehat{\cQ}^b_{-} \right \} =
(\tau_m \varepsilon)^{ab}\,\widehat{\cal T}_m + \epsilon^{ab}
\widehat{L}\,, \qquad \qquad \, \left \{ \widetilde{\cQ}^a_+ \,,
\widetilde{\cQ}^b_{-} \right \} =
(\tau_m \varepsilon)^{ab}\,\widetilde{\cal T}_m + \epsilon^{ab} \widetilde{L}\,,\\
&\, \,\big{[} \widehat{L}\,, \widehat{\cQ}^a_{\pm} \big{]} = \pm
\frac{1}{2} \widehat{\cQ}^a_{\pm}\,, \qquad \qquad \qquad \qquad
\qquad
\big{[} \widetilde{L}\,, \widetilde{\cQ}^a_{\pm} \big{]} = \pm \frac{1}{2} \widetilde{\cQ}^a_{\pm}\,, \\
&\, \, \big{[}\widehat{\cal T}_m\,,\widehat{\cQ}^a_{\pm} \big{]} =
-\frac{1}{2} (\tau_m)^a_{\ b} \widehat {\cQ}^b_{\pm}\,, \qquad
\qquad \qquad \quad \! \big{[}\widetilde{\cal
T}_m\,,\widetilde{\cQ}^a_{\pm} \big{]} = -\frac{1}{2}
(\tau_m)^a_{\ b} \widetilde {\cQ}^b_{\pm}\,,
\end{aligned}
\label{longi-superalgebra}
\end{equation}
with all other not listed commutators vanishing. As claimed above,
this is a $SU(1|2) \times SU(1|2)$  superalgebra. As a side
remark, notice that the $SU(1|2)$ algebra (\ref{SU12}) preserved
by the great $S^2$ loops is just a  diagonal subgroup of the one
we found here.


\section{String solutions}
\label{solutions-appendix}

In this appendix we report the explicit computations of the string solutions in $AdS_5\times S^5$ corresponding to the examples used in the main text.

\subsection{Latitude}
\label{latitude-sol-appendix}

The string solution for the 1/4 BPS latitude was first found in
\cite{Drukker:2005cu,Drukker:2006ga}. Here we reprint the
result in a coordinate system more suited for our present discussion.%
\footnote{Compared to those references, we translated the circle
in the $x_3$ and rescaled it appropriately to fit on $S^3$. We also
replaced $\theta_0\to\pi/2-\theta_0$.}
We use the metric
\begin{equation}
ds^2=\frac{L^2}{z^2}(dz^2+dr^2+r^2 d\phi^2+dx_3^2)
+L^2(d\vartheta^2+\sin^2\vartheta\,d\varphi^2)\,,
\end{equation}
where $(r,\,\phi)$ are radial coordinates in the $(1,\,2)$ plane.
For the latitude at angle $\theta_0$, the boundary of the string
should end along the curve at $r=\sin\theta_0$ and
$x_3=\cos\theta_0$, while on the sphere side of the ansatz it should
end at $\vartheta_0=\pi/2-\theta_0$
see (\ref{scalars_latitude}) and Figure~\ref{latitude-fig}. The
boundary conditions represent motion around both spheres in the
same direction, but with a phase difference of $\pi$.
The string solution will be given
by a constant $x_3$, while in the conformal gauge we may
take the ansatz $z=z(\sigma)$, $r=r(\sigma)$,
$\vartheta=\vartheta(\sigma)$ and $\phi=\varphi+\pi=\tau$. The
solution is given by
\begin{equation}
z=\sin\theta_0\tanh\sigma\,,\qquad
r=\frac{\sin\theta_0}{\cosh\sigma}\,,\qquad
\sin\vartheta=\frac{1}{\cosh(\sigma_0\pm\sigma)}\,.
\label{lat-solution}
\end{equation}
The integration constant $\sigma_0$ is fixed by requiring that at
$\sigma=0$ one has
$\sin\vartheta_0=\cos\theta_0=1/\cosh\sigma_0$. The two
signs in the expression for $\sin\vartheta$ correspond to wrapping the
string either around the north pole of the sphere or around the south pole.

The value of the classical action of the string is
\begin{equation}
{\cal S}=\mp \sqrt\lambda\,\sin\theta_0\, .
\end{equation}
The loop corresponding to the solution wrapping the ``short side'' of the sphere (around the north pole, with the $-$ sign in the expression above) has then a value
\begin{equation}
\vev{W}=e^{\sqrt\lambda\,\sin\theta_0}\, ,
\end{equation}
while the other solution corresponds to an unstable instanton, whose value is exponentially suppressed at large $\lambda$.

\subsection{Two longitudes}
\label{longitudes-sol-appendix}

\begin{figure}
\begin{center}
\includegraphics[width=100mm]{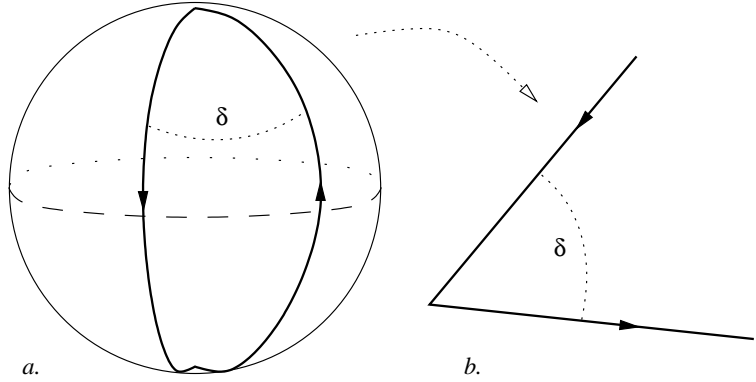}
\parbox{13cm}{
\caption{
The quarter-BPS Wilson loop made of two longitudes (a.) can
be mapped to a stereographic projection the a cusp on the plane (b.).
The scalar couplings (see figure~\ref{longi-fig}~b.) are not altered
and are the natural coupling for a supersymmetric cusp in the plane.
\label{cusp-fig}}}
\end{center}
\end{figure}

The basic idea in finding the string solution for the two longitudes on
the $S^2$ is to observe that a  stereographic projection to the
plane will map this loop to a single cusp at the origin with
an opening angle $\delta$ (see figure~\ref{cusp-fig}).
This will still be $1/4$ BPS and will be of the
type invariant under the $Q$ supercharges \cite{Zarembo:2002an},
therefore it will have trivial expectation value. In that
way our operator is similar to the usual 1/2 BPS circle that is
conformal to the straight line which has trivial value. The
operator on the sphere will
have non-trivial value because of the compactness of the space.

We shall therefore first find the string solution for a single cusp of
angle $\delta$ in the plane  and then we shall conformally transform it to
the interesting system which is compact.

The cusp can be solved by using the conformal symmetry, as was
done in \cite{Drukker:1999zq}. Take the metric on $AdS_3\times S^1$
subspace of $AdS_5\times S^5$ to be
\begin{equation}
ds^2=\frac{L^2}{z^2}\left(dz^2+dr^2+r^2d\phi^2\right)
+L^2 d\varphi^2\,.
\end{equation}
If the cusp is at the origin $r=0$, it is invariant
under rescaling of $r$. This symmetry is then extended
to the string world-sheet, where the $z$ coordinate will have a linear
dependence on $r$.   As world-sheet coordinates we take $r$ and $\phi$. The
ansatz for the other coordinates is
\begin{equation}
z=r\,v(\phi)\,,\qquad
\varphi=\varphi(\phi)\,.
\end{equation}
The Nambu-Goto action is (prime is the derivative with respect to $\phi$)
\begin{equation}
\cS_{NG}=\frac{\sqrt\lambda}{2\pi}\int dr\,d\phi
\frac{1}{r\,v^2}\sqrt{v'^2+(1+v^2)(1+v^2\varphi'^2)}\,.
\end{equation}
The $r$ dependence is trivial and it is easy to find two conserved
quantities, the energy and the canonical momentum conjugate to $\varphi$
\begin{equation}
E=\frac{1+v^2}{v^2\sqrt{v'^2+(1+v^2)(1+v^2\varphi'^2)}}\,,
\qquad
J=\frac{(1+v^2)\varphi'}{\sqrt{v'^2+(1+v^2)(1+v^2\varphi'^2)}}\,.
\label{EJ}
\end{equation}
The BPS condition turns out, not surprisingly, to be $E=|J|$. To derive
it consider the Legendre transform term which should be added to the
action. Using the equations of motion it is
\begin{equation}
\cS_{L.T.}=\frac{\sqrt\lambda}{2\pi}\int dr\,d\phi\,(z\,p_z)'
=\frac{\sqrt\lambda}{2\pi}\int dr\,d\phi
\frac{-2-v'^2-v^2(1+\varphi'^2)}
{r\,v^2\sqrt{v'^2+(1+v^2)(1+v^2\varphi'^2)}}\,.
\end{equation}
Requiring that the total Lagrangian vanishes locally leads to
\begin{equation}
v^4\varphi'^2-1=0\,.
\end{equation}
This can be written in terms of the conserved quantities in (\ref{EJ}) as
$E^2=J^2$.

The equation of motion for $v$ is
\begin{equation}
v'=\frac{1+v^2}{v^2}\sqrt{p^2-v^2}\,,\qquad
p=\frac{1}{E}\,.
\end{equation}
which integrates to
\begin{equation}
\phi=\arcsin\frac{v}{p}-
\frac{1}{\sqrt{1+p^2}}
\arcsin\sqrt{\frac{1+1/p^2}{1+1/v^2}}\,.
\end{equation}
This expression is valid over half the world-sheet, till the midpoint. Beyond
that we should analytically continue to
\begin{equation}
\phi=\pi-\arcsin\frac{v}{p}-
\frac{1}{\sqrt{1+p^2}}
\left(\pi-\arcsin\sqrt{\frac{1+1/p^2}{1+1/v^2}}\right)\,,
\end{equation}
The final value of $\phi$ when $v$ reaches zero again is
\begin{equation}
\delta=\pi\left(1-\frac{1}{\sqrt{1+p^2}}\right)\,.
\label{phi1-p}
\end{equation}

The equation for $\varphi$ is even a bit simpler
\begin{equation}
\varphi'=\pm\frac{1}{v^2}
=\pm\frac{v'}{(1+v^2)\sqrt{p^2-v^2}}\,,
\end{equation}
which integrates to
\begin{equation}
\varphi=\frac{1}{\sqrt{1+p^2}}
\arcsin\sqrt{\frac{1+1/p^2}{1+1/v^2}}\,.
\end{equation}
After going to the second branch the final value is
\begin{equation}
\varphi_1=\frac{\pi}{\sqrt{1+p^2}}\,,
\end{equation}
and indeed $\delta+\varphi_1=\pi$, as should be the case
by the supersymmetric construction of the scalar couplings
(see (\ref{arcs}) and the paragraph thereafter).

As mentioned above, the Nambu-Goto action is equal (up to a sign)
to the total derivative which has to be added, so the full Lagrangian
vanishes
\begin{equation}
\cS_{NG}=\frac{\sqrt\lambda}{2\pi}\int\frac{dr}{r}
\int du\,\frac{-1}{p\,u^2}
=\frac{\sqrt\lambda}{2\pi}\int\frac{dr}{r}
\frac{1}{p\,u_0}\,,
\end{equation}
with $u_0$ a cutoff. Note that for small $u$ the integrand
$1/(rpu_0)\sim 1/z_0$ is the standard divergence. Indeed it cancels
against the Legendre transform.

The next step is to conformally transform to global $AdS$ with metric
\begin{equation}
ds^2=L^2\left[d\rho^2+\sinh^2\rho(d\theta^2+\sin^2\theta d\phi^2)
+d\varphi^2\right]\,,
\end{equation}
by ($\phi$ and $\varphi$ are mapped to themselves)
\begin{equation}
\cosh\rho=\frac{1+z^2+r^2}{2z}\,,\qquad
\sinh\rho\sin\theta=\frac{r}{z}\,.
\end{equation}
This gives the surface
\begin{equation}
\cosh\rho=\frac{1+r^2+r^2v^2}{2rv}\,,\qquad
\sinh\rho\sin\theta=\frac{1}{v}\,.
\end{equation}
The relation between $v$, $\phi$ and $\varphi$ is as before, but
the action will have to be calculated again using a different regularization
that should give the expectation value of the Wilson loop  with two cusps on the
sphere.

Plugging in the solution into the Nambu-Goto action, it may be written
in the following form
\begin{equation}
\begin{aligned}
\cS_{NG}&=\frac{\sqrt{\lambda}}{2\pi}\int\frac{dr}{r}\int d\phi\,
\frac{p(1+v^2)}{v^4}
=\frac{\sqrt{\lambda}}{2\pi}\int\frac{dr}{r}\int dv\,
\frac{p}{v^2\sqrt{p^2-v^2}}
\\&
=\frac{\sqrt{\lambda}}{2\pi}\int d\rho\,d\theta\,
\frac{p\sinh^2\rho\sin\theta}{\sqrt{p^2\sinh^2\rho\sin^2\theta-1}}
\,.
\label{rho-theta-action}
\end{aligned}
\end{equation}
This expression is simple to integrate. For a fixed $\rho$ the variable
$\theta$ varies between the two roots of $\sin\theta\sinh\rho=1/p$,
and then back. Integrating over this variable gives $2\pi\sinh\rho$, so
we are left with the $\rho$ integration between the minimal value, where
$\sinh\rho=1/p$ and a cutoff $\rho_0$ at large $\rho$
\begin{equation}
\cS_{NG}=\sqrt\lambda\int d\rho\,\sinh\rho
=\sqrt\lambda\left(\cosh\rho_0-\sqrt{1+\frac{1}{p^2}}\right)\,.
\label{div-action}
\end{equation}

One may be tempted to simply throw away the divergent $\cosh\rho_0$ term,
but some more care is actually required to proceed. As we noted before, the range of the $\theta$
integration for fixed $\rho_0$ is not $2\pi$, but roughly
\begin{equation}
2\pi-\frac{4}{p\sinh\rho_0}\,.
\end{equation}
So this gives the possibility of some finite corrections left over from the
divergent piece. The precise prescription for getting a
finite value for the Wilson loop expectation value was given in
\cite{Drukker:1999zq}. It is defined in the Poincar\'e patch, where
one can resort to considerations on the near horizon limit of D3-branes.
The divergence in the bulk action is canceled by a boundary term
which is a Legendre transform of the six coordinates orthogonal to the
brane. In global $AdS$ this translates to
\begin{equation}
\cL_\text{boundary}=-\coth\rho_0\,p_\rho
=-\coth\rho_0\,\rho'\,\frac{\delta\cL_{NG}}{\delta\rho'}\,,
\end{equation}
where $\rho'$ is the derivative of $\rho$ with respect to the world-sheet
coordinate orthogonal to the boundary. In the limit of large $\rho_0$ we
can replace $\coth\rho_0\to1$.

To evaluate it in practice one has to reintroduce $\rho'$ into
(\ref{rho-theta-action}), where it was set to one, leading to the expression
\begin{equation}
\begin{aligned}
\cS_\text{boundary}
&=-\frac{\sqrt{\lambda}}{2\pi}\int d\theta\,
\frac{\sqrt{p^2\sinh^2\rho_0\sin^2\theta-1}}
{p\sinh^2\rho_0\sin\theta}
\left[\sinh^2\rho_0(1+\sin^2\theta(\partial_\theta\phi)^2)
+(\partial_\theta\varphi)^2\right]\\
&=-\frac{\sqrt{\lambda}}{2\pi}\int d\theta\,\sin\theta\,
\frac{p^2\sinh^2\rho_0(\sinh^2\rho_0\sin^2\theta+1)-\cosh^2\rho_0}
{p(\sinh^2\rho_0\sin^2\theta+1)\sqrt{p^2\sinh^2\rho_0\sin^2\theta-1}}
\,.
\end{aligned}
\end{equation}
The first term in the numerator cancels part of the denominator giving
the same integral over $\theta$ as in (\ref{rho-theta-action}),
which is equal to $2\pi\sinh\rho_0$.
The second term, with $\cosh^2\rho_0$ in the numerator integrates to a
finite answer such that the final result for the boundary term is
\begin{equation}
\cS_\text{boundary}\simeq-\sqrt{\lambda}
\left(\sinh\rho_0-\frac{\coth\rho_0}{p\sqrt{1+p^2}}\right)\,.
\end{equation}

Combining this with the bulk action (\ref{div-action}), the divergences
indeed cancel and we get the final answer for the action of the string
dual to the two-longitudes Wilson loop
\begin{equation}
\cS
=-\frac{p}{\sqrt{1+p^2}}\,\sqrt{\lambda}
=-\frac{\sqrt{\lambda\, \delta(2\pi-\delta)}}{\pi}\,.
\label{longi-action}
\end{equation}
In the last equality we used (\ref{phi1-p}) to represent $p$
in terms of $\delta$.

\subsubsection*{Non-BPS case}
For completion we consider here the case of the general
non-supersymmetric cusp in the plane with opening angle $\delta$
and {\em arbitrary} jump in the scalar coupling $\varphi_1$.
This calculation is not used in the main text, as this loop is not
BPS, but it was left unsolved in \cite{Drukker:1999zq} and is
a simple generalization of the BPS case.

In the supersymmetric case the ratio of the two conserved charges
$J$ and $E$ in (\ref{EJ}) was $\pm1$. In the non-supersymmetric
case it is still simple and we denote it by $q$
\begin{equation}
q\equiv\frac{J}{E}=v^2\varphi'\,.
\end{equation}
Using this we find the differential equation for $v$
\begin{equation}
v'^2=\frac{1+v^2}{v^4}[p^2+(p^2-q^2)v^2-v^4]\,,\qquad
p=\frac{1}{E}\,.
\label{v-eqn}
\end{equation}
This is an elliptic equation. To see that define
\begin{equation}
\zeta=\sqrt{\frac{v^2(1+b^2)}{b^2(1+v^2)}}\,,\qquad
b^2=\frac{1}{2}\left(p^2-q^2+\sqrt{(p^2-q^2)^2+4p^2}\right)\,.
\end{equation}
Then $\zeta$ satisfies
\begin{equation}
\zeta'^2=\frac{p^2}{b^2}\left(1-\frac{1+b^2}{b^2\zeta^2}\right)^2
(1-\zeta^2)(1- k^2\zeta^2)\,,
\qquad
k^2=\frac{b^2(p^2-b^2)}{p^2(1+b^2)}\,.
\end{equation}
Therefore the relation between $\zeta$ and $\phi$ is given in
terms of incomplete elliptic integrals of the first and third kind
$F$ and $\Pi$ with argument $\arcsin \zeta$ and modulus
 $k$
\begin{equation}
\phi=\frac{b}{p\sqrt{1+b^2}}
\left[F(\arcsin \zeta; k)
-\Pi\left(\frac{b^2}{1+b^2},\arcsin \zeta;k\right)\right]\,.
\end{equation}
At the boundary $v=0$ so also $\zeta=0$. It reaches a maximal value $\zeta=1$
beyond which another copy of the surface continues with
\begin{equation}
\phi=\frac{b}{p\sqrt{1+b^2}}
\left[2K(k)-2\Pi\left(\frac{b^2}{1+b^2};k\right)
-F(\arcsin \zeta; k)
+\Pi\left(\frac{b^2}{1+b^2},\arcsin \zeta;k\right)\right]\,.
\end{equation}
The final value of $\phi$ when we reach the boundary again is twice the
complete elliptic integrals
\begin{equation}
\delta=\frac{2b}{p\sqrt{1+b^2}}
\left[K(k)-\Pi\left(\frac{b^2}{1+b^2};k\right)\right]\,.
\end{equation}

Integrating $\varphi$ leads to an even simpler expression in terms of
elliptic integrals of the first kind
\begin{equation}
\varphi=\int d\phi\,\frac{q}{v^2}
=\frac{q\,b}{p\sqrt{1+b^2}}
F(\arcsin \zeta;k)\,.
\end{equation}
The final value of $\varphi$ is again related to the complete integral
\begin{equation}
\varphi=2\frac{q\,b}{p\sqrt{1+b^2}}
K(k)\,.
\end{equation}

Then we can calculate the classical action
\begin{equation}
\begin{aligned}
\cS&=\frac{\sqrt{\lambda}}{2\pi}\int dr\,d\phi\,
\frac{p}{r}\frac{1+v^2}{v^4}
\\
&=\frac{\sqrt{\lambda}}{2\pi}\int\frac{dr}{r}
\frac{\sqrt{1+b^2}}{b}
\left[-\frac{\sqrt{(1-\zeta^2)(1-k^2\zeta^2)}}{\zeta}
+\left[F(\arcsin \zeta;k)-E(\arcsin \zeta;k)\right]\right]\,,
\end{aligned}
\end{equation}
where $E$ denotes an elliptic integral of the second kind. The
right hand side should be evaluated at the two boundaries where
$\zeta=0$ (on the two branches). The result is
\begin{equation}
\cS_{NG}=\frac{\sqrt{\lambda}}{2\pi}\int\frac{dr}{r}
\frac{\sqrt{1+b^2}}{b}\left[\frac{2}{\zeta_0}
+2\left[K(k)-E(k)\right]\right]\,.
\end{equation}
Here $\zeta_0$ is a cutoff at small $\zeta$, so the first term is equal to
\begin{equation}
\frac{\sqrt\lambda}{2\pi}\int dr\frac{2\sqrt{1+b^2}}{b\, r \zeta_0}
=\frac{\sqrt\lambda}{2\pi}\int dr\frac{2}{z_0}\,,
\end{equation}
where $z_0$ is a cutoff on $z$, and this is the standard divergence
for the two rays making the cusp. The divergence is canceled as usual by
a boundary term.

\subsection{Toroidal loops}
\label{torus-appendix}

We describe now the toroidal loops introduced in Section~\ref{hopf-section}
and Section~\ref{torus-subsection}. We perform the calculation in the
general case where the radii of the loops $r_1$ and $r_2$ are independent
of the periods $k_1$ and $k_2$ along the two cycles of the torus.
To focus on the case of the latitude on the Hopf base discussed in
 Section~\ref{hopf-section}, one should simply set
\begin{equation}
\sin{\theta\over 2}=\sqrt{\frac{k_2}{k_1+k_2}}\,.
\end{equation}

Consider a doubly-periodic motion on $S^3$
\begin{equation}
x_1=\sin\frac{\theta}{2}\sin k_1 t\,,\quad
x_2=\sin\frac{\theta}{2}\cos k_1 t\,,\quad
x_3=\cos\frac{\theta}{2}\sin k_2 t\,,\quad
x_4=\cos\frac{\theta}{2}\cos k_2 t\,,
\end{equation}
where $\theta$ is one of the Euler angles, while the other two
angles are given by
\begin{equation}
\phi=(k_1+k_2)t\,,\qquad
\psi=(k_2-k_1)t\,.
\end{equation}
The scalar couplings for these loops are simple
\begin{equation}
\begin{aligned}
\frac{1}{2}\,\sigma_1^R
&=\frac{1}{2}(k_1+k_2)\sin\theta\,\cos(k_2-k_1)t\,dt\,,\\
\frac{1}{2}\,\sigma_2^R
&=\frac{1}{2}(k_1+k_2)\sin\theta\,\sin(k_2-k_1)t\,dt\,,\\
\frac{1}{2}\,\sigma_3^R
&=\left(k_2\cos^2\frac{\theta}{2}-k_1\sin^2\frac{\theta}{2}
\right)dt\,.
\end{aligned}
\end{equation}
This is just a periodic motion, as in the case of the latitude on the great
$S^2$.

It is possible to find the minimal surface representing this Wilson loop in
$AdS_5\times S^5$ using the techniques of \cite{Drukker:2005cu}.
There it was shown how to calculate a general periodic Wilson loop,
but the example of motion on a torus was not done explicitly.

One first notices that the $AdS_5$ and $S^5$ parts of the $\sigma$-model
completely decouple. In principle the two systems may be coupled because of the
Virasoro constraint, which should be satisfied on the combined system only.
All the examples in \cite{Drukker:2005cu} where this occurred were the
correlation functions of two loops. Here we have a single loop and in this
case the Virasoro constraint is indeed satisfied
independently on both sides.

The solution to the equations of motion on the $S^5$ side are like in the
latitude on the great $S^2$ example (\ref{lat-solution})
\begin{equation}
\sin\vartheta=\frac{1}{\cosh[(k_2-k_1)(\sigma_0\pm\sigma)]}\,,\qquad
\varphi=(k_2-k_1)\tau\,.
\end{equation}
The sign choice corresponds to a surface wrapping the northern or
southern hemisphere and the integration constant $\sigma_0$ is
chosen so that at $\sigma=0$ it reaches the boundary value
\begin{equation}
\sin\vartheta_0=\frac{1}{\cosh[(k_2-k_1)\sigma_0]}=
\frac{(k_1+k_2)\sin\theta}{2\sqrt{k_1^2\sin^2\frac{\theta}{2}
+k_2^2\cos^2\frac{\theta}{2}}}\,.
\end{equation}
The action for the string will be the sum of $AdS_5$ part and of the $S^5$
part. The latter is just the area of the part of the sphere covered by the string
(taking $k_2\geq k_1$)
\begin{equation}
\begin{aligned}
\cS_{S^5}&=\left(k_2-k_1\right)\left(1\pm
\frac{k_2\cos^2\frac{\theta}{2}-k_1\sin^2\frac{\theta}{2}}
{\sqrt{k_1^2\sin^2\frac{\theta}{2}+k_2^2\cos^2\frac{\theta}{2}}}
\right)\sqrt{\lambda}\\
&=\left(k_2-k_1
\pm\sqrt{k_1^2\sin^2\frac{\theta}{2}+k_2^2\cos^2\frac{\theta}{2}}
\mp\frac{k_1k_2}
{\sqrt{k_1^2\sin^2\frac{\theta}{2}+k_2^2\cos^2\frac{\theta}{2}}}
\right)\sqrt{\lambda}\,.
\end{aligned}
\label{torus-S5-action}
\end{equation}
The sign choice again corresponds to the two possible wrappings of $S^2$.

To solve the $AdS_5$ part it is convenient to write it as
a hypersurface in flat six-dimensional Minkowski space
\begin{equation}
-Y_0^2+Y_1^2+Y_2^2+Y_3^2+Y_4^2+Y_5^2=-L^2\,.
\end{equation}

Now let us define the coordinates $r_0$, $r_1$, $r_2$, $v$, $\phi_1$
and $\phi_2$ by
\begin{equation}
\begin{aligned}
&Y_0=Lr_0\cosh v\,,\qquad
&Y_5=Lr_0\sinh v\,,
\\
&Y_1=Lr_1\cos\phi_1\,,\qquad
&Y_2=Lr_1\sin\phi_1\,,
\\
&Y_3=Lr_2\cos\phi_2\,,\qquad
&Y_4=Lr_2\sin\phi_2\,.
\end{aligned}
\end{equation}
Those coordinates satisfy the constraint $-r_0^2+r_1^2+r_2^2=-1$, and
the metric of the embedding flat Minkowski space is
\begin{equation}
ds^2=L^2\left(-dr_0^2+r_0^2dv^2
+dr_1^2+r_1^2d\phi_1^2
+dr_2^2+r_2^2d\phi_2^2\right)\,.
\end{equation}

The relevant ansatz for our system of periodic motion on $T^2$ is
\begin{equation}
r_i=r_i(\sigma)\,,\qquad
v=v(\sigma)\,,\qquad
\phi_1=k_1\tau+\alpha_1(\sigma)\,,\qquad
\phi_2=k_2\tau+\alpha_2(\sigma)\,.
\end{equation}
Furthermore we can set $\alpha_1$, $\alpha_2$ and $v$ to be constants,
leaving only the action for $r_0$, $r_1$, and $r_2$
\begin{equation}
\cS_{AdS_5}=\frac{L^2}{4\pi\alpha'}\int d\sigma\,d\tau
\big[-r_0'^2+r_1'^2+r_2'^2+r_1^2k_1^2+r_2^2k_2^2
+\Lambda\left(-r_0^2+r_1^2+r_2^2+1\right)\big]\,.
\end{equation}
Here $\Lambda$ is a Lagrange multiplier.

The equations of motion for $r_0$, $r_1$ and $r_2$ are
\begin{equation}
r_0''=\Lambda r_0\,,\qquad
r_1''=(k_1^2+\Lambda)r_1\,,\qquad
r_2''=(k_2^2+\Lambda)r_2\,.
\end{equation}
It is simple to find the first integral of motion, it is the diagonal
component of the $AdS_5$ contribution to the stress-energy tensor
\begin{equation}
-r_0'^2+r_1'^2+r_2'^2-k_1^2r_1^2-k_2^2r_2^2=0\,.
\end{equation}
Using the Virasoro constraint, the classical action is twice the kinetic piece
\begin{equation}
\cS_{AdS_5}=2\cS_{AdS_5}^{kinetic}
=\frac{\sqrt\lambda}{2\pi}\int d\sigma\,d\tau
\left(r_1^2k_1^2+r_2^2k_2^2\right)\,.
\end{equation}

The other integrals of motion are
\begin{equation}
\begin{aligned}
I_0=&r_0^2-\frac{1}{k_1^2}(r_0r_1'-r_1r_0')^2
-\frac{1}{k_2^2}(r_0r_2'-r_2r_0')^2\,,
\\
I_1=&r_1^2-\frac{1}{k_1^2}(r_0r_1'-r_1r_0')^2
+\frac{1}{k_1^2-k_2^2}(r_1r_2'-r_2r_1')^2\,.
\end{aligned}
\end{equation}
We can define $I_2$ in a similar fashion, but it is not an independent
integral, since $-I_0+I_1+I_2=-1$.

We know that the range of the world-sheet coordinate $\sigma$
is infinite (from the $S^5$ part of the solution), and that for large
$\sigma$ both $r_1$ and $r_2$ vanish (as well as their derivatives),
while $r_0\to1$. From this we easily conclude that the integration
constants are $I_0=1$ and $I_1=I_2=0$.

To solve these equations we define the coordinates $\zeta_1$ and $\zeta_2$ which
are the roots of the equation
\begin{equation}
\frac{r_0^2}{\zeta_i^2}
-\frac{r_1^2}{\zeta_i^2-k_1^2}
-\frac{r_2^2}{\zeta_i^2-k_2^2}
=0\, ,
\end{equation}
and we find
\begin{equation}
r_0=\frac{\zeta_1\zeta_2}{k_1k_2}\,,\qquad
r_1=\sqrt{\frac{(\zeta_1^2-k_1^2)(\zeta_2^2-k_1^2)}{k_1^2(k_2^2-k_1^2)}}\,,
\qquad
r_2=\sqrt{\frac{(\zeta_1^2-k_2^2)(\zeta_2^2-k_2^2)}{k_2^2(k_1^2-k_2^2)}}\,.
\end{equation}
The integrals of motion $I_0$ and $I_1$ lead to the equations
\begin{equation}
\zeta_1'=\pm\frac{(\zeta_1^2-k_1^2)(\zeta_1^2-k_2^2)}{\zeta_1^2-\zeta_2^2}\,,\qquad
\zeta_2'=\pm\frac{(\zeta_2^2-k_1^2)(\zeta_2^2-k_2^2)}{\zeta_1^2-\zeta_2^2}\,.
\label{torus-z'}
\end{equation}
The ratio of those two equations is then simple to integrate. If we
assume without loss of generality that $k_1<k_2$, then it turns out that
for our system we can take $k_1\leq \zeta_1\leq k_2\leq \zeta_2$ and in the
first equation in (\ref{torus-z'}) there should be the negative sign while
in the second the positive one.
The solution is given in terms of a constant $c$
\begin{equation}
k_1\,\hbox{arctanh}\,\frac{\zeta_1}{k_2}
-k_2\,\hbox{arccoth}\,\frac{\zeta_1}{k_1}
+k_1\,\hbox{arccoth}\,\frac{\zeta_2}{k_2}
-k_2\,\hbox{arccoth}\,\frac{\zeta_2}{k_1}
=c\,.
\end{equation}
or
\begin{equation}
\left(\frac{(\zeta_1-k_1)(\zeta_2+k_1)}{(\zeta_1+k_1)(\zeta_2-k_1)}\right)^{k_2}
\left(\frac{(k_2+\zeta_1)(\zeta_2-k_2)}{(k_2-\zeta_1)(\zeta_2+k_2)}\right)^{k_1}
=C\,.
\label{torus-z1z2-integral}
\end{equation}
Note that this solution is valid for any torus. The radii $\sin\theta/2$
and $\cos\theta/2$
are encoded in the asymptotic values of $r_1$ and $r_2$ whose ratio
should approach $\tan(\theta/2)$.
In terms of the $\zeta$'s, this corresponds to one of them ($\zeta_1$)
approaching the constant
$k_1k_2/\sqrt{k_1^2\sin^2(\theta/2)+k_2^2\cos^2(\theta/2)}$,
while $\zeta_2$ diverges.

So our solution has $\zeta_1$ starting at this constant near the boundary
of $AdS_5$ and decreasing to $k_1$, while $\zeta_2$ will start at infinity and
decrease to $k_2$. The constant $C$ in (\ref{torus-z1z2-integral}) is
\begin{equation}
\left(\frac{k_2-\sqrt{k_1^2\sin^2\frac{\theta}{2}+k_2^2\cos^2\frac{\theta}{2}}}
{k_2+\sqrt{k_1^2\sin^2\frac{\theta}{2}+k_2^2\cos^2\frac{\theta}{2}}}\right)^{k_2}
\left(\frac{\sqrt{k_1^2\sin^2\frac{\theta}{2}+k_2^2\cos^2\frac{\theta}{2}}+k_1}
{\sqrt{k_1^2\sin^2\frac{\theta}{2}+k_2^2\cos^2\frac{\theta}{2}}-k_1}\right)^{k_1}.
\end{equation}
Is is not easy to solve for
the $\zeta$'s (or $r$'s) in terms of $\sigma$, but that turns out not
to be necessary. The action can be evaluated without that
\begin{equation}
\begin{aligned}
\cS_{AdS_5}&=\frac{\sqrt\lambda}{2\pi}\int d\sigma\,d\tau\,
(-r_0'^2+r_1'^2+r_2'^2)
\\&
=\sqrt\lambda\int d\sigma\,\left(
\frac{\zeta_1'^2(\zeta_2^2-\zeta_1^2)}{(\zeta_1^2-k_1^2)(k_2^2-\zeta_1^2)}
+\frac{\zeta_2'^2(\zeta_2^2-\zeta_1^2)}{(\zeta_2^2-k_1^2)(\zeta_2^2-k_2^2)}\right)
\\&
=-\sqrt\lambda\left(
\int_{\frac{k_1k_2}{\sqrt{k_1^2\sin^2\frac{\theta}{2}+k_2^2\cos^2\frac{\theta}{2}}}}^{k_1} d\zeta_1
+\int_{\infty}^{k_2} d\zeta_2\right)
\\&
\simeq-\sqrt{\lambda}\left(k_1+k_2-\frac{k_1k_2}{\sqrt{k_1^2\sin^2\frac{\theta}{2}+k_2^2\cos^2\frac{\theta}{2}}}
\right)
\,.
\end{aligned}
\end{equation}
In the last expression the divergence was removed.

Together with the $S^5$ part (\ref{torus-S5-action}) one gets the total action
\begin{equation}
\cS=\left(-2k_1
\pm\sqrt{k_1^2\sin^2\frac{\theta}{2}+k_2^2\cos^2\frac{\theta}{2}}
+(1\mp1)\frac{k_1k_2}
{\sqrt{k_1^2\sin^2\frac{\theta}{2}+k_2^2\cos^2\frac{\theta}{2}}}
\right)\sqrt{\lambda}\,.
\label{torus-total-action}
\end{equation}


\section{Almost complex structure for $S^2$ and $S^6$}
\label{S6app}

In this appendix we provide an alternative, more geometrical
understanding of the origin of the almost complex structure ${\cal
J}$. The main clue comes from observing that our string solutions
satisfy
\be
x^2+z^2=1\,,
\label{emb}
\ee
and therefore reside inside an
$AdS_4\times S^2$ subspace of $AdS_5\times S^5$ . It is then
natural to look for an almost complex structure on this subspace.
To understand how this observation can help we note that we could
rewrite equation (\ref{emb}) as $x^\mu x^\mu+z^4 y^i y^i=1$, which,
up to a $z$ factor, is analogous to the equation of a 6-sphere
embedded in $(x^\mu,y^i)$. It will be therefore insightful to
review how we can construct an almost complex structure on $S^6$
in $\mathbb{R}^7$.  To understand how to proceed let us begin with
a simpler case, which is the construction of an almost complex
structure on $S^2$. This structure is by definition a linear
endomorphism on the tangent space of the sphere which satisfies
\be J: TS^2\rightarrow TS^2,\qquad J^2=-1. \ee
In terms of the usual embedding of the sphere in $\mathbb{R}^3$
let us consider the following linear operator
\be\label{acss2}
J=\left(%
\begin{array}{ccc}
  0 & -x_3 & x_2 \\
  x_3 & 0 & -x_1 \\
  -x_2 & x_1 & 0 \\
\end{array}%
\right). \ee This $J$ defines an almost complex structure on
$S^2$. To see that we first observe that $J$ is a well defined map
on $TS^2$ because for any $\vec p=(p_1,p_2,p_3)$ in the tangent
space of $\vec x=(x_1,x_2,x_3)$ we have $J(\vec p)\cdot{\vec
x}=0$. This says that $J(\vec p)$ is orthogonal to $\vec x$ and
therefore $J$ maps tangent vectors into tangent vectors. For $J$
to be an almost complex structure it remains to prove that it
squares to minus the identity, and indeed \be
J^2 (p)=\left(%
\begin{array}{c}
  -p_1+x_1\, x\cdot p \\
  -p_2+x_2\, x\cdot p \\
  -p_3+x_3\, x\cdot p\\
\end{array}%
\right) =-\left(%
\begin{array}{c}
  p_1 \\
  p_2 \\
  p_3\\
\end{array}%
\right).
\label{S2-J^2}
\ee
 Note that the action of $J$ on $\vec p$ can be simply
thought of as the cross product $\vec x \times \vec p$ .

Let us try to extend this construction. It is a fact that the only
spheres which admit an almost complex structure are $S^2$
(in which case $J$ is also integrable) and
$S^6$.
\footnote{It is
widely believed, but not proved, that $S^6$ does not admit a
complex structure.}
 The construction of an almost complex structure for the
latter case can be carried over in analogy to what done for the unit
2-sphere if we work with the octonion algebra $\mathbb{O}$. An
octonion element
 can be written  as
\be
\label{oct}{\bf x}=x_0+x_1 {\bf e}_1 +x_2
{\bf e}_2+x_3 {\bf e}_3+x_4 {\bf e}_4+x_5 {\bf e}_5+x_6 {\bf
e}_6+x_7 {\bf e}_7
\ee
where the algebra generators satisfy
\be
{\bf e}_i^2=-1,\qquad {\bf e}_i {\bf e}_j=-{\bf e}_j\, {\bf e}_i.
\ee
We can think of $S^6$ as the hypersurface $|x|=1$ with $x\in
\rm{Im}\, \mathbb{O}$, the imaginary octonions being obtained by
setting $x_0=0$. We will see that $S^6$, when considered as the
set of unit norm imaginary octonions, inherits an almost complex
structure from the octonion multiplication \cite{calabi}.

If we want to construct an almost complex structure on $S^6$ using
the analogy with $S^2$ we need to define a cross product. Luckily
a cross product between two vectors, satisfying all the usual
assumptions, exists only in dimensions 3 and 7. The cross product
between two octonions ${\bf x}$ and ${\bf y}$ is defined as \be
{\bf x}\times {\bf y}=\half({\bf x}{\bf y}-{\bf y}{\bf x} )={\rm
Im}\left({\bf x}{\bf y}\right) \ee where ${\bf x}{\bf y}$ is the
non-commutative and non-associative octonion product. If we work
with imaginary octonions the cross product reduces to the ordinary
octonion multiplication. The claim is that an almost complex
structure $J$ can be constructed as \be J={x}\times {p}\,,\qquad
{x}\in S^6,\qquad { p}\in T_{x}S^6 \ee where ${
x}=(x_1,x_2,x_3,x_4,x_5,x_6,x_7)$ and ${
p}=(p_1,p_2,p_3,p_4,p_5,p_6,p_7)$ are thought of as imaginary
octonions. Using a particular choice\footnote{There is not a
universal choice for the octonion multiplication table. The one
used here has been chosen to highlight the similarities with the
almost complex structure $\cJ$ relevant to the discussion of the
string solutions.} for the multiplication table one gets the
following matrix \be J=\left(\begin{array}{ccccccc}
 0 &-x_7 & x_6 & x_5 & -x_4 & -x_3 & x_2 \\
 x_7 & 0 & -x_5 & x_6 & x_3 & -x_4 & -x_1 \\
 -x_6 & x_5 & 0 & x_7 & -x_2 & x_1 & -x_4 \\
 -x_5 & -x_6 & -x_7 & 0 & x_1 & x_2 & x_3 \\
 x_4 & -x_3 & x_2 & -x_1 & 0 & x_7 & -x_6 \\
 x_3 & x_4 & -x_1 & -x_2 & -x_7 & 0 & x_5 \\
 -x_2 & x_1 & x_4 & -x_3 & x_6 & -x_5 & 0
\end{array}\right)
\ee
In complete analogy to the $S^2$ case we can show that $J$
defines linear endomorphism on the tangent space and that
$J^2({p})= -{p}$ for any tangent vector ${p}$. This proves we have
constructed an almost complex structure on the unit 6-sphere.
Using a notation similar to (\ref{notnot}) we
can write the matrix $J^{i}_{\  j}$ as
\be
\label{notation}
J^{i}_{\  j}=J^{i}_{\  j;\,k}\,x^k. \ee

Note that up to $z$ factors, $J$ coincides with the almost complex
structure ${\cal J}$ associated to the Wilson loops, see
(\ref{acsx4}), after the relabeling $ x_5\rightarrow
-y_1,\,x_6\rightarrow -y_2,\,x_7\rightarrow -y_3$.
The corresponding fundamental two-form reads%
\footnote{$dx_{\mu\nu}=dx_\mu\wedge dx_\nu$ and
$dx_{\mu\nu\rho}=dx_\mu\wedge dx_\nu\wedge dx_\rho$.}
\bea
\label{2form}
&& J=\half\,J_{MN}\,dx^M\wedge dx^N\nonumber\\
&&=x_1(dx_{72}+dx_{36}+dx_{45})+x_2(dx_{17}+dx_{53}+dx_{46})+x_3(dx_{61}+dx_{25}+dx_{47})\nonumber\\
&&+x_4(dx_{51}+dx_{62}+dx_{73})+x_5(dx_{14}+dx_{32}+dx_{67})+x_6(dx_{13}+dx_{24}+dx_{75})\nonumber\\
&&+x_7(dx_{21}+dx_{34}+dx_{56}).
\eea
Note that, as was the case
for ${\cal J}$, this two-form is not closed but rather we have
\be
\label{associative}
dJ=3(dx_{172}+dx_{136}+dx_{145}+dx_{325}+dx_{246}+dx_{347}+dx_{567}).
\ee
This form is the associative three-form $\phi$ preserved by
the $G_2$ group. The explanation for the appearance of $\phi$ in
this context is that $G_2\subset SO(7)$ is the automorphism group
of the octonions.\footnote{Also note that $S^6=G_2/SU(3).$} The
reason for which $dJ\neq 0$ is the well known fact that $S^6$ is
not K\"ahler.


\section{2-dimensional YM in the WML $\xi=-1$ gauge}
\label{WML-appendix}

In this appendix we present an explicit computation in the $\xi=-1$
generalized Feynman gauge with WML prescription for the
two-dimensional near-flat limit discussed in Section \ref{near-flat-subsection}.
Since we know that the non-interacting graphs from our Wilson loops
in four dimensions in the Feynman gauge agree with the 2-dimensional
propagators in this gauge, we turned to the first interacting graphs,
which appear at order $\lambda^2$.

While we were not able to find agreement between the interacting graphs
in four dimensions and two dimensions for a general loop, we present
the calculation here nonetheless in the hope that it would aid in
future explorations of the subject. To get some concrete results
we focused on the
one case where the interacting graphs were calculated in four
dimensions---the circular loop. In this special example
the two propagators in the $\xi=-1$ gauge sum up to
the (single) propagator in the light-cone gauge, hence the ladder diagrams
in the two gauges are equal. In the light-cone gauge there are
no interactions and therefore in our gauge the interaction graphs
for the circle should all cancel, which we indeed verify.

We start by deriving the Feynman rules in this generalized gauge in the near-flat limit. The Euclidean action reads
\bea
L=\frac{1}{g_{2d}^2}\left[\frac{1}{4}\left(F_{rs}^a\right)^2+\frac{1}{2\xi}\left(\partial_r A^{a,r}\right)^2+\partial_r b^a\left(D^r c\right)^a\right]\, ,
\eea
where $r,s=1,2$ and
\be
F^a_{rs}=\partial_{[r}A^a_{s]}+f^{abc}A^b_r\,A^c_s\, , \qquad \left(D_r c\right)^a=\partial_r c^a+f^{abc}A^b_r\, c^c\, .
\ee
Choosing the gauge $\xi=-1$ and using the light-cone coordinates $x^\pm=\frac{1}{2}(x^1\pm i x^2)$ (so that the metric is $g_{+-}=2$) the action becomes
\bea
L&=&\frac{1}{g_{2d}^2}\bigg[-\frac{1}{4}(\partial_+A_-^a)^2-\frac{1}{4}(\partial_-A_+^a)^2-b^a\partial_+\partial_- c^a
\cr
&&\hskip .8cm + \frac{1}{4}f^{abc}(\partial_+A^a_- -\partial_- A^a_+)A^b_-A^c_+ -\frac{1}{8}f^{abc}f^{ade}A^b_+A^c_-A^d_+A^e_-
\cr
&&\hskip .8cm +\frac{1}{2}f^{abc}(\partial_+ b^a) A_-^b c^c+\frac{1}{2}f^{abc}(\partial_- b^a) A_+^b c^c\bigg]\, .
\eea
The propagators for the gauge fields in the WML prescription are then
\bea
&& \Delta^{ab}_{++}(x,y)\equiv \delta^{ab}\Delta_{++}(x,y)=\delta^{ab}\frac{g_{2d}^2}{2\pi}\frac{x^--y^-}{x^+-y^+}\, , \cr && \Delta^{ab}_{--}(x,y)\equiv  \delta^{ab}\Delta_{--}(x,y)=\delta^{ab}\frac{g_{2d}^2}{2\pi}\frac{x^+-y^+}{x^--y^-}\, ,
\eea
where the normalization is fixed by requesting
\be
\frac{1}{2g_{2d}^2}\partial^2_{x^-} \Delta_{++}(x,y)=\delta^2(x-y)\, ,\label{norma}
\ee
and similarly for $\Delta_{--}$.
For the ghosts one has
\be
\Delta_{gh}^{ab}(x,y)\equiv \delta^{ab}\Delta_{gh}(x,y)=-\delta^{ab}\frac{g_{2d}^2}{4\pi}\log (x-y)^2\, .
\ee
The vertices can be easily read off from the action.

\subsection{Three-point graphs}

We now write down the interacting graphs starting with the ones with an internal 3-vertex. On the loop there can be two $A_+$'s and one $A_-$ or two $A_-$'s and one $A_+$. These two cases are one the complex conjugate of the other so it is sufficient to compute only one of them, say the first one, which we denote $\Sigma^{(3)}_{++-}$. Expanding the action $e^{-S}$ to first order and the Wilson loop to third order, one obtains after performing all the Wick contractions
\bea
&& \Sigma^{(3)}_{++-}=-\frac{i^3}{3!N}\frac{1}{4g_{2d}^2}\frac{iN(N^2-1)}{4}\oint d\tau_1 d\tau_2 d\tau_3\, \varepsilon(\tau_1\tau_2\tau_3)\int d^2y\,\times \cr
&& \hskip 1cm \times \bigg\{\dot x^+_1 \dot x^+_2 \dot x^-_3\Delta_{--}(y,x_3)\Big[\Delta_{++}(y,x_2)\partial_{y^-}\Delta_{++}(y,x_1)-\Delta_{++}(y,x_1)\partial_{y^-}\Delta_{++}(y,x_2)\Big]\cr && \hskip 3cm -(1\leftrightarrow 3)-(2\leftrightarrow 3)\bigg\}\, .\label{mercedes}
\eea
Here we have used that $\Tr(T^aT^bT^c)f^{abc}=\frac{i}{4}N(N^2-1)$ and the symbol $\varepsilon(\tau_1\tau_2\tau_3)$ enforces the path ordering through the antisymmetrization of $\tau_1$, $\tau_2$, and $\tau_3$.

We now proceed with the integration over $y$. The first term in curly brackets can be explicitly written (up to the $\dot x^+_1 \dot x^+_2 \dot x^-_3$ structure and the constant factors in the propagators which we do not include) as
\bea
&& \int d^2 y\,
\frac{(y^+-x^+_3)(x^-_1-x^-_2)}{(y^--x^-_3)(y^+-x^+_1)(y^+-x^+_2)}=\cr
&& \hskip 1cm = \frac{x^-_1-x^-_2}{x^+_1-x^+_2}\int
d^2y\left(\frac{x^+_1-x^+_3}{(y^--x^-_3)(y^+-x^+_1)}-\frac{x^+_2-x^+_3}{(y^--x^-_3)(y^+-x^+_2)}\right)\,
. \label{inty} \eea It is convenient to parametrize the position of
the vertex as $y^\pm\equiv \frac{\rho}{2}e^{\pm i\phi}$.  The
integrals in $\phi$ are of the type \be
\int_0^{2\pi}\frac{d\phi}{\left(e^{-i\phi}-a\right)\left(e^{i\phi}-b\right)}=\frac{2\pi}{ab-1}\left[\vartheta(|a|-1)-\vartheta(1-|b|)\right]\,
, \label{phi-int}
\ee
where $a,b\in \mathbb{C}$ and $\vartheta$ is
the step function. This identity can be easily proven starting from
\be
\int_0^{2\pi}\frac{d\phi}{e^{i\phi}-a}=-\frac{2\pi}{a}\vartheta(|a|-1)\,
. \ee
After integrating over $\phi$, equation (\ref{inty}) becomes
\bea
&& 8\pi\frac{(x^-_1-x^-_2)(x^+_1-x^+_3)}{x^+_1-x^+_2}\int_0^R
\frac{\rho\, d\rho}{4 x^-_3
x^+_1-\rho^2}\left[\vartheta(r(\tau_3)-\rho)-\vartheta(\rho-r(\tau_1))\right]-(1\leftrightarrow
2)\, , \cr &&
\eea
where we have introduced an IR cutoff $R$ and
parametrized $x^\pm_i\equiv \frac{r(\tau_i)}{2}e^{\pm i \tau_i}$. For convenience,
we will use in the following the shorthand notation $r_i \equiv r(\tau_i)$.
The integral above can be easily performed and yields \be
4\pi\frac{(x^-_1-x^-_2)(x^+_1-x^+_3)}{x^+_1-x^+_2}
\log\left( \frac{R^2 - 4 x^-_3 x^+_1}{r_1^2+r_3^2-2 r_1 r_3 \cos\tau_{13}}\right)-(1\leftrightarrow 2)\,
,\label{intyr} \ee
where we have also introduced the notation $\tau_{ij}\equiv \tau_i-\tau_j$. Including all the prefactors in equation (\ref{mercedes}) and
summing over the permutations yields the final result
\bea
&&\Sigma^{(3)}_{++-}=-\frac{\lambda^2}{3! 32 \pi^2}\oint d\tau_1 d\tau_2 d\tau_3\, \varepsilon(\tau_1\tau_2\tau_3)
\bigg[\dot x^+_1 \dot x^+_2 \dot x^-_3 \frac{x^-_1-x^-_2}{x^+_1-x^+_2}\Big( (x^+_1-x^+_2) \log R^2 + \cr
&& + (x^+_3-x^+_1) \log (x_3-x_1)^2 + (x^+_2-x^+_3) \log (x_2-x_3)^2 \Big)
-(1\leftrightarrow 3)-(2\leftrightarrow 3) \bigg], \cr &&
\eea
where we have expanded at large $R$ and neglected terms of order $1/R^2$. Adding the complex conjugate of this expression gives the total contribution of the 3-vertex graphs for a general curve.

We can now specialize to the case of the circular loop $x^\pm =
\frac{1}{2}e^{\pm i\tau}$ (for simplicity we take a circle of unit
radius, one could reinsert an arbitrary radius at the end by
dimensional analysis). In this case, the above expression yields
\bea && \Sigma^{(3)}_{++-}=-\frac{\lambda^2}{3! 256 \pi^2}\oint
d\tau_1d\tau_2 d\tau_3\,
\varepsilon(\tau_1\tau_2\tau_3)\Big[(\sin\tau_{21}+\sin\tau_{32}+\sin\tau_{13})
\log R^2+\cr && \hskip 1cm +
\sin\tau_{12}\log(2-2\cos\tau_{12})+\sin\tau_{23}\log(2-2\cos\tau_{23})+
\sin\tau_{31}\log(2-2\cos\tau_{13})\Big]\, ,\cr &&
\label{circleintyr} \eea Since the expression in square brackets
is totally antisymmetric in $\tau_1,\tau_2$, and $\tau_3$, one can
choose a fixed ordering of the $\tau$'s, say
$\tau_1\ge\tau_2\ge\tau_3$, and multiply by $3!$. The finite terms
not containing the $\log R^2$ integrate to zero and the final
result is $\Sigma^{(3)}_{++-}=\frac{\lambda^2}{32}\log R$. The
total contribution of the three-point interaction graphs in the
case of the circle is then \be
\Sigma^{(3)}=\Sigma^{(3)}_{++-}+\Sigma^{(3)}_{--+}=\frac{\lambda^2}{16}\log
R\, . \label{Sigma3circle} \ee

\subsection{Self-energy graphs}

We now compute the gluon self-energy graphs. We need to consider the 1-loop corrections to the $\Sigma^{(2)}_{++}$  and  $\Sigma^{(2)}_{+-}$ graphs and their complex conjugates. These graphs receive contributions from both gauge fields and ghosts running in the loop and are obtained by expanding the Wilson loop to quadratic order in the gauge fields.

We start with the $\Sigma^{(2)}_{++}$ graph. The ghost contribution reads
\bea
&&
\Sigma^{(2)}_{++}(ghost)=\frac{1}{2}\frac{i^2}{N}\left(\frac{1}{2g_{2d}^2}\right)^2\left(\frac{g_{2d}^2}{4\pi}\right)^2\left(\frac{g_{2d}^2}{2\pi}\right)^2\frac{-N(N^2-1)}{2}\times \cr &&
\hskip 1cm \times \int_{\tau_1\ge\tau_2}d \tau_1 d\tau_2\int d^2 y\, d^2 w\,
\frac{\dot x^+_1\dot x^+_2}{(y^--w^-)^2}\left\{\frac{(y^--x^-_1)(w^--x^-_2)}{(y^+-x^+_1)(w^+-x^+_2)}+(1\leftrightarrow 2)\right\}\, ,\label{gh++}\cr &&
\eea
where the first factor of $1/2$ comes from the Taylor expansion of $e^{-S}$.
The gauge field running in the loop contributes with three graphs: One graph with a 4-vertex and two graphs with two 3-vertices. In the first one of these two graphs with 3-vertices the propagators in the loop are a $\Delta_{++}$ and a $\Delta_{--}$, whereas in the second one they are two $\Delta_{--}$'s.

We find that the seagull graph is given by the following expression
\bea
&& \Sigma^{(2)}_{++}(seagull)=-\frac{i^2}{N}\left(-\frac{1}{8 g_{2d}^2}\right)\left(\frac{g_{2d}^2}{2\pi}\right)^3N(N^2-1)\times \cr
&& \hskip 1cm \times\oint_{\tau_1\ge \tau_2}d\tau_1 d\tau_2\int d^2y\,  \dot x^+_1 \dot x^+_2\frac{(y^+-y^+)(y^--x^-_1)(y^--x^-_2)}{(y^--y^-)(y^+-x^+_1)(y^+-x^+_2)}\, ,\label{seagull}
\eea
where we used the formal expression $(y^+-y^+)/(y^--y^-)$ to indicate the propagator in the limit of coincident points.

The graph with internal $\Delta_{++}$ and  $\Delta_{--}$ propagators reads
\bea
&&
\frac{1}{2}\frac{i^2}{N}\left(\frac{1}{4g_{2d}^2}\right)^2\left(\frac{g_{2d}^2}{2\pi}\right)^4\frac{N(N^2-1)}{2}\int_{\tau_1\ge\tau_2}d \tau_1 d\tau_2\int d^2 y\, d^2 w\, \dot x^+_1\dot x^+_2
\times \cr &&
\hskip .6cm \times
\bigg\{
 \frac{x^-_1-x^-_2}{y^--w^-}\left(\frac{1}{(y^+-x^+_1)(w^+-x^+_2)}-(1\leftrightarrow 2)\right)+
 \cr && \hskip 2.5cm {}+
\frac{y^+-w^+}{y^--w^-}\left(\frac{(y^--x^-_1)(w^--x^-_2)}{(y^+-x^+_1)(w^+-x^+_2)}+(1\leftrightarrow 2)\right)\partial_{y^-}\partial_{w^-}\left(\frac{y^--w^-}{y^+-w^+}\right)\bigg\}\, .\label{A++}\cr &&
\eea
The second graph with two $\Delta_{--}$ propagators gives a term which exactly cancels the ghost contribution equation (\ref{gh++}) and another term which is equal to the last term in equation (\ref{A++}) except that the factor
\be
\partial_{y^-}\partial_{w^-}\left(\frac{y^--w^-}{y^+-w^+}\right)
\ee
is replaced by its complex conjugate. Let us write these two terms more explicitly
\be
\partial_{y^-}\partial_{w^-}\left(\frac{y^--w^-}{y^+-w^+}\right)+c.c.=-\partial^2_{y^-}\left(\frac{y^--w^-}{y^+-w^+}\right)+c.c.=-8\pi \delta^2(y-w)\, ,
\ee
where we have used equation (\ref{norma}) and its complex conjugate. This term containing the $\delta$ function cancels then the seagull contribution (\ref{seagull}).

Similarly for $\Sigma^{(2)}_{+-}$  one finds that the ghost contribution is given by
\bea
&&
\Sigma^{(2)}_{+-}(ghost)=\frac{i^2}{N}\left(\frac{1}{2g_{2d}^2}\right)^2\left(\frac{g_{2d}^2}{4\pi}\right)^2\left(\frac{g_{2d}^2}{2\pi}\right)^2\frac{-N(N^2-1)}{2}\times \cr &&
\hskip 1cm \times \int_{\tau_1\ge\tau_2}d \tau_1 d\tau_2\int d^2 y\, d^2 w\,
\frac{\dot x^+_1\dot x^-_2(y^--x^-_1)(w^+-x^+_2)}{(y^+-w^+)(y^--w^-)(y^+-x^+_1)(w^--x^-_2)}\, .\label{gh+-}\cr &&
\eea
As for the gluons running in the loop, now only the graph with two 3-vertices,  one $\Delta_{++}$ and a $\Delta_{--}$
contributes (there is no seagull graph contributing to $\Sigma^{(2)}_{+-}$). This is given by
\bea
&&
\Sigma^{(2)}_{+-}(gluon)=\frac{i^2}{N}\left(\frac{1}{4g_{2d}^2}\right)^2\left(\frac{g_{2d}^2}{2\pi}\right)^4\frac{N(N^2-1)}{2}\times \cr &&
\hskip 1cm \times \int_{\tau_1\ge\tau_2}d \tau_1 d\tau_2\int d^2 y\, d^2 w
\frac{\dot x^+_1\dot x^-_2(y^+-x^+_2)(w^--x^-_1)}{(y^+-w^+)(y^--w^-)(y^+-x^+_1)(w^--x^-_2)}\, .\label{A+-}\cr &&
\eea
Putting together all the pieces one obtains
\bea
&&
\Sigma^{(2)}_{++}+\Sigma^{(2)}_{+-}=\frac{i^2}{N}\left(\frac{1}{4g_{2d}^2}\right)^2\left(\frac{g_{2d}^2}{2\pi}\right)^4\frac{N(N^2-1)}{2} \int_{\tau_1\ge\tau_2}d \tau_1 d\tau_2\int d^2 y\, d^2 w\,  \times \cr &&
\hskip .8cm \times
\bigg\{
\frac{\dot x^+_1\dot
x^+_2}{2}\frac{x^-_1-x^-_2}{y^--w^-}\left(\frac{1}{(y^+-x^+_1)(w^+-x^+_2)}-(1\leftrightarrow
2)\right)+ \cr && \hskip .8cm +\dot x^+_1\dot
x^-_2\left(\frac{y^--x^-_1}{(y^--w^-)(y^+-x^+_1)(w^--x^-_2)}-\frac{y^+-x^+_2}{(y^+-w^+)(y^+-x^+_1)(w^--x^-_2)}\right)
\bigg\}\, .\label{Sigma2tot}\cr && \eea
Adding the complex conjugate
of this expression gives the total contribution of the self-energy
graphs.

We start by evaluating the first term in equation
(\ref{Sigma2tot}), corresponding to $\Sigma^{(2)}_{++}$.
As before, we use polar coordinates for the
integration over the internal vertices, by defining
$y^{\pm}=\frac{\rho}{2} e^{\pm i\phi}$ and $w^{\pm}=\frac{\xi}{2}
e^{\pm i\psi}$. The generic loop is parameterized as $x_i^{\pm} = \frac{r_i}{2} e^{\pm i \tau_i}$, where
$r_i \equiv r(\tau_i)$.
Computing first the integrals over $\phi$ and $\psi$ with the help of (\ref{phi-int}), we get
\bea
&&\int d^2 y\, d^2 w\, \frac{1}{(y^--w^-)(y^+-x^+_1)(w^+-x^+_2)}= \cr
&& = 16 \pi^2 \int_0^R \int_0^R  \frac{\rho \xi\, d\rho\, d\xi}{x_1^+ \xi^2 - x_2^+ \rho^2}
\Big[ \vartheta(\rho-\xi) -\vartheta(\xi-r_2) \Big] \Big[ \vartheta(\xi^2-r_2 \rho)-\vartheta(\rho-r_1)\Big]\cr
&& = -16 \pi^2 \bigg[\int_{r_2}^{R} d\xi \int_0^{\xi} d\rho + \int_{r_1}^{R} d\rho \int_0^{\rho} d\xi
- \int_{r_1}^{R} d\rho \int_{r_2}^{R} d\xi  \bigg] \frac{\rho \xi}{x_1^+ \xi^2 - x_2^+ \rho^2}\, ,
\label{int++}
\eea
where $R$ is the large distance cutoff. The remaining integrals can be easily performed. Expanding at large $R$, one finds
that quadratic divergences cancel and the final result for the integral in equation (\ref{int++}) is
\bea
16 \pi^2 (x_1^- - x_2^-) \bigg( \log R^2 + 1 - \log (r_1^2+r_2^2 -2 r_1 r_2 \cos\tau_{12})\bigg)
 + { \cal O} \left(\frac{1}{R^2}\right)\,.
\eea
Including all the prefactors in equation (\ref{Sigma2tot}) as well as the contribution obtained by exchanging $x_1$ and $x_2$,
we thus obtain
\bea
\Sigma^{(2)}_{++} = -\frac{\lambda^2}{32 \pi^2} \int_{\tau_1 \ge \tau_2} d\tau_1 d\tau_2\,
\dot x^+_1 \dot x^+_2 (x^-_1-x^-_2)^2 \bigg( \log R^2 + 1 - \log (x_1-x_2)^2 \bigg)\,.
\label{Sigma++}
\eea

We integrate now the second term in equation (\ref{Sigma2tot}), which corresponds to
$\Sigma^{(2)}_{+-}$. We proceed as before by first integrating over $\phi$ and $\psi$ using
identities analogous to (\ref{phi-int}), and then we integrate over the radial directions
$\rho$ and $\xi$ with an IR cutoff $R$. After expanding at large $R$ the final result
for the integrals on the internal vertices is
\bea
&& \int d^2 y\, d^2 w\,
\left(\frac{y^--x^-_1}{(y^--w^-)(y^+-x^+_1)(w^--x^-_2)}-\frac{y^+-x^+_2}{(y^+-w^+)(y^+-x^+_1)(w^--x^-_2)}\right)=\cr
&& = 8\pi^2 \bigg[ R^2 -(r_1^2 + r_2^2 -2 r_1 r_2 \cos\tau_{12}) \log R^2 + \cr
&& ~~~~~~
+ (r_1^2 + r_2^2 -2 r_1 r_2 \cos\tau_{12}) \log (r_1^2 + r_2^2 -2 r_1 r_2 \cos\tau_{12})
+ 6 x_1^- x_2^+ -r_1^2 -r_2^2 \bigg]\,.\cr &&
\eea
The quadratic divergence appearing here cancels out for a general curve once we sum the contribution of the complex conjugate graph $\Sigma^{(2)}_{-+}$. Indeed, the $R^2$ term is then proportional to
\bea
\int_{\tau_1 \ge \tau_2} d\tau_1 d\tau_2\, \left( \dot x^+_1\dot x^-_2 + c.c. \right) =
\frac{1}{2} \int_{\tau_1 \ge \tau_2} d\tau_1 d\tau_2\, \dot x_1 \cdot \dot x_2 = 0\,.
\eea
Including the prefactors in (\ref{Sigma2tot}), we thus get
\be
\begin{aligned}
&\Sigma^{(2)}_{+-} + \Sigma^{(2)}_{-+} = -\frac{\lambda^2}{64 \pi^2}
\int_{\tau_1 \ge \tau_2} d\tau_1 d\tau_2\,
\dot x^+_1\dot x^-_2 \Big[-(x_1-x_2)^2 \log R^2 +
\\&\hskip 3cm {}+
(x_1-x_2)^2 \log (x_1-x_2)^2
+\, 6 x_1^- x_2^+ -r_1^2 -r_2^2 \Big] + c.c.
\label{Sigma+-}
\end{aligned}
\ee

We now specialize to the circle $x_i^{\pm} = \frac{1}{2} e^{\pm i\tau_i}$. From
 (\ref{Sigma++}) we readily obtain
\bea
\Sigma^{(2)}_{++} + \Sigma^{(2)}_{--} &=& -\frac{\lambda^2}{128 \pi^2}
\int_0^{2\pi} d\tau_1 \int_0^{\tau_1} d\tau_2
(1-\cos\tau_{12}) \bigg( \log R^2 + 1 - \log (2-2\cos\tau_{12}) \bigg)\cr
&=& -\frac{\lambda^2}{32} \log R \, ,
\eea
while (\ref{Sigma+-}) yields
\bea
&& \Sigma^{(2)}_{+-} + \Sigma^{(2)}_{-+} = -\frac{\lambda^2}{64 \pi^2}
\int_0^{2\pi} d\tau_1 \int_0^{\tau_1} d\tau_2
\bigg[\frac{3}{4} - \cos\tau_{12}+\\
&& \hskip 2.7cm {}+  \cos\tau_{12}(1-\cos\tau_{12})
\Big(-\log R^2 + \log (2-2\cos\tau_{12})\Big)
 \bigg] = -\frac{\lambda^2}{32} \log R \, .
\nonumber
\eea
Recalling the contribution of the 3-vertex (\ref{Sigma3circle}), we see that for the circle the sum of the interacting graphs at this order vanishes as expected
\bea
\Sigma^{(2)}+\Sigma^{(3)} = 0\,.
\eea


\end{document}